\documentclass{aa}
\usepackage[varg]{txfonts}
\pdfoutput=1
\usepackage{graphicx}
\usepackage{newtxtext,newtxmath}
\usepackage{xcolor}
\RequirePackage[colorlinks=true,linkcolor=blue,citecolor=blue,urlcolor=blue]{hyperref}

\begin{document} 
\title{An Atlas of Gas Motions in the TNG-Cluster Simulation: \\from Cluster Cores to the Outskirts}

\titlerunning{Gas Motions in TNG-Cluster}

\author{Mohammadreza Ayromlou
      \inst{1}\thanks{\email{ayromlou@uni-heidelberg.de}}
      \and Dylan Nelson\inst{1}
      \and Annalisa Pillepich\inst{2}
      \and Eric Rohr\inst{2}
      \and Nhut Truong\inst{3,4}
      \and \\Yuan Li\inst{5}
      \and Aurora Simionescu\inst{6,7,8}
      \and Katrin Lehle\inst{1}
      \and Wonki Lee\inst{9,10}
      }

\institute{Universit{\"a}t Heidelberg, Zentrum f{\"u}r Astronomie, ITA, Albert-Ueberle-Str. 2, 69120 Heidelberg, Germany \label{1}
\and Max-Planck-Institut f\"{u}r Astronomie, K\"{o}nigstuhl 17, 69117 Heidelberg, Germany \label{2}
\and NASA Goddard Space Flight Center, Greenbelt, MD 20771, USA \label{3}
\and Center for Space Sciences and Technology, University of Maryland, 1000 Hilltop Circle, Baltimore, MD 21250, USA \label{4}
\and Department of Physics, University of North Texas, Denton, TX, United States \label{5}
\and SRON Netherlands Institute for Space Research, Niels Bohrweg 4, NL-2333 CA Leiden, the Netherlands \label{6}
\and Leiden Observatory, Leiden University, P.O. Box 9513, NL-2300 RA Leiden, the Netherlands \label{7}
\and Kavli Institute for the Physics and Mathematics of the Universe (WPI), University of Tokyo, Kashiwa 277-8583, Japan \label{8}
\and Yonsei University, Department of Astronomy, Seoul, Republic of Korea \label{9}
\and Harvard-Smithsonian Center for Astrophysics, 60 Garden St., Cambridge, MA 02138, USA \label{10}}

\date{}
 
\abstract
{Galaxy clusters are unique laboratories for studying astrophysical processes and their impact on halo gas kinematics. Despite their importance, the full complexity of gas motion within and around these clusters remains poorly known. This paper is part of a series presenting first results from the new TNG-Cluster simulation, a suite comprising 352 high-mass galaxy clusters including the full cosmological context, mergers and accretion, baryonic processes and feedback, and magnetic fields. Studying the dynamics and coherence of gas flows, we find that gas motions in galaxy cluster cores and intermediate regions are largely balanced between inflows and outflows, exhibiting a Gaussian distribution centered at zero velocity. In the outskirts, even the net velocity distribution becomes asymmetric, featuring a double peak where the second peak reflects cosmic accretion. Across all cluster regions, the resulting net flow distribution reveals complex gas dynamics. These are strongly correlated with halo properties: at a given total cluster mass, unrelaxed, late-forming halos with less massive black holes and lower accretion rates exhibit a more dynamic behavior. Our analysis shows no clear relationship between line-of-sight and radial gas velocities, suggesting that line-of-sight velocity alone is insufficient to distinguish between inflowing and outflowing gas. Additional properties, such as temperature, can help break this degeneracy. A velocity structure function (VSF) analysis indicates more coherent gas motion in the outskirts and more disturbed kinematics towards halo centers. In all cluster regions, the VSF shows a slope close to the theoretical models of Kolmogorov ($\sim 1/3$), except within 50 kpc of the cluster centers, where the slope is significantly steeper. The outcome of TNG-Cluster broadly aligns with observations of the VSF of multiphase gas across different scales in galaxy clusters, ranging from $\sim 1\,\rm kpc$ to Megaparsec scales.}


\maketitle


\section{Introduction}

Currently, the $\Lambda$CDM model is the most observationally supported theoretical framework describing our Universe. Observations of the cosmic microwave background (CMB) infer matter perturbations in the early Universe. These are the seeds for the formation of observed large-scale structure \citep{Spergel2003First,komatsu2011seven,planck2015_xiii}. In the hierarchical structure formation paradigm, dark matter halos emerge at the peaks of initial density perturbations \citep{white1978core}. As these halos evolve, they grow through both mergers with other halos and smooth accretion of matter from their surroundings \citep{white1991galaxy,Lacey1994Merger}. Within these evolving potential wells, baryonic matter -- initially in the form of gas -- radiates energy, cools, and condenses at the centers of the halos, setting the stage for star formation and, eventually, the assembly of galaxies \citep[see][for a full discussion]{mo2010galaxy}.

The highest density peaks in the matter distribution are destined to become the most massive virialized structures in the Universe: galaxy clusters. These structures, composed of dark matter, gas, stars, super-massive black holes (SMBHs), and hundreds to thousands of individual galaxies, are laboratories for studying a wide range of astrophysical processes responsible for galaxy formation and evolution \citep[see][]{Kravtsov2012Formation}.

Unlike lower mass halos, the most-massive clusters are observed to have a baryon fraction consistent with the cosmic baryon fraction, $\Omega_{\rm b}/\Omega_{\rm m}\sim 0.16$ \citep{giodini2009stellar,chiu2018baryon,Eckert2021Feedback}. This is broadly true also according to hydrodynamical cosmological simulations, suggesting that clusters are closed systems. In particular, the `closure radius' of clusters, within which all baryons associated with dark matter are found, is equal or smaller than the virial radius \citep{Ayromlou2023Feedback}. However, these facts do not imply that galaxy clusters are static systems. In fact, observational and theoretical studies have shown that galaxy clusters are dynamically evolving structures \citep{Voit2005Tracing,Nagai2007Effects,Arnaud2010Universal}. The motions of material within clusters arise due to the interplay of several physical processes.

The gas in galaxy clusters, known as the intracluster medium (ICM), is heated to extreme temperatures by gravitational collapse, forming a hot, diffuse medium that plays a key role in their evolution \citep{Ghirardini2019Universal}. In some clusters, the ICM can cool and condense at the cluster center, leading to cooling flows that can trigger star formation and affect the evolution of the central galaxy \citep{Fabian1994Cooling,Petersen2003HighResolution}.

Major mergers between galaxy clusters are violent and transformative events. They can reshape the structure of the clusters, heat the ICM, and are key drivers of cluster evolution \citep{Sarazin2002Physics,Markevitch2007Shocks,Lokas2023Merging}. Additionally, smaller groups of galaxies can fall into larger clusters, contributing to the growth and evolution of the cluster over time \citep{Tully1987Nearby}. Galaxy clusters also continuously accrete individual galaxies, smooth gas and dark matter from their surroundings, which further contributes to their growth and evolution \citep{Kravtsov2012Formation}.

In addition, SMBH feedback from active galactic nuclei (AGN) is believed to play a crucial role in regulating the cooling of the ICM and preventing catastrophic cooling flows in cluster centers \citep{croton2006many}. AGN feedback is effective in quenching star formation in massive cluster galaxies, including the central or brightest, and maintaining the observed low star formation rates in these galaxies \citep{Kauffmann2003Host,McNamara2007Heating}. In doing so it drives strong, high-velocity jets or outflows into the surrounding medium \citep{Cicone2014Massive}.

All of these processes shape not only cluster galaxies, but also the gas within and in the outskirts of clusters. This gas constitutes the majority of the baryon budget of the halo, and its properties and dynamics are complex. In particular, the kinematics of the hot ICM gas is the focus of this paper.

Observationally, the kinematics of the ICM are challenging to study \citep{Simionescu2019Constraining}. In its pioneering study, \cite{Hitomi2018Atmospheric} probed the gas dynamics in the Perseus cluster, reporting measurements of the gas motions within the central 100 kpc of the cluster, using X-ray emission lines. They found that the gas velocity dispersion is very low ($\sim 100 \, \rm{km/s}$) and mostly uniform, except near the central AGN and the north-western ghost bubble ($\sim 200 \, \rm{km/s}$), which are regions where the AGN injects energy into the gas \citep[see also][for simulation work]{Mohapatra2022Velocity}. A velocity gradient also exists across the cluster core, consistent with sloshing of the gas caused by a past merger event \citep[see also][]{Hitomi2016quiescent}.

Beyond the velocity dispersion, the dynamics of the ICM as a function of scale can be quantified with the velocity structure function (VSF). For example, \cite{Ganguly2023nature} show that the Abell 1795 cluster exhibits a flattening of the VSF on small scales, consistent with Kolmogorov predictions, but this behavior is observed only in regions distant from the SMBH. This finding emphasizes the significant role that SMBHs may play in the small-scale kinematics of the ICM. \cite{Gatuzz2023Measuring} measure the VSFs of three galaxy clusters: these also generally exhibit power-law slopes out to $\sim$\,Mpc scales, indicative of turbulent motions. For Virgo, they observe a flattening in the VSF, which they associate with a turbulent driving scale of $\sim 10-20$\,kpc.

Studying colder gas phases, \cite{Li2020Direct} investigate the kinematics of multi-phase $\rm H\alpha$ filaments in the centers of three nearby galaxy clusters: Perseus, Abell 2597, and Virgo, using high-resolution optical data. On scales smaller than a few tens of kiloparsecs, the VSF reveals multi-scale turbulence. Notably, VSF features correlate with the sizes of jet-inflated bubbles, suggesting that AGN feedback drives turbulence in these clusters. In addition, they compare with X-ray measurements of line widths and surface brightness fluctuations from \cite{Hitomi2018Atmospheric}, finding broad agreement between the kinematics of cold filaments and the hot ICM.

In the realm of simulations, numerous studies have addressed galaxy clusters and the kinematics of their gas \citep[see, e.g.,][]{Yang2016How,Cui2018Three, Mohapatra2019Turbulence}. Utilizing idealized hydrodynamic simulations, \cite{Zuhone2016Simulating} examined a minor merger between a large cluster with a cool core and a smaller gasless subcluster, resulting in cold fronts within the cluster core. They generated synthetic X-ray spectra from these simulations, convolved with the Astro-H Soft X-ray Spectrometer (SXS) responses, to investigate the effects of projection and non-Gaussian line shapes on gas velocity measurements. With respect to major mergers, \cite{Biffi2022Velocity} studied a major merger of two galaxy clusters using hydrodynamical simulations with the AREPO code \citep{springel2010pur}. They explored the possibility of detecting multiple velocity components or velocity gradients by generating mock X-ray spectra \citep[see also][]{Vazza2017Turbulence}. Additionally, \cite{Ehlert2021Connecting} explored the impact of AGN jets on hot gas and magnetic fields in galaxy clusters. They found that AGN jets create bubbles of hot gas that rise buoyantly, disrupting cluster turbulence but only locally affecting the gas velocity field \citep[see also][]{Beckmann2022AGN}. Their analysis of the gas uplifted by the AGN jets revealed fast, coherent outflows with low velocity dispersion \citep[see also][]{Chen2019Jets}. They also noted that projected velocity distributions exhibit complex structures, complicating the interpretation of observations.

\cite{Wang2021NonKolmogorov} investigated the properties of turbulence in both hot and cold phases of the ICM in galaxy cluster simulations. They found that cold gas forms filaments that gravitate toward the cluster center and exhibit a VSF with a slope steeper than the 1/3 predicted by \cite{Kolmogorov1941Local}. In contrast, the hot gas displayed a VSF slope close to 1/2 in hydrodynamic scenarios. They posited that the turbulence in the hot phase could be driven by a combination of AGN jet stirring and filament motion, facilitated by magnetic fields.

In this work, we investigate the kinematics of the gas in galaxy clusters using the new TNG-Cluster simulation. Focusing on $z=0$, we cover a halo mass range of $14.3<\log_{10} (M_{\rm 200c}/{\rm M_{\odot}})<15.4$, exploring the gas from small kpc scales in the core of the cluster to large Mpc scales in the cluster outskirts.

This paper, together with a series of companion papers, collectively offer an initial overview of the TNG-Cluster simulation suite, its scientific scope, and its prospects for future research. \textcolor{blue}{Nelson et al. (submitted)} introduce the simulation framework and discuss the general properties of the ICM. \textcolor{blue}{Pillepich et al. (in preparation)} delve into both the global and spatially-resolved characteristics of the ICM, encompassing phenomena such as pressure waves and feedback-induced bubbles and perturbations. \textcolor{blue}{Rohr et al. (submitted)} examine the circumgalactic medium (CGM) surrounding cluster satellites, focusing on their observable signatures. \textcolor{blue}{Lehle et al. (submitted)} provide a complete analysis of the cool-core and non-cool-core populations within TNG-Cluster, linking them to cluster properties. \textcolor{blue}{Lee et al. (submitted)} identify systems undergoing mergers and recover radio relics, the diversity and features of which align with existing observations. Lastly and most relevant to this study, in \textcolor{blue}{Truong et al. (submitted)} we investigate the velocity dispersion of cores Perseus-like clusters, anticipated to be observed in the XRISM mission.

This paper is structured as follows. In Section \ref{sec: Methodology}, we describe the TNG-Cluster simulation and the methodology used in this paper. In Section \ref{sec: results}, we present our main results in 3D simulation space, including the initial findings on the kinematics of gas in and around galaxy clusters, as well as the coherence of gas motion from small to large scales, as revealed through the analysis of the velocity structure function (VSF) of gas. Section \ref{sec: obs_kinematics} is dedicated to the observable kinematics of the gas in galaxy clusters, where we compare our results with various observations. Finally, we summarize our findings and conclude our study in Section \ref{sec: summary}.


\section{Methodology}
\label{sec: Methodology}

\subsection{The TNG-Cluster simulation suite}
\label{subsubsec: TNG-Cluster}

The TNG-Cluster simulation\footnote{\url{www.tng-project.org/cluster}} is a suite of 352 zoom simulations of massive galaxy clusters with $\log_{10}(M_{\rm 200c}/{\rm M_{\odot}})\gtrsim 14.2$ chosen from a $\sim 1\,\rm Gpc$ parent box. These simulations are specifically designed to study the physics of galaxy clusters. This project extends the previous IllustrisTNG simulations \citep[TNG hereafter;][]{nelson18a,pillepich2018First,springel2018first,marinacci2018first,naiman2018first}, particularly TNG300, to the high-mass regime. The TNG-Cluster simulation adopts the same numerical methodology and physical galaxy formation model as the other TNG simulations. It also maintains the relatively high resolution of TNG300-1 with $m_{\rm gas} \sim 10^7$\,M$_\odot$ (see \textcolor{blue}{Nelson et al. submitted} for details).

The TNG model is implemented within the moving-mesh \textsc{AREPO} code \citep{springel2010pur} and solves the equations governing gravity and magnetohydrodynamics \citep{pakmor2011magnetohydrodynamics,pakmor2013simulations}. It is a comprehensive physical model for galaxy formation that includes radiative cooling of gas, star formation from cold gas, stellar evolution, and stellar feedback \citep{pillepich2018Simulating}. It also models SMBHs, including their seeding, merging, and feedback \citep{weinberger17}. 

The TNG simulation utilizes a real-time halo identification algorithm and inserts a supermassive black hole with a fixed mass into the halo as soon as the halo's mass crosses a predetermined value. The SMBH then grows by accreting matter and merging with other SMBHs. The accretion process is simulated using the Bondi accretion formula \citep[][]{Bondi1952spherically,Bondi1944mechanism}, which takes into account various factors like the mass of the SMBH, local gas density, temperature, and the SMBH's relative velocity to the surrounding gas. In the model, AGN feedback is the key factor that quenches star formation in massive galaxies such as those located at the cluster centers. The SMBH feedback can function in either thermal or kinetic modes, but not both simultaneously. In the thermal mode, energy is consistently added to the adjacent gas cells, elevating their temperature and affecting their properties. On the other hand, when the accretion rate drops—typically in massive galaxies—the feedback switches to the kinetic mode, injecting momentum and kinetic energy into neighboring gas cells. This can generate non-isotropic outflows, depending on the specific conditions of the nearby gas \citep{weinberger17, nelson2019First, Pillepich2021X-ray}. However, these outflows do not significantly alter the gas distribution in galaxy clusters on a large scale, unlike what we previously found in galaxy groups and Milky Way-like halos \citep{Ayromlou2023Feedback}. Magnetic fields in galaxy clusters are also taken into account in our simulations. Due to a mix of turbulent and small-scale dynamo processes, magnetic fields are self-consistently amplified from a nearly negligible primordial seed field in the initial conditions to their values across different cosmic times.

The TNG-Cluster simulation adopts the same $\Lambda$CDM cosmology as TNG, with cosmological parameters consistent with Planck observations of the cosmic microwave background \citep{planck2015_xiii}. The cosmological parameters used are $\Omega_{\text{m}} = 0.3089$, $\Omega_{\Lambda} = 0.6911$, $H_0 = 67.74 \, \text{km/s/Mpc}$, $n_{\text{s}} = 0.9667$, $\sigma_8 = 0.8159$, and $\Omega_{\text{b}} = 0.0486$.

We identify halos with the friends-of-friends (FoF) algorithm \citep{Davis1985TheEvolution} as linked groups of particles. Within these halos, substructures, termed subhalos, are identified using the \textsc{Subfind} algorithm \citep{springel2001populating}. Each halo contains one central subhalo, typically the most massive substructure within that halo, while the remaining subhalos are categorized as satellites.

\subsection{Analysis: radial profiles and 2D projections}
\label{subsec: radial_profiles_and_2D_histograms}

We measure the radial profiles of several physical quantities, from the halo center to $\sim 1.5 R_{\rm 200c}$. In doing so we consider all cells/particles in the simulation volume, and do not restrict to e.g. the FoF halo member cells/particles. For each halo, we compute the weighted mean of a given property $X$ at a fixed halocentric distance as $\langle X \rangle = \sum_i X_i \, w_i / \sum_i w_i$ where $w_i$ is the weight assigned to each gas cell, which we take as mass, density, or X-ray emissivity. In the latter case we adopt a simple estimator for bolometric X-ray luminosity \citep{Navarro1995Simulations}, as a function of gas density and temperature. This approach accounts for free-free bremsstrahlung emission from hot gas with $T > 10^6 \, \rm K$ which dominates in high-mass clusters.

When deriving line-of-sight observables in 2D, we project along the z-axis of the simulation box. We also integrate along the full simulation box in the line-of-sight direction, in order to incorporate clustering and local projection effects.

The physical properties we consider are defined as:

\begin{itemize}
    \item \textbf{Radial Velocity:} The radial component of the velocity vector of each gas cell in the rest frame of the halo. Positive and negative values indicate outflows and inflows, respectively.
    \item \textbf{Line-of-Sight Velocity:} The z-axis component of the velocity vector of each gas cell in the rest frame of the halo. Positive and negative values indicate redshift and blueshift, respectively.
    \item \textbf{Velocity Dispersion:} The standard deviation of the radial or line-of-sight velocity for all gas cells within a given bin.
    \item \textbf{Mass Inflow and Outflow Rates:} Measured from the instantaneous gas velocities as $\dot{M} = \sum_i m_i \, v_{\rm i} / \Delta r_i$ where $m_i$ is the mass of each gas cell, $v_{\rm i}$ is its radial or line of sight velocity, and $\Delta r$ is the thickness of the radial bin. Positive and negative values correspond to outflows and inflows, respectively.
\end{itemize}

\subsection{Classification of galaxy clusters}
\label{subsec: halo_classification}

We categorize halos based on several key attributes: halo mass (or radius), halo radial zone, relaxedness (state of relaxation), formation redshift, SMBH mass, and SMBH accretion rate. We define each of these as follows:

\begin{itemize}
    \item \textbf{Halo mass:} We characterize halo size with $R_{\rm 200c}$ (or $R_{\rm 500c}$), the halocentric radius enclosing a total matter density of 200 (or 500) times the critical density of the Universe. The mass within this radius, $M_{\rm 200c}$ (or $M_{\rm 500c}$), is the halo mass.
    \item \textbf{Halo radial zone:} We sometimes partition halos into three distinct radial zones: the core, the intermediate region, and the outskirts. The core is defined as the region within $0.2 , R_{\rm 200c}$ of the halo center. The intermediate region is centered around $\sim 0.5 , R_{\rm 200c}$, and the outskirts are defined as the volume at $R_{\rm 200c}$. Each of these zones has a thickness of $0.2 \, R_{\rm 200c}$.
    \item \textbf{Relaxedness:} We combine two criteria, and classify a halo as relaxed if and only if it satisfies both criteria.
    \begin{enumerate}
        \item \textit{Distance criterion:} The separation between the halo center of mass and its most bound particle. A halo is labelled relaxed if this distance is less than $0.1 \, R_{\rm 200c}$; otherwise, it is unrelaxed.
        \item \textit{Central subhalo mass criterion:} The mass fraction of the central subhalo relative to the total halo mass. A halo is considered relaxed if this ratio exceeds $0.85$; otherwise, it is unrelaxed.
    \end{enumerate}
    \item \textbf{Formation redshift:} The redshift at which the halo reaches half of its present-day mass. Halos are categorized as early- or late-forming, with the threshold at $z\sim0.5$, corresponding to the median formation redshift of TNG-Cluster halos.
    \item \textbf{SMBH mass to halo mass ratio:} The mass of the central SMBH relative to the total halo mass. We split halos into two subsets at the median mass ratio for a given mass bin.
    \item \textbf{SMBH accretion rate:} The instantaneous accretion rate of the central SMBH. Halos are divided into two subsets at the median accretion rate of the SMBHs of all halos at fixed halo mass bin.
\end{itemize}

\subsection{Velocity structure function}
\label{subsec: methods_vsf}

The velocity structure function (VSF) is a robust two-point statistical measure to quantify the coherence and turbulence of gas motions as a function of separation distance, denoted as $\Delta r$ \citep{Kolmogorov1941Local}. The VSF is calculated as:
\begin{equation}
    \label{eq: VSF}
    S_1(\Delta r) = \frac{\sum_{i,j} w_i w_j \left| \vec{v}_i - \vec{v}_j \right|^p}{\sum_{i,j} w_i w_j},
\end{equation}
where $p$ represents the order of the VSF, and $p=1$ for the first order VSF. For the 3D (theoretical) VSF, $\vec{v}$ is the velocity vector gas. For the 1D (mock observational) VSF, $\vec{v}$ is the line-of-sight velocity. To compute the VSF, we initially bin the space in either 3D or 2D, depending on whether we are dealing with theoretical or observational VSF measurements. We then calculate the average velocity for each bin ($\vec{v}$ in the formula above). When computing the average velocity of each bin, we employ mass, density, or X-ray bolometric luminosity weighting. We have verified that altering the bin size by a factor of 5 does not significantly affect our VSF results. The sole criterion for selecting the bin size is that it should be smaller than the minimum separation scale of interest. To ensure maximum consistency and caution, we always choose our bins to be at least four times smaller than the minimum separation scale.

In addition to the bins from which we measure the VSF, the VSF itself can be weighted. In the equation above, this is denoted by $w_i$ and $w_j$ as weights of the $i$-th and $j$-th gas cells or bins, respectively. Throughout this paper, we employ $w_i = w_j = 1$, except for the case of the VSF of the line-of-sight velocity in comparison with observations, where we explore various choices.


\section{Gas kinematics according to TNG-Cluster}
\label{sec: results}

\begin{figure*}
    \centering
    \includegraphics[width=0.7\textwidth]{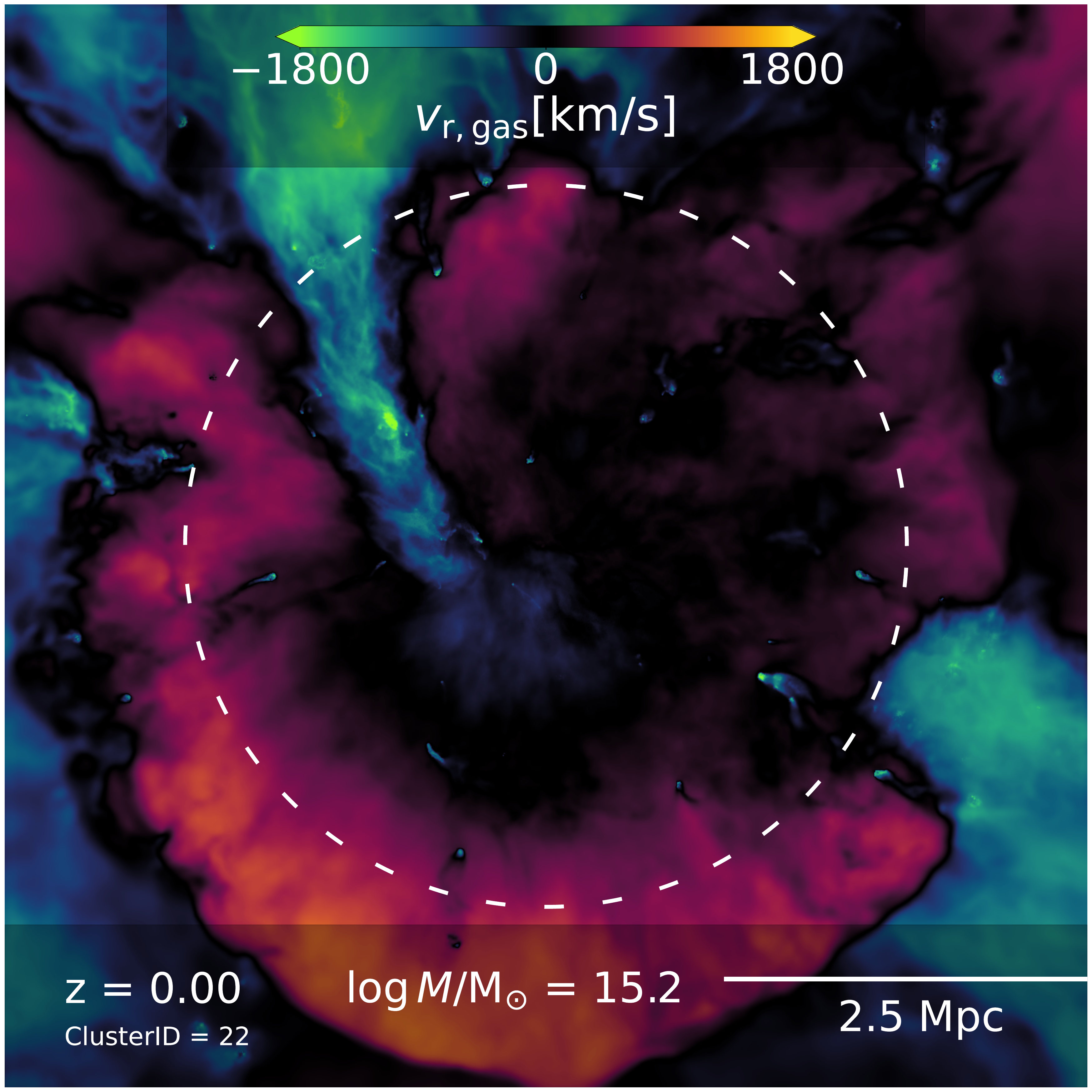}
    \includegraphics[width=0.33\textwidth]{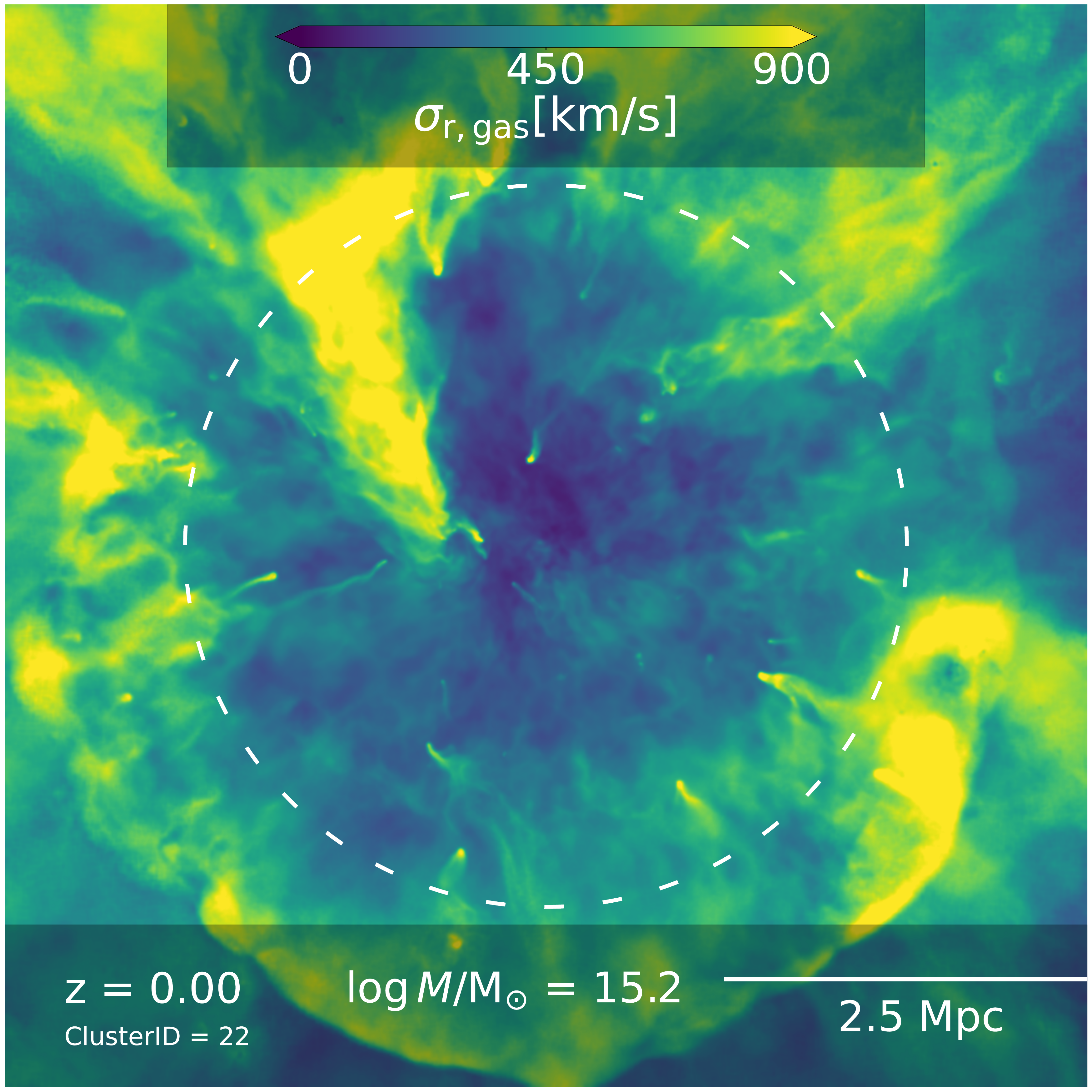}
    \includegraphics[width=0.33\textwidth]{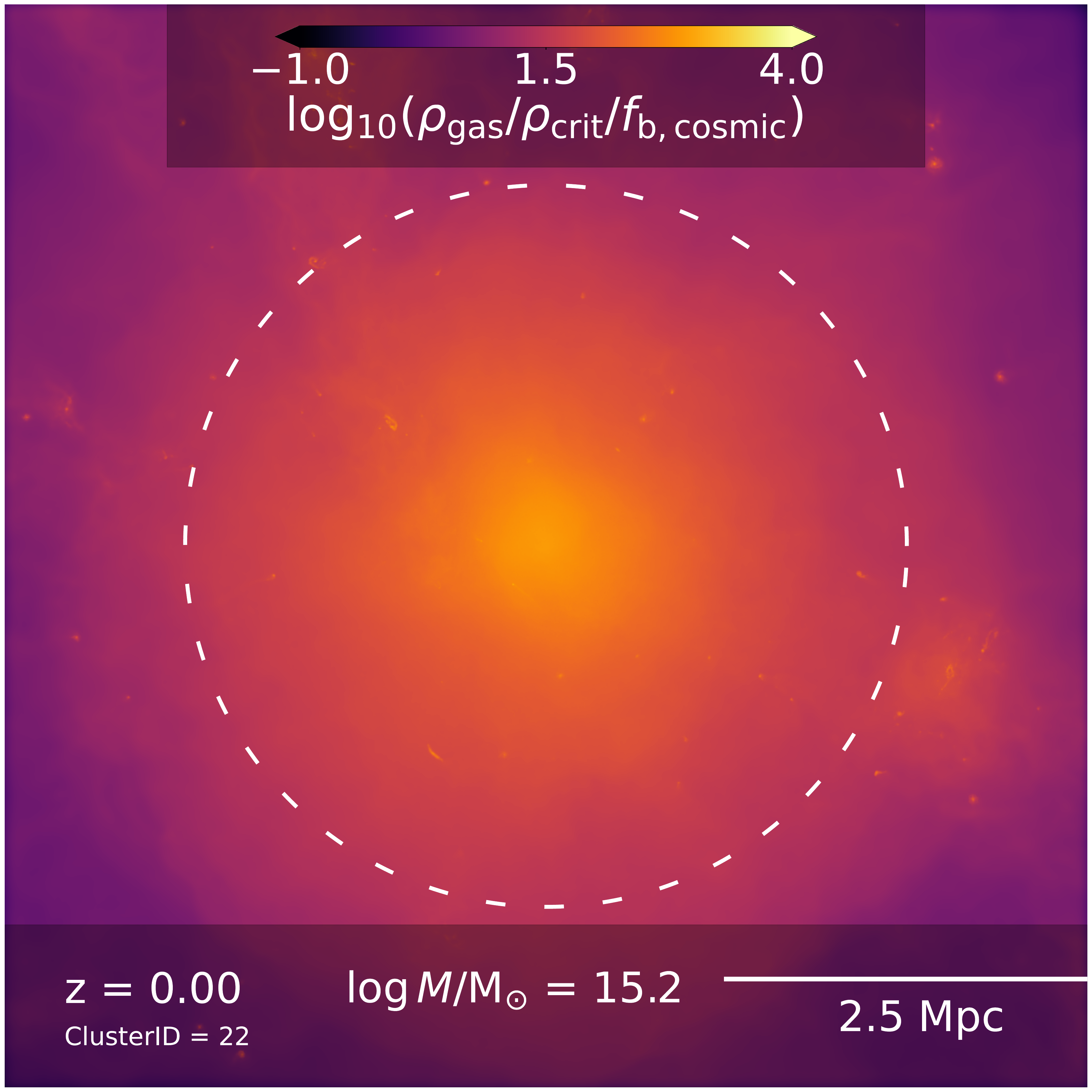}
    \includegraphics[width=0.33\textwidth]{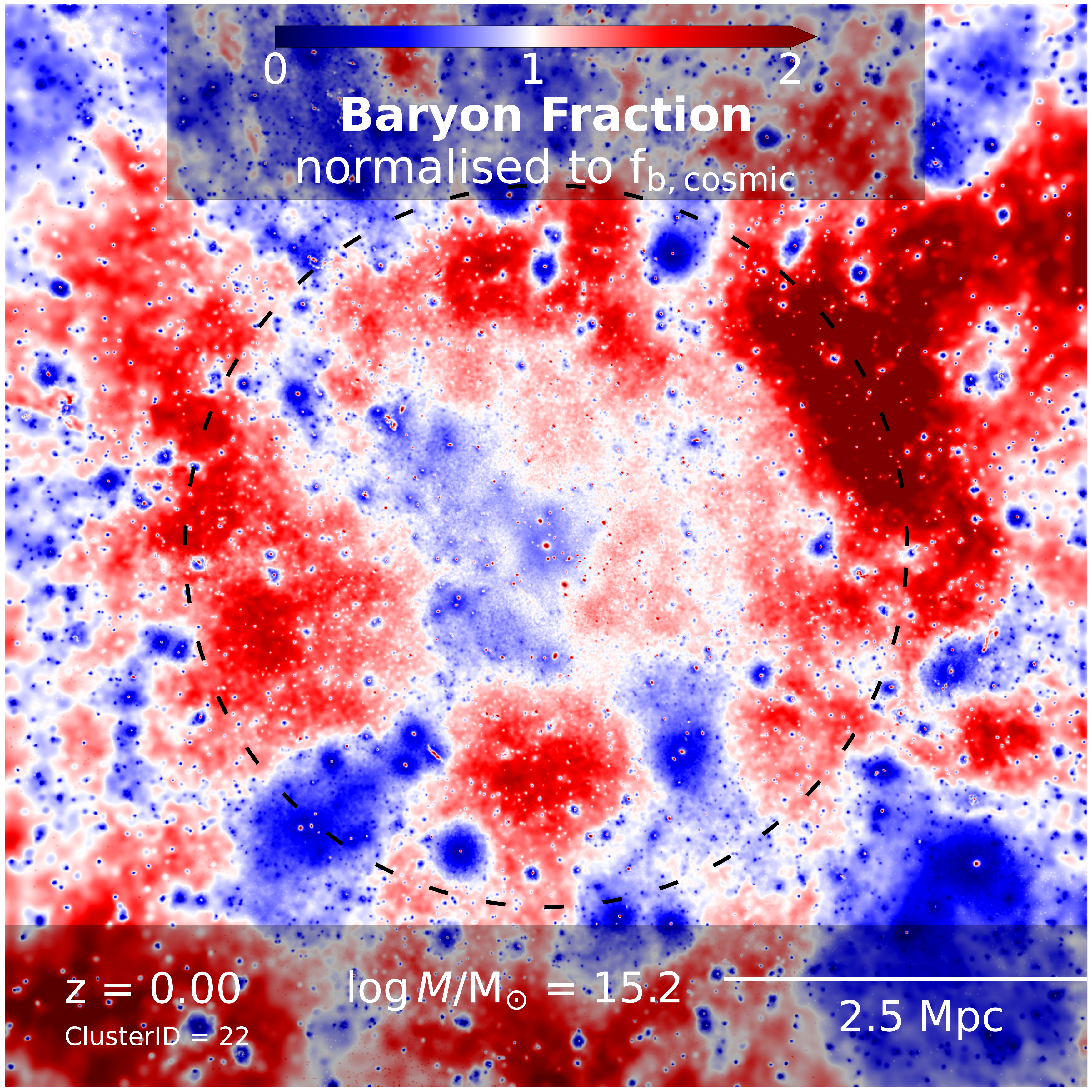}
    \caption{Gas kinematics in and around a massive halo of the TNG-Cluster simulation with $M\sim 1.6\times 10^{15} {\rm M_{\odot}}$ at $z=0$. The halo virial radius (white/black circle) and velocity are $R_{200} \sim 2.5 \, \rm Mpc$ and $V_{200} \sim 1800 \, \rm km/s$, respectively. The primary panel illustrates the radial velocity of the gas, weighted by X-ray luminosity. The trio of smaller panels below detail the dispersion of the radial velocity of the gas weighted by X-ray luminosity (left), the density of the gas normalized to the critical density of the Universe (middle), and the fraction of baryons normalized to the cosmic baryon fraction (right), all for the same object. The projection depth across all panels is set at $1.5 \, R_{\rm 200c}$. The halo is closed, meaning all of its baryons are confined within its boundary (the closure radius, $R_{\rm c}$, smaller than $R_{200}$). In contrast, its kinematics is in a state of significant evolution.}
\label{Fig: vrad_box}
\end{figure*}

\subsection{Gas flows in and out of galaxy clusters}
We begin by visualizing a massive galaxy cluster at $z = 0$ in the TNG-Cluster simulation, with a halo mass $M_{\text{200c}} \approx 1.6 \times 10^{15} \, M_{\odot}$. Fig. \ref{Fig: vrad_box} provides an overview of the key scientific phenomena discussed in this paper. It depicts the mass-weighted gas radial velocity in the top panel, the gas radial velocity dispersion in the bottom left panel, the gas density distribution in the bottom middle panel, and the baryon fraction in the bottom right panel. In all cases, the images extend to a radial distance of $1.5 R_{\text{200c}}$, where $R_{\text{200c}}$ is denoted by the white circles and is approximately $2.5 \, \text{Mpc}$. We consider a slice-like projection depth along the line of sight of $1.5 \, R_{\text{200c}}$. The halo has formed half of its mass by $z \approx 0.64$ and is in a relaxed state, as over 85\% of its mass resides within the central subhalo, and it is currently not undergoing major mergers. The BCG of the cluster contains a massive central SMBH with a mass of $M_{\text{SMBH}} \approx 4 \times 10^{10} \, M_{\odot}$.

This cluster is dynamically evolving at a significant rate. The main panel reveals complex gas kinematics within and around this massive halo. The gas motions vary substantially from the core to the outskirts of the cluster. Both inflowing (depicted in green, negative velocity) and outflowing (depicted in orange, positive velocity) gas motions are relatively slow near the cluster center. These velocities increase in the intermediate regions of the halo and continue to rise towards the outskirts, even exceeding the halo's virial velocity, which is marked at the edges of the colorbar as $\sim 1800 \, \text{km/s}$.

The radial velocity dispersion (bottom left panel) demonstrates that even within small regions of approximately one pixel ($\sim 10 \, \text{kpc}$), the gas motion is not coherent. This suggests that the gas dynamics within the cluster are influenced by, and may correlate with, a variety of factors which we discuss below. We will also delve into the coherency of gas motions through a detailed analysis of the velocity structure function later in Section \ref{subsec: velocity structure function}.

The gas density (lower center panel) is highest in the core and decreases with increasing radius. However, it is not spherically symmetric, particularly towards the outskirts of the cluster. Similarly, the baryon fraction, shown in the lower right panel, peaks red at the very center of the halo, indicating the presence of the brightest cluster galaxy (BCG). The blue region surrounding the halo center represents a high concentration of dark matter and a low gas fraction, while other blue clumps correspond to satellites. This halo is `closed' and contains all the baryonic mass associated with its dark matter within the virial radius \citep{Ayromlou2023Feedback}. The red regions dominated by baryonic matter and the blue regions dominated by dark matter effectively cancel each other out. As a result, the measured closure radius $R_{\text{c}}$ is smaller than the halo size $R_{\text{200c}}$. However, the radial velocity, as shown in the top panel, reveals a distinct pattern of inflowing (green) and outflowing (orange) gas. Clearly, even though galaxy clusters contain their full baryon budget, they are not static systems \citep[see also][]{Mitchell2022Baryonic}.


\subsection{Inflows, outflows and velocity dispersion}
\label{subsec: kinematics}

\begin{figure}
    \centering
    \includegraphics[width=0.46\textwidth]
    {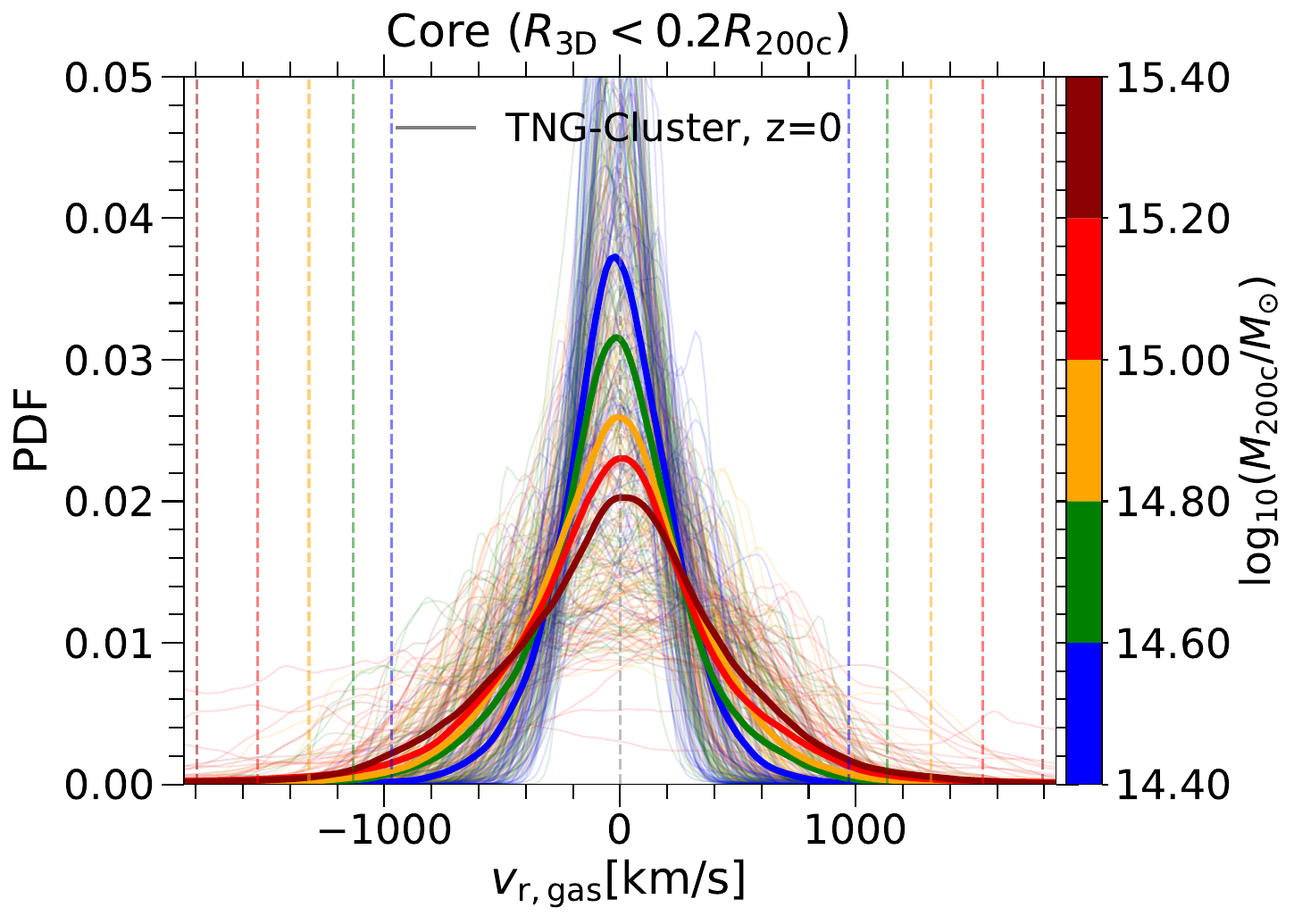}
    \includegraphics[width=0.46\textwidth]{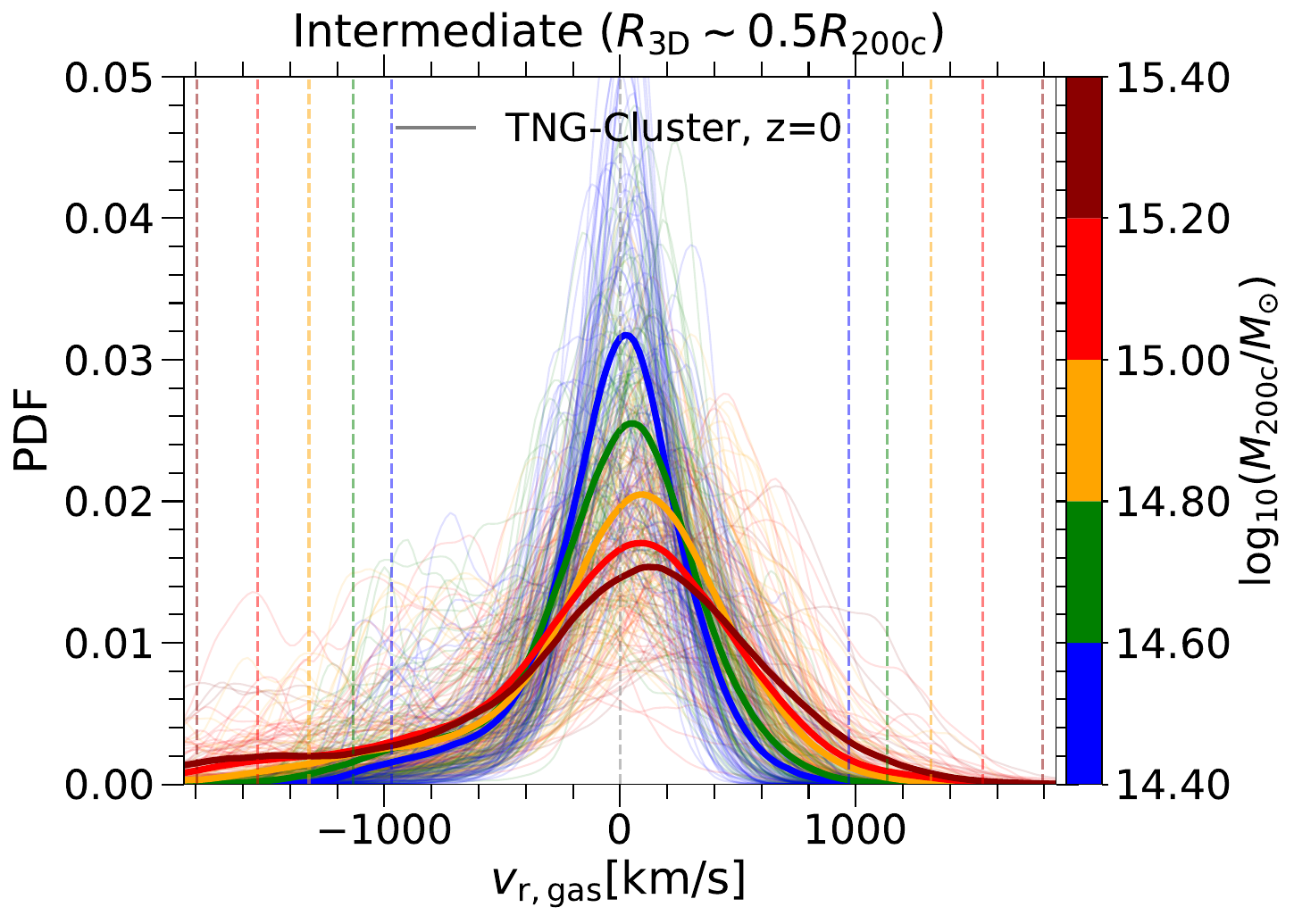}
    \includegraphics[width=0.46\textwidth]{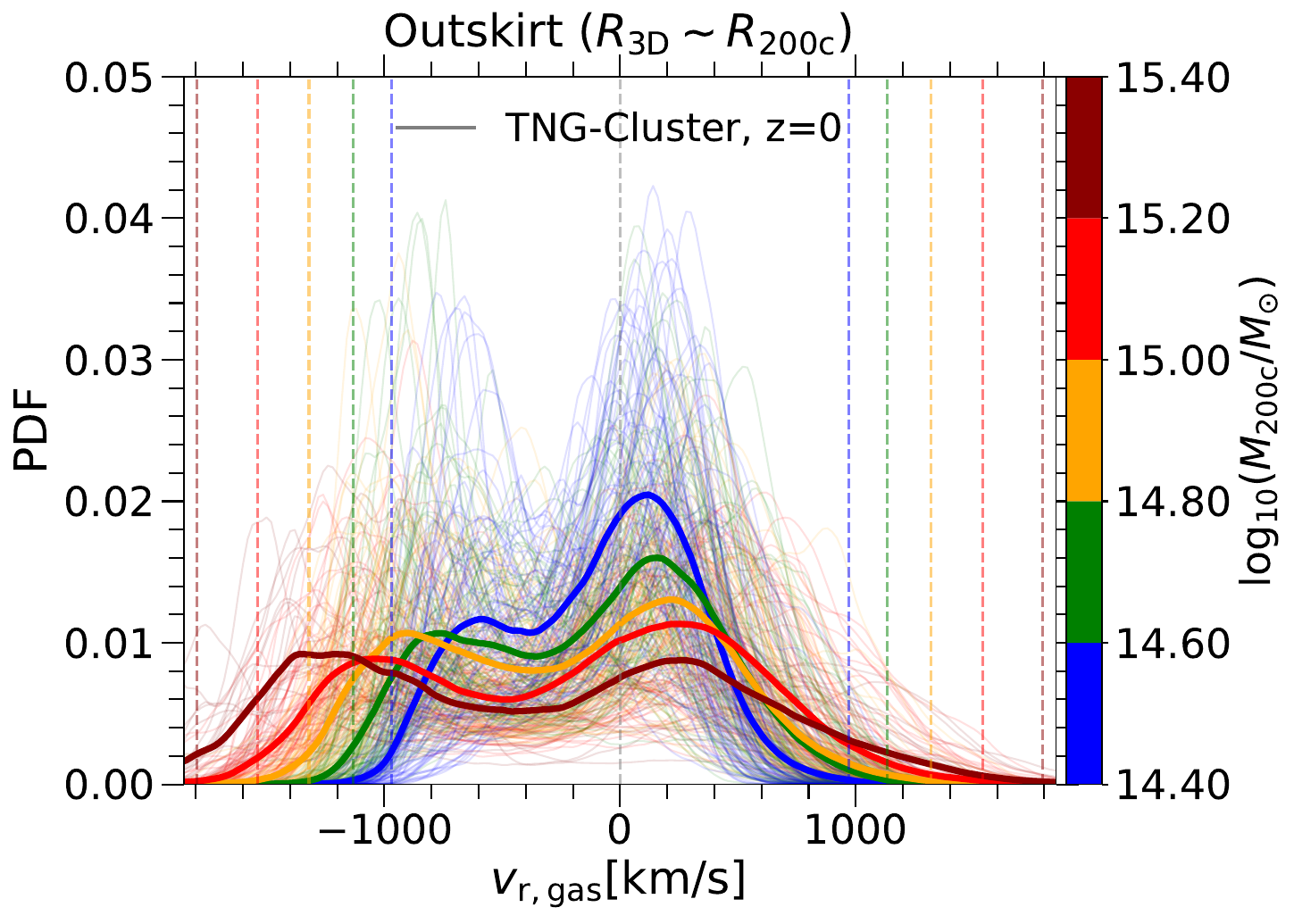}
    \caption{The normalized distributions of gas radial velocities in three different regions of TNG-Cluster halos at $z=0$. The top, middle, and bottom panels represent the cluster cores, intermediate regions, and outskirts, respectively. Each color corresponds to a different halo mass bin. The thick solid lines depict the mean profile, while the thin lines represent individual clusters. The colored dashed vertical lines indicate the virial velocity for each mass bin. The gas radial velocity distribution exhibits both inflows and outflows. It is symmetric and follows a Gaussian distribution in the core but becomes asymmetric in the outskirts, featuring a clear second peak at negative velocities.}
\label{Fig: vel_hist1d}
\end{figure}

\begin{figure}
    \centering
    \includegraphics[width=1\columnwidth]{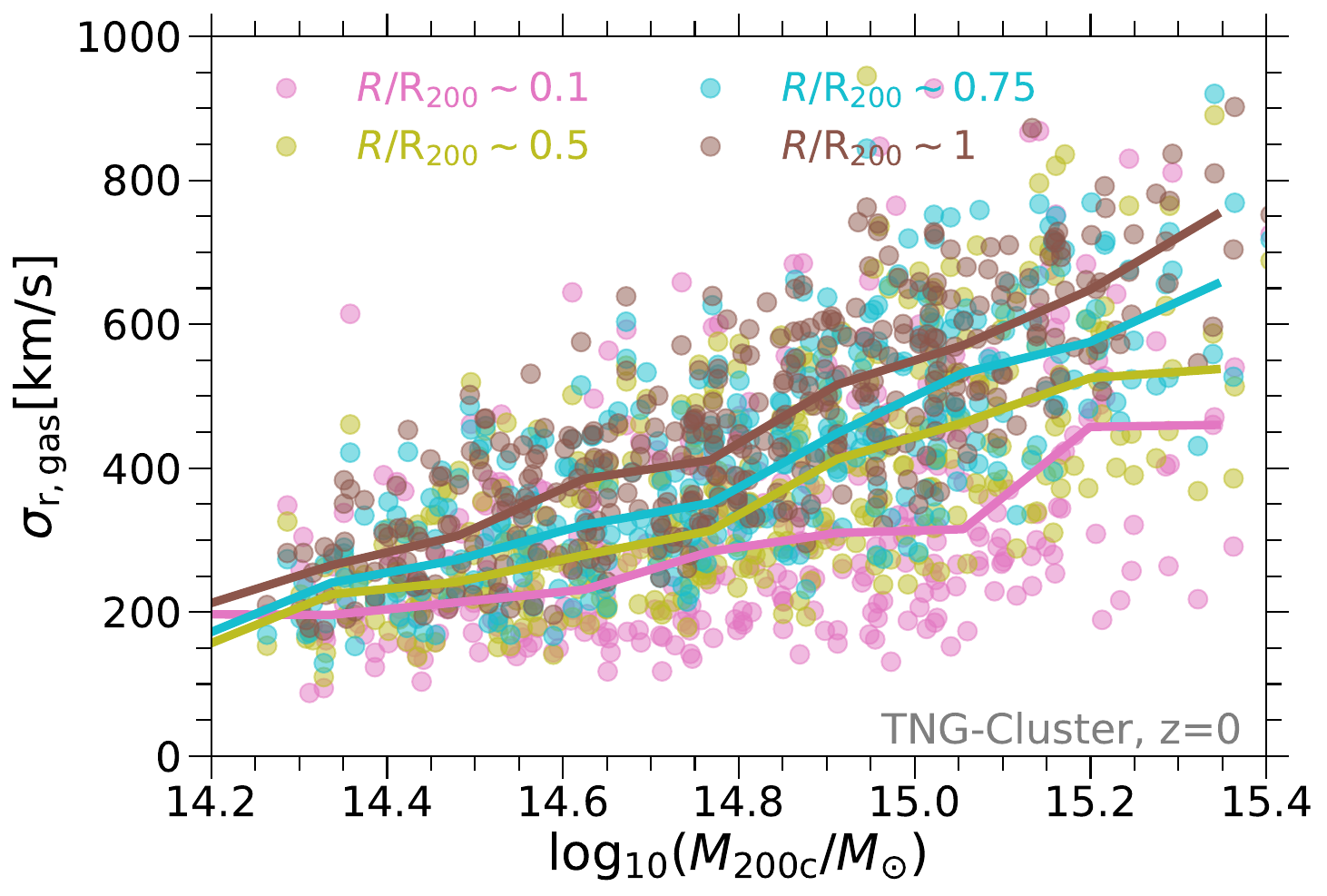}
    \caption{The radial velocity dispersion of ICM gas as a function of the halo mass. The four colors correspond to four different halocentric radii. Individual markers show the complete TNG-Cluster sample at $z=0$, while lines show the median relations at each of the four radii.}
\label{Fig: vrad_disp}
\end{figure}

In Fig. \ref{Fig: vel_hist1d}, we quantify the mass weighted distribution of gas radial motion in three different halo zones (three panels) and for different halo masses (colors). Overall, the PDFs exhibit a wide range of velocities within each region, ranging from stationary gas to gas moving at speeds exceeding $1000 \, \text{km/s}$. All 352 individual halos are shown with thin lines, while the four thick colored lines show mean distributions, stacking in five halo mass bins.

In the core ($R_{\text{3D}} < 0.2 \, R_{\text{200c}}$), the average distributions (thick lines) are symmetric and follow a Gaussian distribution, with a mean at $v_{\text{r, gas}} \sim 0$. This suggests that the net flow of gas in the core is approximately zero. In the intermediate regions ($R_{\text{3D}} \sim 0.5 \, R_{\text{200c}}$), the distributions remain relatively symmetric on average, but their peaks shift towards positive values, depending on the halo mas. However, as shown in the top left panel of Fig. \ref{Fig: profiles_main}, the net velocity remains close to zero. Nevertheless, individual halos show significant variation, with some exhibiting strong outflows and others showing strong inflows. The presence of these asymmetries in the velocity distribution of massive clusters, that are also seen in the CGM of much less massive halos \citep[e.g.,][]{Lochhaas2020Properties,Fielding2020First}, indicates that the ICM is dynamically evolving, similar to its lower mass counterparts.

In the outskirts ($R_{\text{3D}} \sim R_{\text{200c}}$), the velocity distributions are highly asymmetric and exhibit a double peak. The left peak (negative values) arises from cosmic accretion onto the halo at these scales. This aligns with the radial velocity profiles presented in Fig. \ref{Fig: profiles_main}. The cosmic accretion peak happens at higher velocities in more massive halos, a trend which follows the halo virial velocities (dashed vertical lines). Furthermore, the PDFs in the outskirts are significantly broader than those in the core and intermediate regions, indicating less coherent gas motions in the outskirts. We will investigate this in detail in Section \ref{subsec: velocity structure function}.

There is also a clear trend with halo mass: the higher the mass, the lower the amplitude of the peak and the broader its width. However, there is considerable halo-to-halo variation. Individual halos, represented by thin lines, can exhibit velocity distributions that deviate significantly from the average within their respective halo mass bins. This variation is strongly correlated with both the properties of the clusters and other attributes of the gas itself, as we discuss below.

In Fig. \ref{Fig: vrad_disp} we show the intrinsic, three-dimensional, radial velocity dispersion of gas, measured within radial bins i.e. spherical shells. We plot the entire TNG-Cluster sample at $z=0$, as a function of halo mass, for four different radii extending from the core (pink) to intermediate regions (yellow, teal), to the outskirts (brown). The ICM radial velocity dispersion is a strong function of mass, and is larger for more massive halos, increasing by a factor of $\sim 2-4$ from our least to most massive clusters. Simultaneously, $\sigma_{\rm r}$ increases with halocentric distance, and is smallest in the core.

\subsection{Radial and mass dependence of 3D gas kinematics}
\label{subsubsec: radial and mass dependence}

\begin{figure*}
    \centering
    \includegraphics[width=0.49\textwidth]{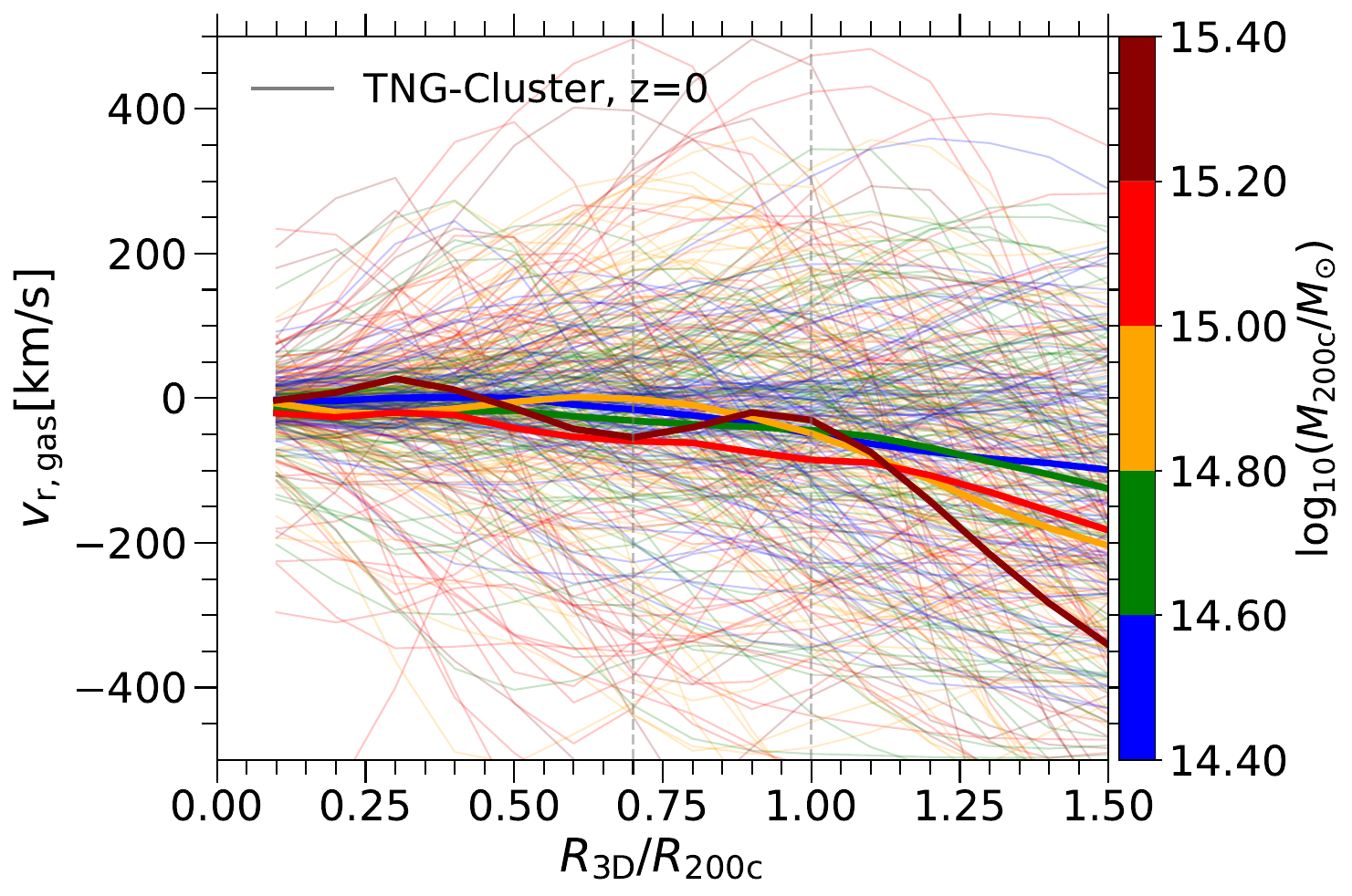}
    \includegraphics[width=0.49\textwidth]{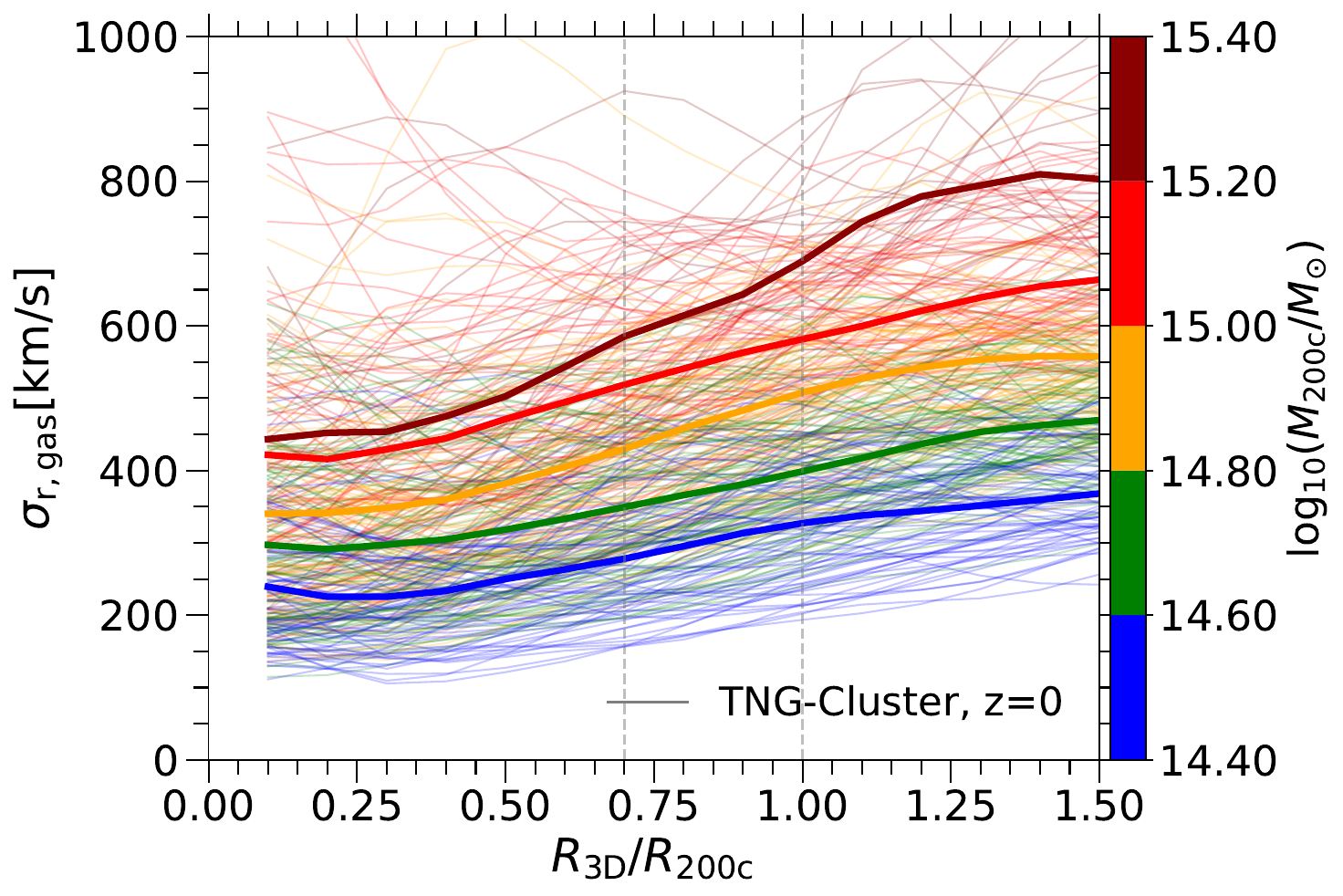}
    \includegraphics[width=0.49\textwidth]{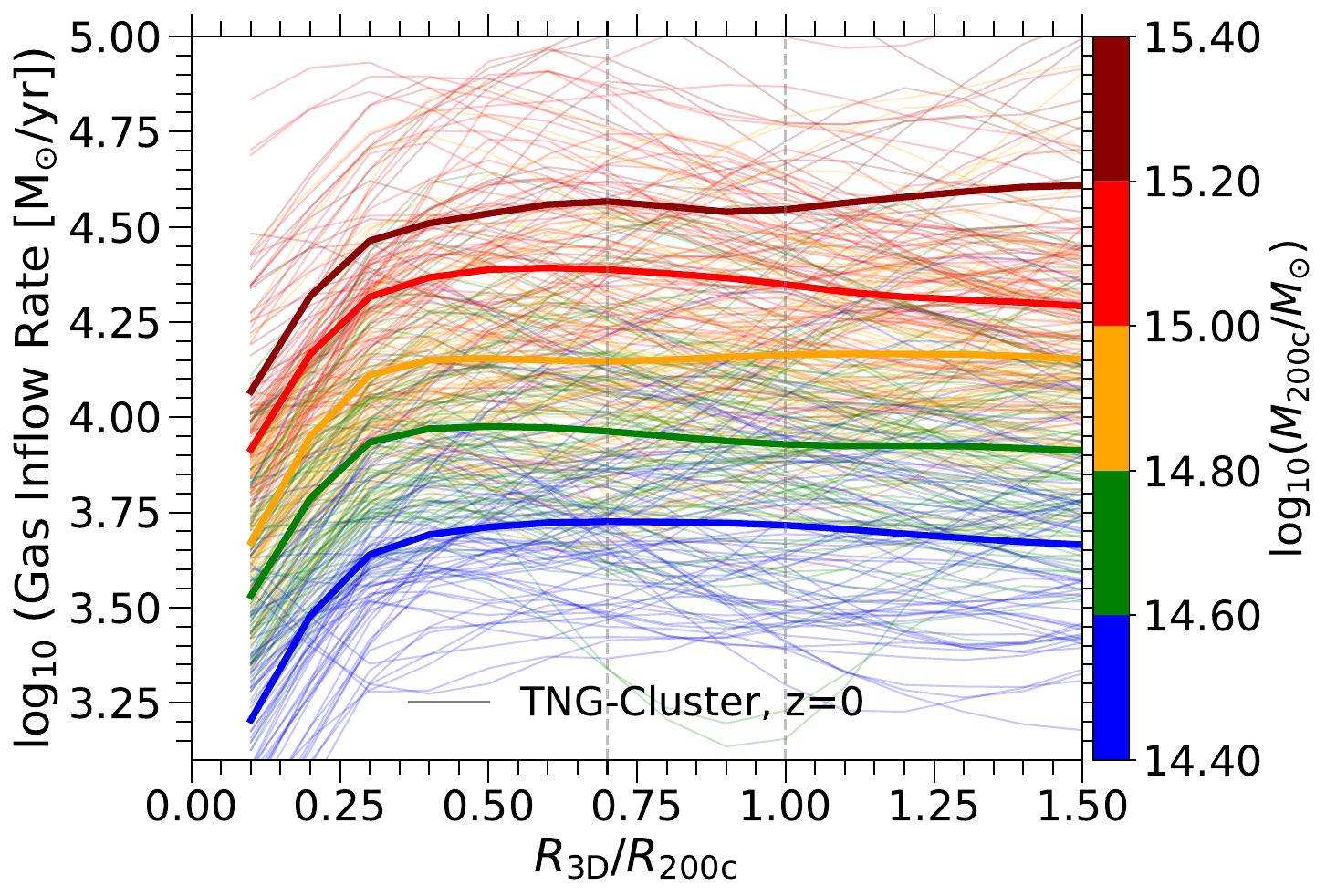}
    \includegraphics[width=0.49\textwidth]{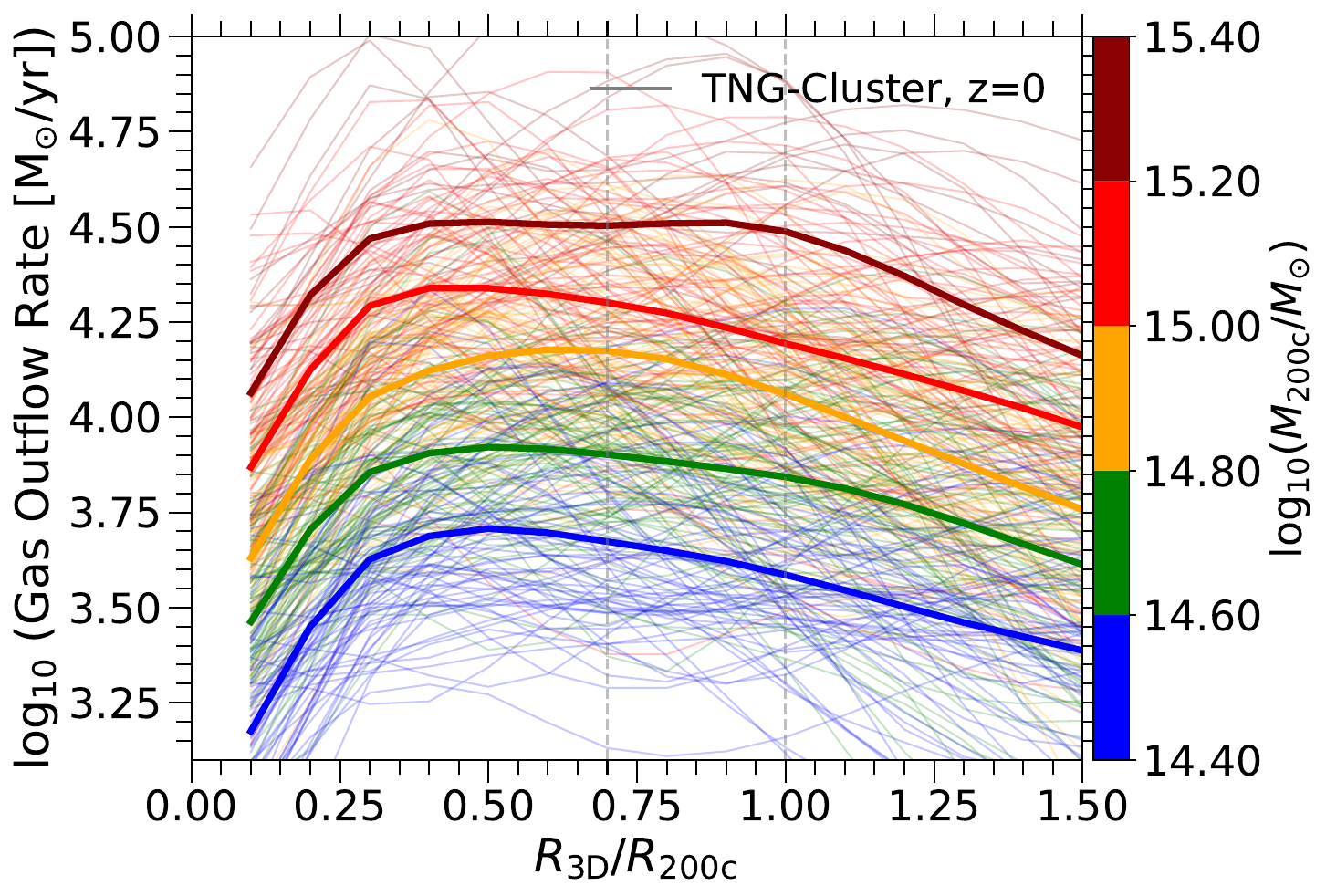}
    \caption{Radial profiles of gas properties across the full TNG-Cluster sample of galaxy clusters at $z=0$. Each color denotes a different halo mass: thick solid lines depict the mean profile, while the thin lines represent individual clusters. The panels, from top left to bottom right, show the radial velocity, radial velocity dispersion, gas inflow rate, and gas outflow rate. On average, gas is inflowing (with negative radial velocity) in cluster outskirts, reaching quasi-equilibrium (zero radial velocity) in cluster cores. Similarly, on average, the net gas flow across all stacked clusters is near zero. However, this is not the case for individual clusters, which can exhibit strong inflows and/or outflows of high velocity and/or high velocity dispersion gas at a variety of radii.}
\label{Fig: profiles_main}
\end{figure*}

In Fig. \ref{Fig: profiles_main}, we present the radial and mass trends of various gas properties from the cluster center to $1.5 \, R_{\text{200c}}$ for clusters of differing masses, as denoted by the color of each line. The top-left panel shows the radial velocity of the gas, indicating that, on average, gas is inflowing in the outskirts (negative radial velocity) and reaches a quasi-equilibrium state (zero radial velocity) in the cores of clusters. This suggests that the net flow of the gas is approximately zero near the core and shifts towards inflow-dominated as the halocentric distance increases. However, the thin lines for individual clusters reveal that this general trend does not necessarily represent the gas kinematics for a specific cluster. There is considerable scatter and variation among halos. The radial velocity of the gas can be positive (outflowing) or negative (inflowing) at any given radius, and its magnitude can vary significantly. This scatter is strongly correlated with the properties and evolutionary state of the halo, as we discuss below.

To understand the variation of velocities at different scales, the top-right panel of Fig. \ref{Fig: profiles_main} displays the radial velocity dispersion of the gas. In this analysis, we measure the velocity dispersion within radial bins (i.e., spherical shells). The radial velocity dispersion is highest in the outskirts and decreases towards the core of the cluster, highlighting the greater asymmetry of gas motions in the outskirts. The amplitude of the velocity dispersion is also higher in more massive clusters, as indicated by the color of the lines. This is partly due to the stronger gravitational forces in more massive clusters, given that both the overall velocity dispersion and $ \rm V_{200c}$ increase with halo mass. However, there is noticeable scatter among individual clusters, reflecting the unique gas kinematics in each halo. Therefore, while case studies of individual clusters can provide valuable insights into the gas kinematics of specific halos, a statistical study of a large sample of clusters is essential for understanding overarching trends and the impact of halo properties on gas kinematics. This is one of the key objectives of both the TNG-Cluster project and this paper.

The bottom panels of Fig. \ref{Fig: profiles_main} display the radial profiles of the gas inflow and outflow rates. The gas inflow rate (bottom left panel) is lowest in the cluster core and monotonically increases towards the outskirts. This trend is expected, as gas is accreted from the cosmic web, and the accretion rate diminishes towards the cluster center. Beyond a certain halocentric distance, this trend flattens, indicating a relatively constant accretion through the halo. More massive halos have higher inflow rates at fixed halocentric distances, reflecting their stronger gravitational pull. In contrast, the outflow rate does not monotonically increase with halocentric distance. Between the cluster center and $\sim 0.3-0.5 R_{200c}$, the outflow rate rises with increasing halocentric distance but then declines towards the outskirts. This suggests that the impact of physical processes leading to outflows weakens considerably towards the halo outskirts. As the average radial velocity of outflows keeps increasing with halocentric distance (see Fig. \ref{Fig: profiles_vrad}), this indicates the lower amount of gas moving outwards, as opposed to slower outflows, leading to lower gas outflow rates. For both inflows and outflows, there is notable scatter among individual halos, further confirming the variation in gas kinematics from one halo to another. As we discuss below, this halo-to-halo variation is strongly correlated with the properties and evolutionary history of each halo.

\subsection{Connecting gas motions to cluster properties}
\label{subsec: impact of galaxy and halo properties}

\begin{figure*}
    \centering
    \includegraphics[width=0.40\textwidth]{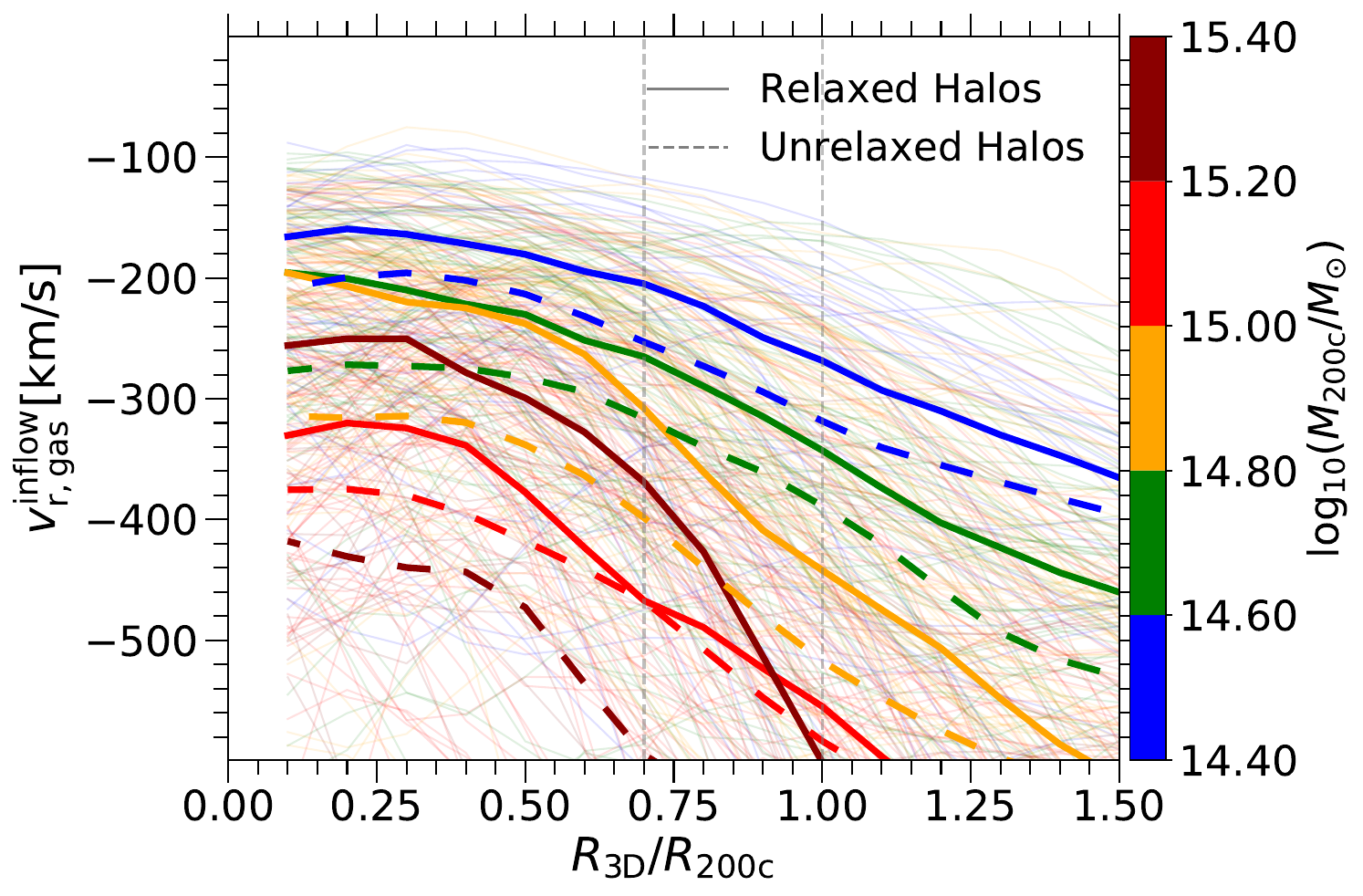}\hspace{5mm}
    \includegraphics[width=0.40\textwidth]{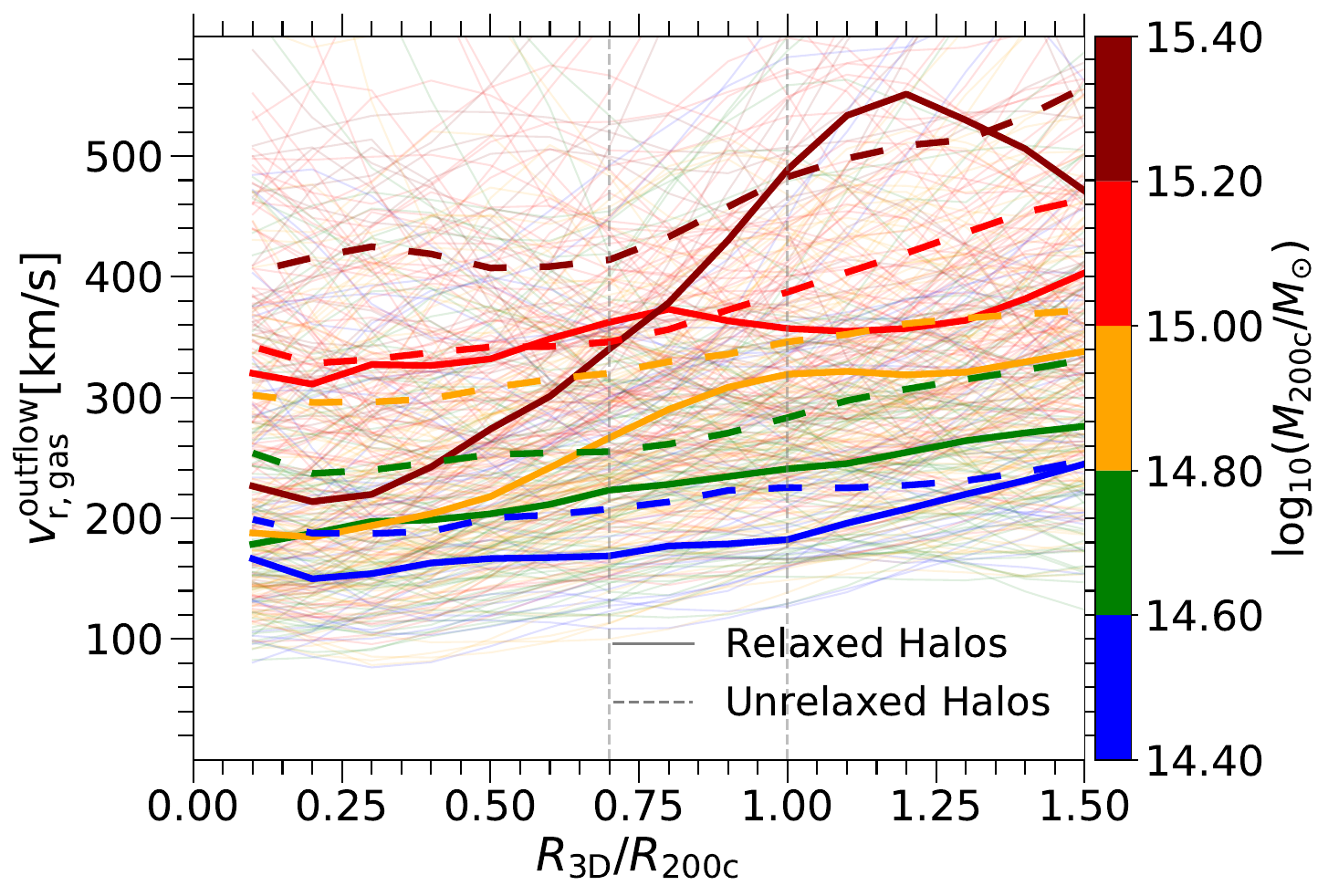}
    \includegraphics[width=0.40\textwidth]{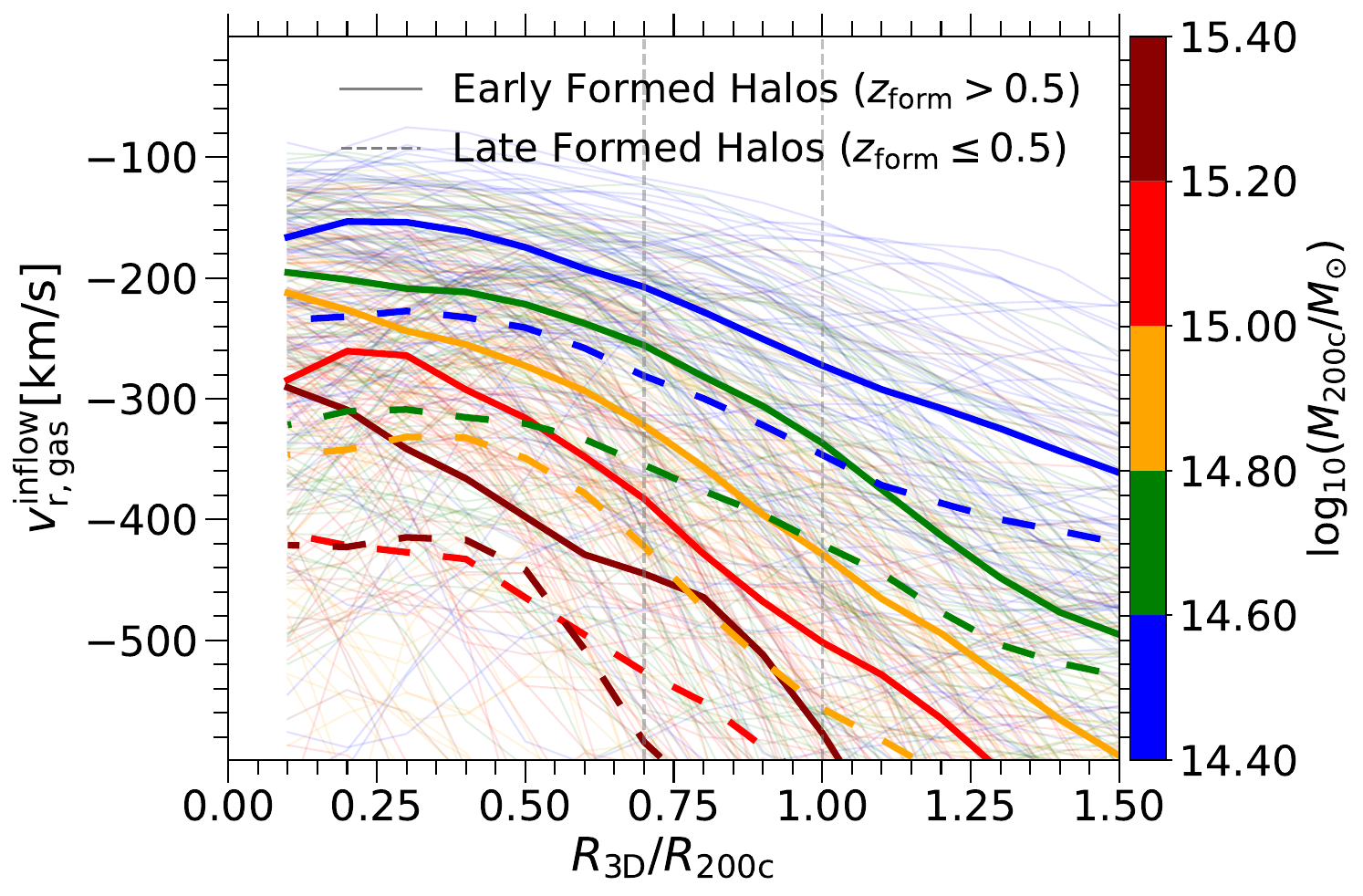}\hspace{5mm}
    \includegraphics[width=0.40\textwidth]{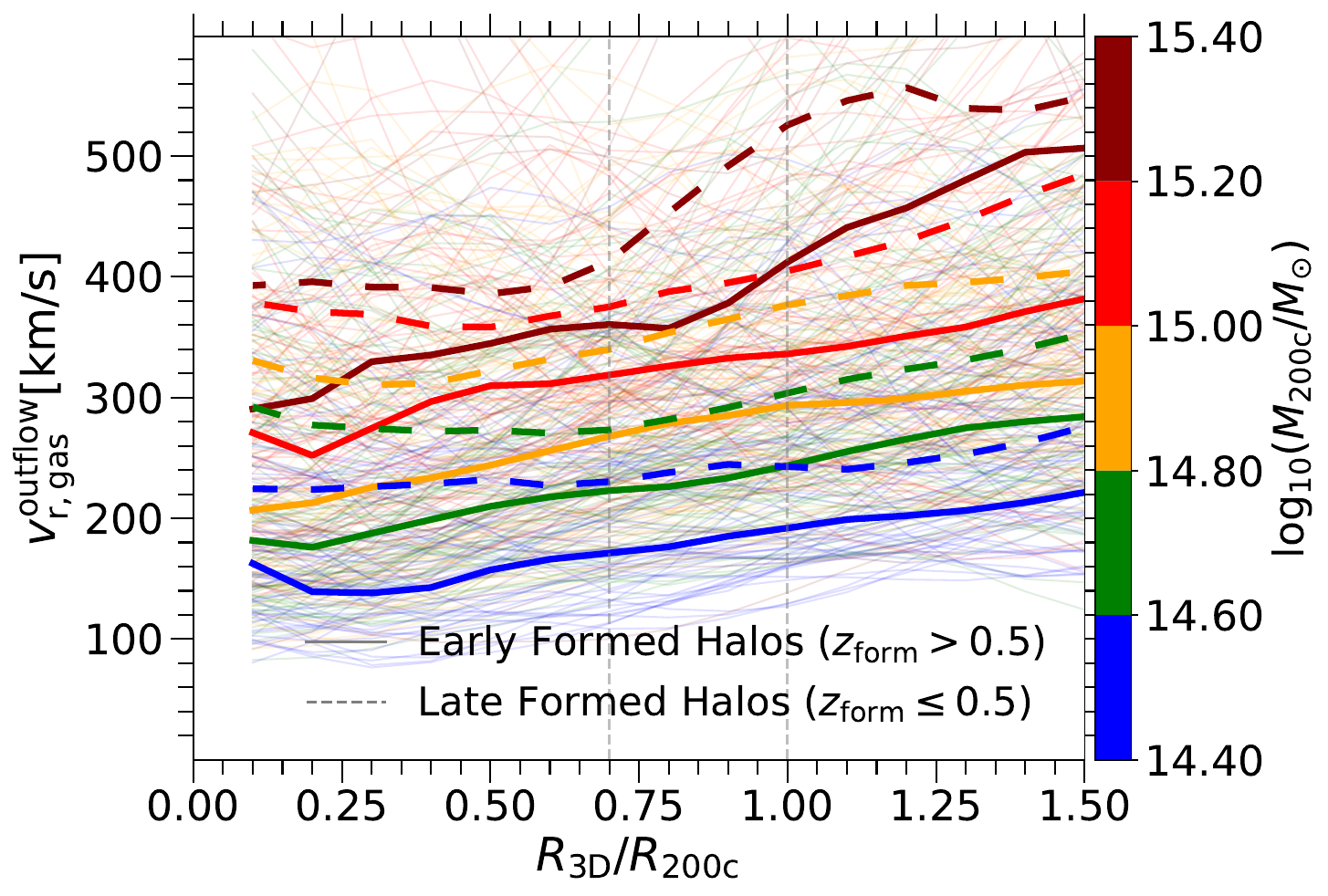}
    \includegraphics[width=0.40\textwidth]{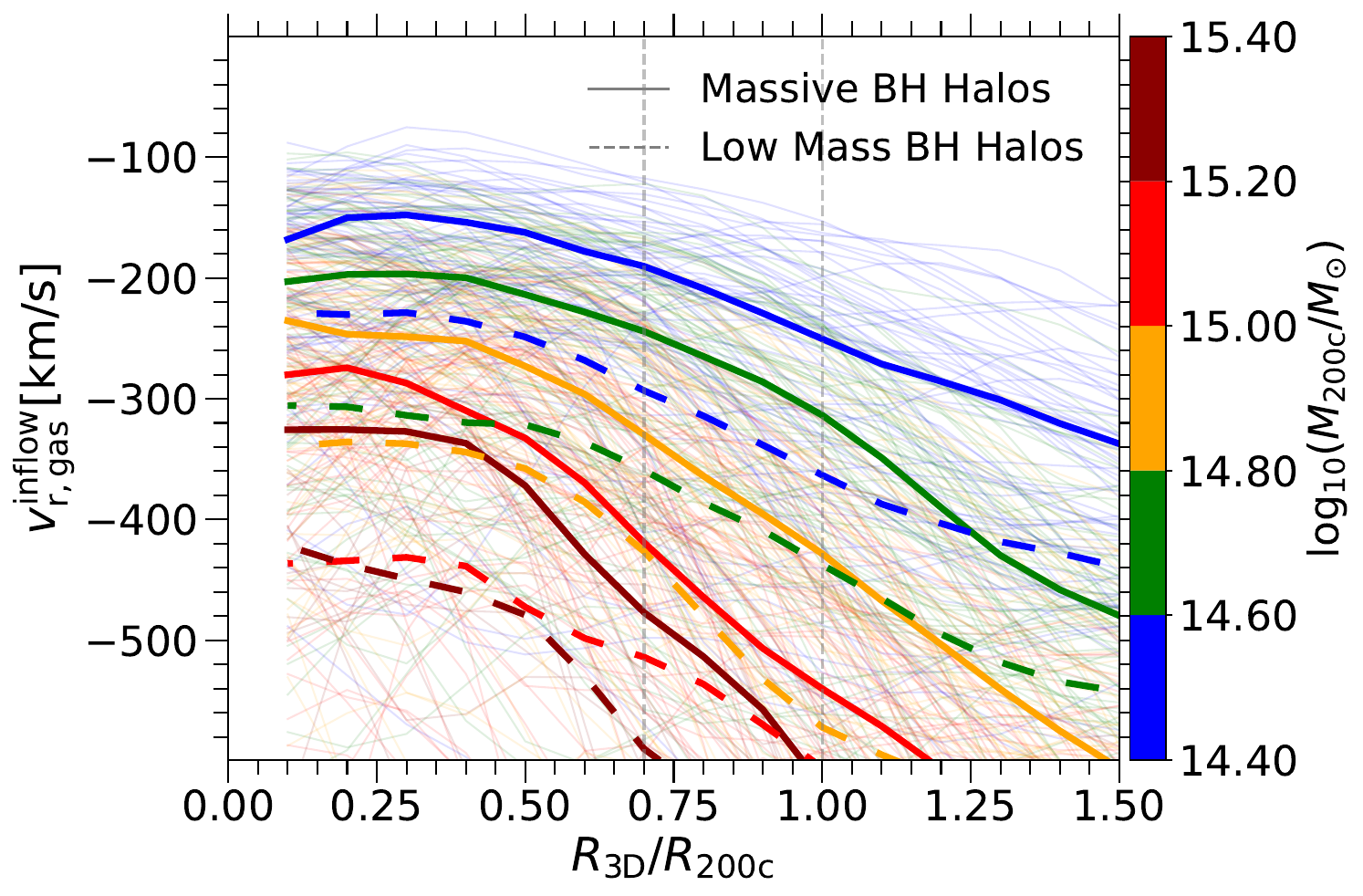}\hspace{5mm}
    \includegraphics[width=0.40\textwidth]{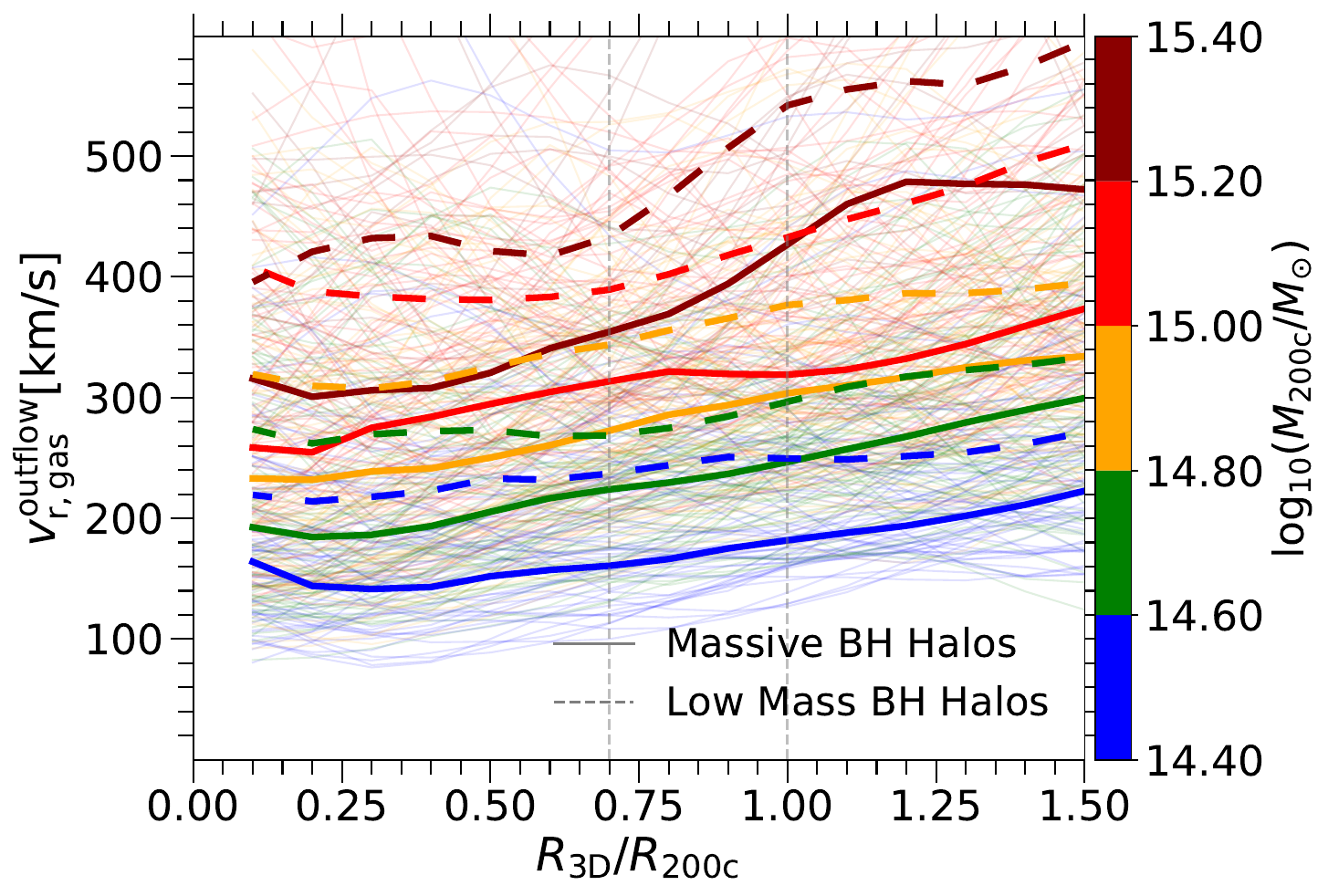}
    \includegraphics[width=0.40\textwidth]{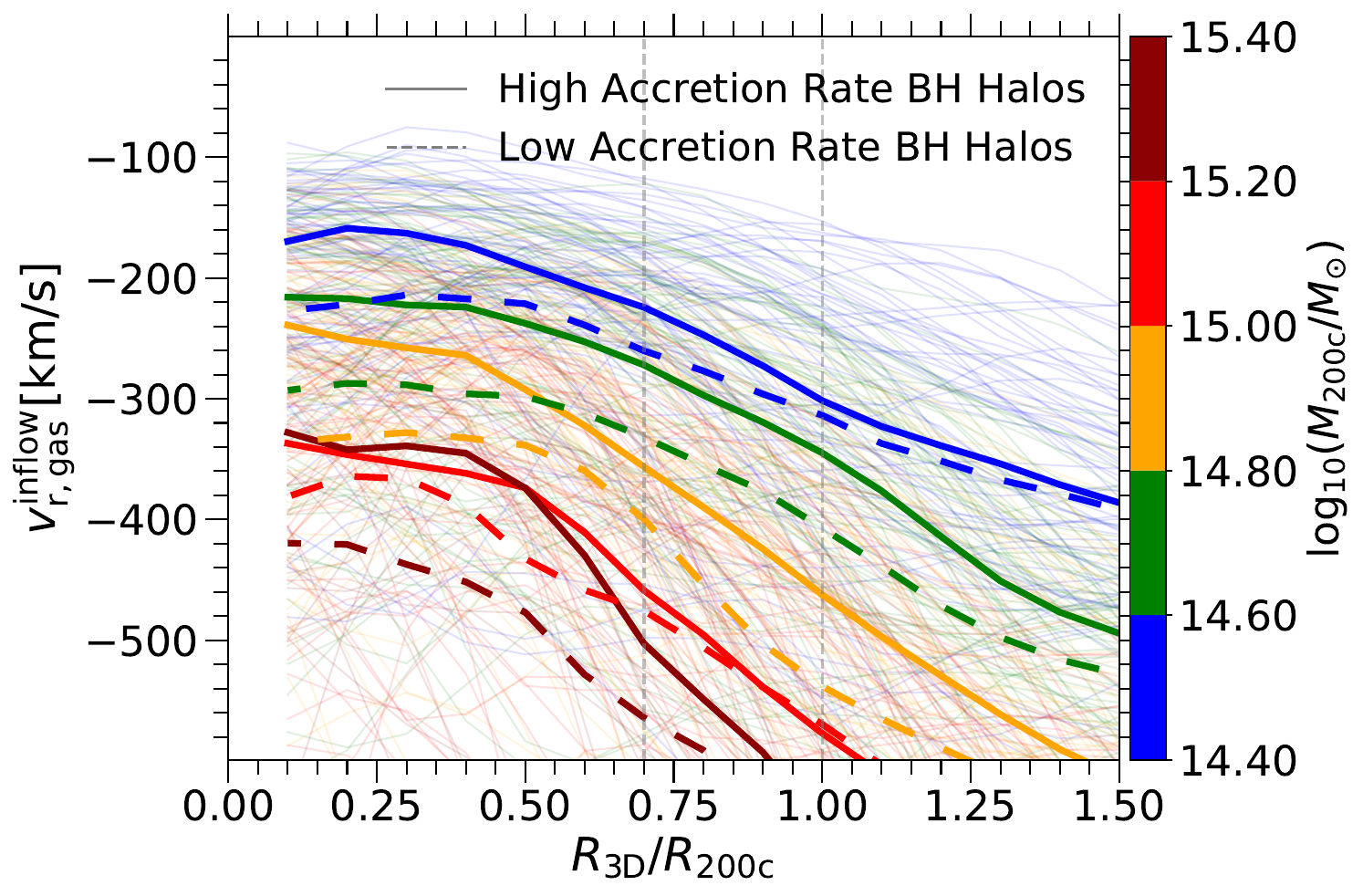}\hspace{5mm}
    \includegraphics[width=0.40\textwidth]{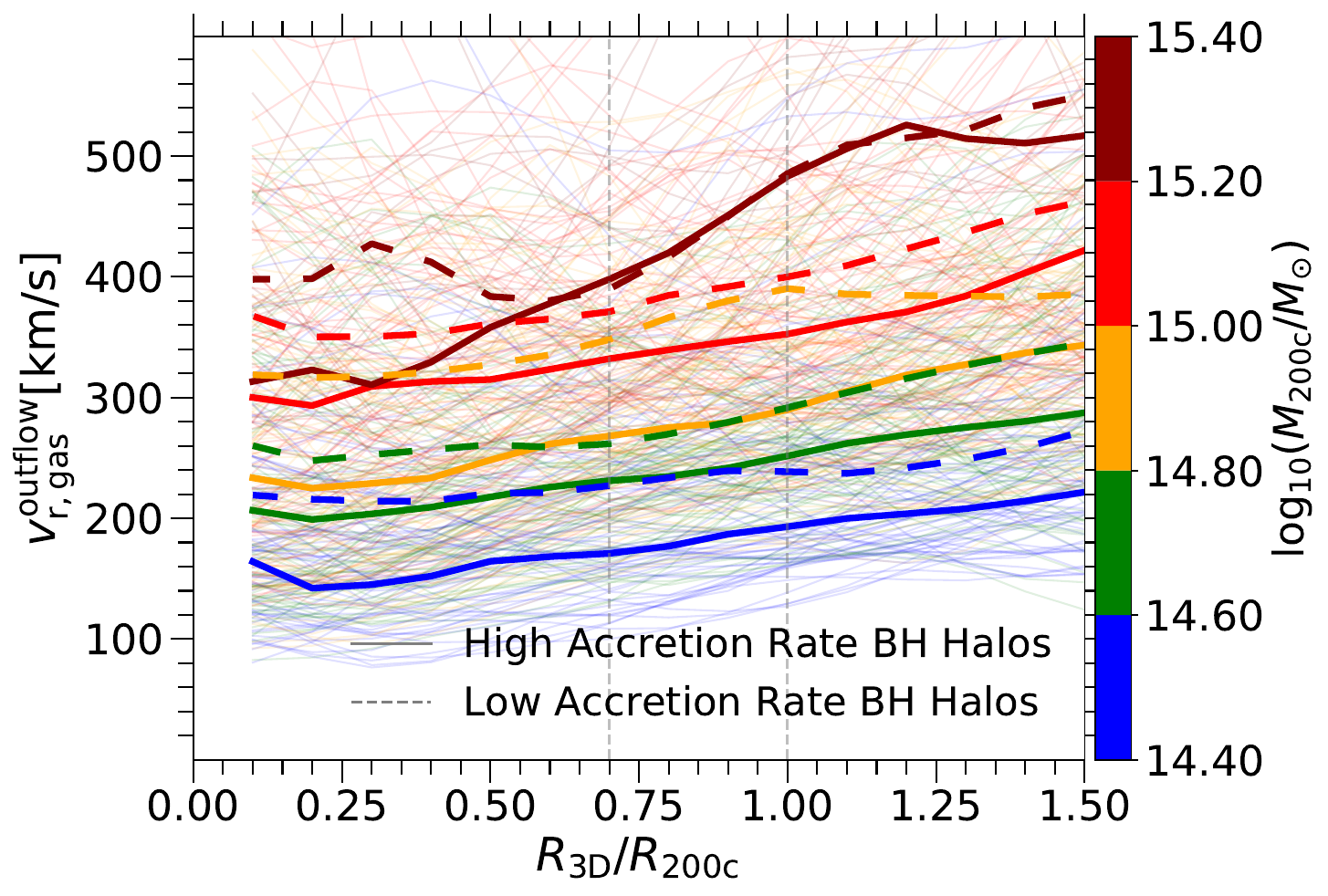}
    \caption{Radial profiles of the inflowing gas radial velocity (left column) and the outflowing gas radial velocity (right column) for halos classified based on their relaxedness (first row), formation time (second row), SMBH mass (third row), and SMBH accretion rate (fourth row). Each color corresponds to a different halo mass bin. The thick solid lines depict the mean profile, while the thin lines represent individual clusters. The figure emphasizes the significant correlations between halo properties and the kinematics of gas.}
\label{Fig: profiles_vrad}
\end{figure*}

In Fig. \ref{Fig: profiles_vrad}, we present the radial profiles of inflowing (left column) and outflowing (right column) gas radial velocity. Halos are split into subsets based on four physical parameters in the four rows. Our aim is to explore potential correlations between gas motion and the properties of the host galaxy clusters.

The top row splits halos based on their state of relaxation. We see that unrelaxed halos have faster inflows (more negative radial velocities) and faster outflows (more positive radial velocities). This trend is consistent across all mass ranges (indicated by colors) and nearly all halocentric radii. However, the average velocity difference between relaxed and unrelaxed objects can vary from a few tens to a few hundreds of km/s, depending on the halo mass and halocentric radius. This suggests a strong correlation between the gas kinematics and the dynamical state of the halo. Unrelaxed halos, which are typically undergoing mergers and are dynamically evolving, have strong flows driven by these mergers, and their gas motions are not in equilibrium. In contrast, relaxed halos are in a quasi-equilibrium state, and their gas motions are more balanced. This is further confirmed by the velocity dispersion of the gas, which is higher in unrelaxed halos, as we discuss later in the paper (see Figs.\ref{Fig: vLOS_disp},\ref{Fig: profiles_vrad_disp}).

In the second row, we sub-divide halos based on their formation time. Late-forming halos, which are more dynamically evolving, exhibit faster inflows and outflows compared to early-forming halos. This indicates that halo formation history impacts ICM kinematics. On average, late-forming halos have undergone more recent merger events, leading to more dynamic gas motions. In contrast, early-forming halos have been in a quasi-equilibrium state for a longer period, resulting in more stable gas motions.

The third row splits halos based on the mass of the SMBHs in their central subhalos. For halos of a fixed mass (colors), those with less massive SMBHs exhibit faster inflows and outflows compared to those with more massive SMBHs. Interestingly, radial velocities increase with increasing halo mass, but at a fixed halo mass, they decrease with increasing SMBH mass. This is because more massive SMBHs reside within more relaxed halos and those with earlier formation times. Consequently, the impact of SMBH mass on gas kinematics is partially obscured by other, stronger correlations. We discuss the cross-correlation between halo properties in more detail in Appendix \ref{app: halo correlations}.

Finally, the fourth row classifies halos based on the accretion rate onto their SMBHs. Halos with more rapidly accreting SMBHs (represented by solid lines) generally exhibit slower inflows and outflows compared to those hosting SMBHs with slower accretion rates. This trend is strongly dependent on halocentric distance: the closer the gas is to the halo center, the more significant the trend becomes. This reflects the strong coupling of SMBH feedback with the halo center, while its direct influence diminishes towards the outskirts of the halo.

We note that the trends discussed above are, in most cases, significantly more pronounced in our most massive halo mass bin ($15.2<\log_{10} M_{\rm 200c} / \rm M_{\odot} < 15.4$). This bin contains 22 galaxy clusters, making the mean profiles generally robust. However, we do not rule out the possibility that the trends may be magnified due to the small number of clusters in this bin.

\begin{figure}
    \centering
    \includegraphics[width=0.46\textwidth]{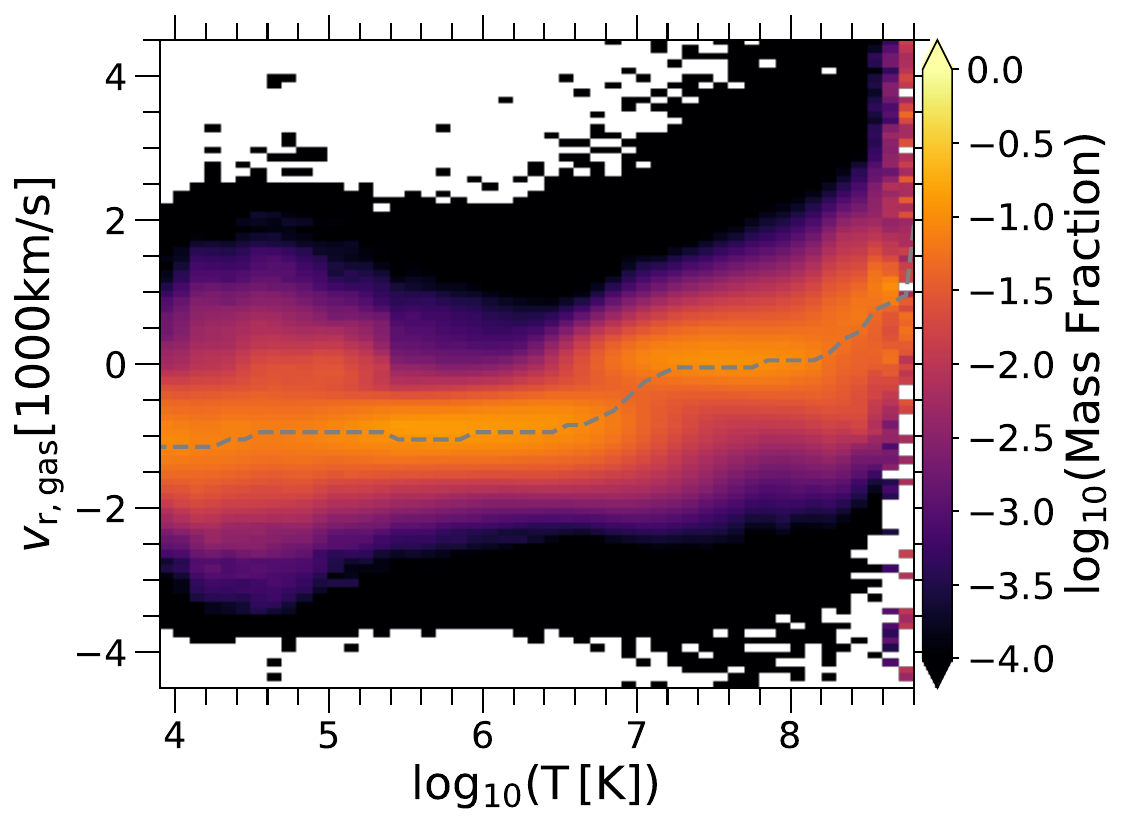}
    \includegraphics[width=0.46\textwidth]{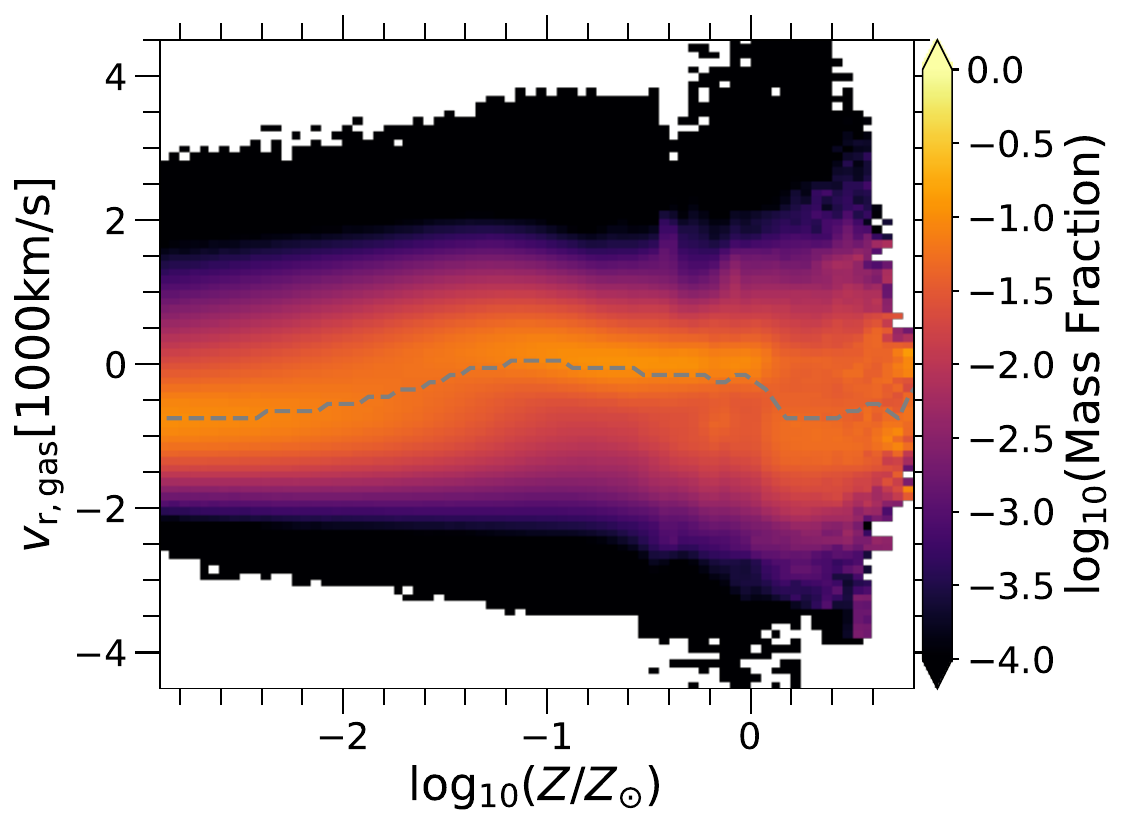}
    \includegraphics[width=0.46\textwidth]{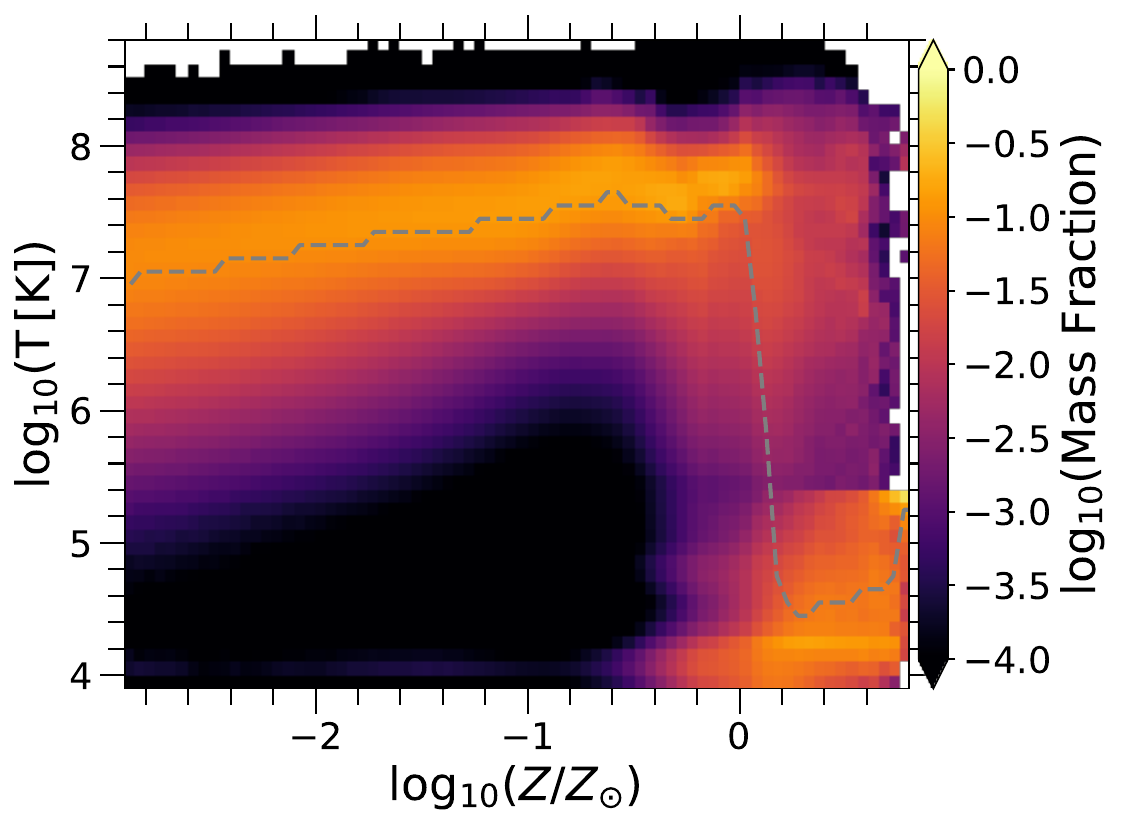}
    \caption{The connection between gas kinematics, temperature, and metallicity. Each panel shows the 2D histogram (i.e. distribution of gas mass) in the plane of radial velocity versus: temperature (top), and metallicity (middle). The bottom panel shows the cross correlation of temperature and metallicity. The colors show the conditional fraction of mass at fixed x-axis value, i.e. each column is normalized such that the sum equals unity. Gas velocity and temperature are connected: cool gas is typically inflowing while hot gas is typically outflowing.}

\label{Fig: 2dhists}
\end{figure}

We further explore the relationship between the gas radial velocity and other gas properties, such as temperature and metallicity, in Fig. \ref{Fig: 2dhists}. In all panels, the colors represent the conditional fraction of mass at a fixed value on the x-axis; that is, each column is normalized so that the sum equals one. In this manner, each column serves as a histogram, where we have stacked data from all TNG-Cluster halos at $z=0$.

The top panel shows the correlation between gas radial velocity and temperature. In both mass fractions (indicated by colors) and the running median (depicted by the dashed line) we see that the majority of cool gas, with $\rm T < 10^6 \, K$, is inflowing. Conversely, hot gas with $\rm T \geq 10^6 \, K$ exhibits both inflowing and outflowing components. The median radial velocity of hot gas is near zero, but there is a significant scatter, ranging from hundreds to thousands of km/s. Nevertheless, even the median radial velocity of extremely hot gas with $\rm T \geq 10^8 \, K$ is positive, indicating that this phase in mostly outflowing.

The middle panel of Fig. \ref{Fig: 2dhists} presents the relationship between gas radial velocity and its metallicity. A correlation is evident: both low-metallicity gas ($\rm \log_{10}Z/Z_{\odot} < -1$) and very high-metallicity gas ($\rm \log_{10}Z/Z_{\odot} > 0$) are more likely to be inflowing. As depicted in the bottom panel of the same figure, low-metallicity gas is predominantly hot, while high-metallicity gas is generally cool. This aligns with the idea that low-metallicity gas is accreted from the cosmic web, whereas high-metallicity gas may be ejected from galaxies. Gas with intermediate metallicity ($\rm -1 < \log_{10}Z/Z_{\odot} < 0$) is typically hot and exhibits a median radial velocity close to zero, indicating a balance between inflowing and outflowing material. However, the scatter indicates the complexity of velocity-metallicity relations. This intermediate-metallicity gas is the most abundant component of galaxy clusters.

\begin{figure*}
    \centering
    \includegraphics[width=0.49\textwidth]{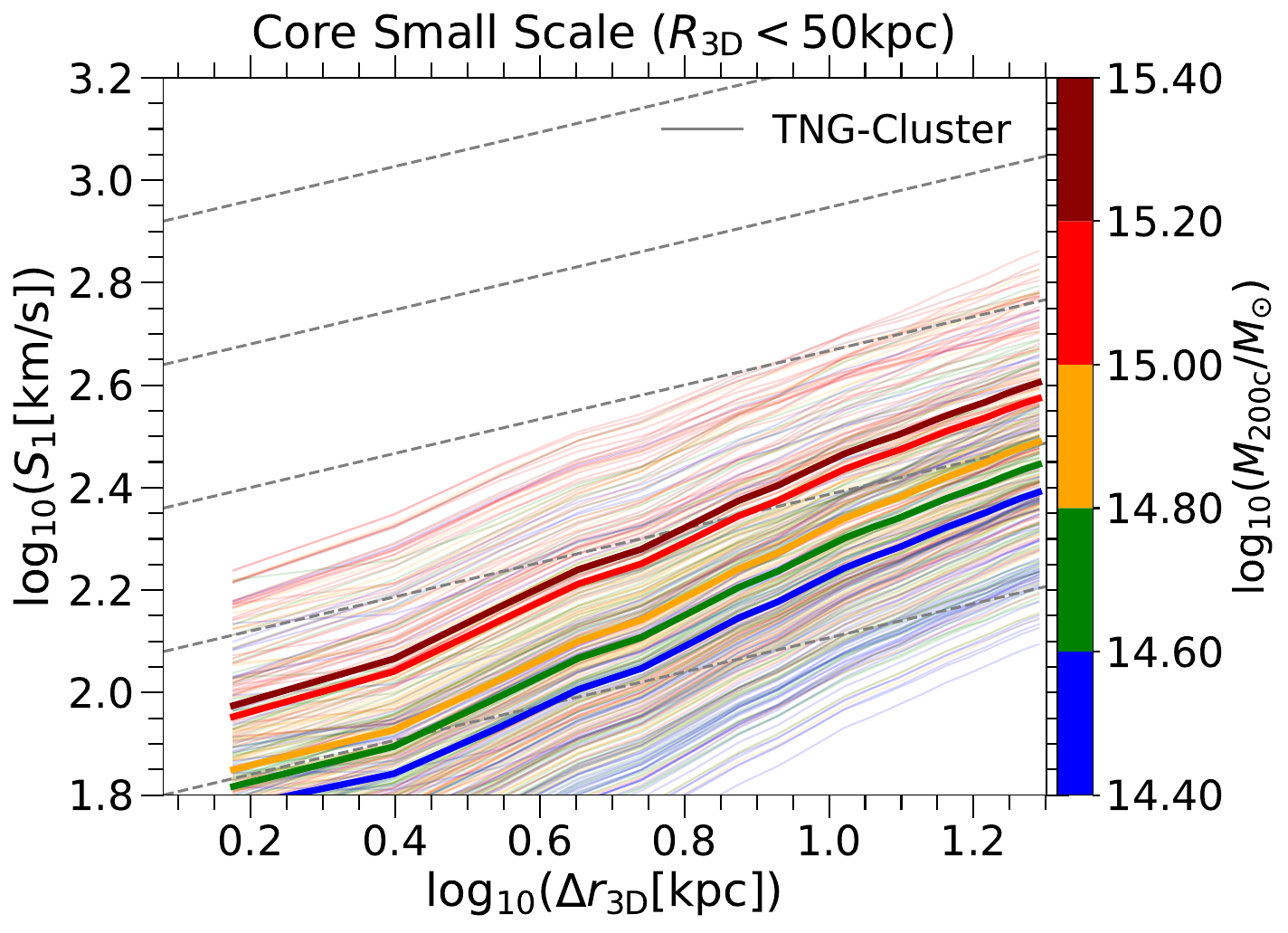}
    \includegraphics[width=0.49\textwidth]{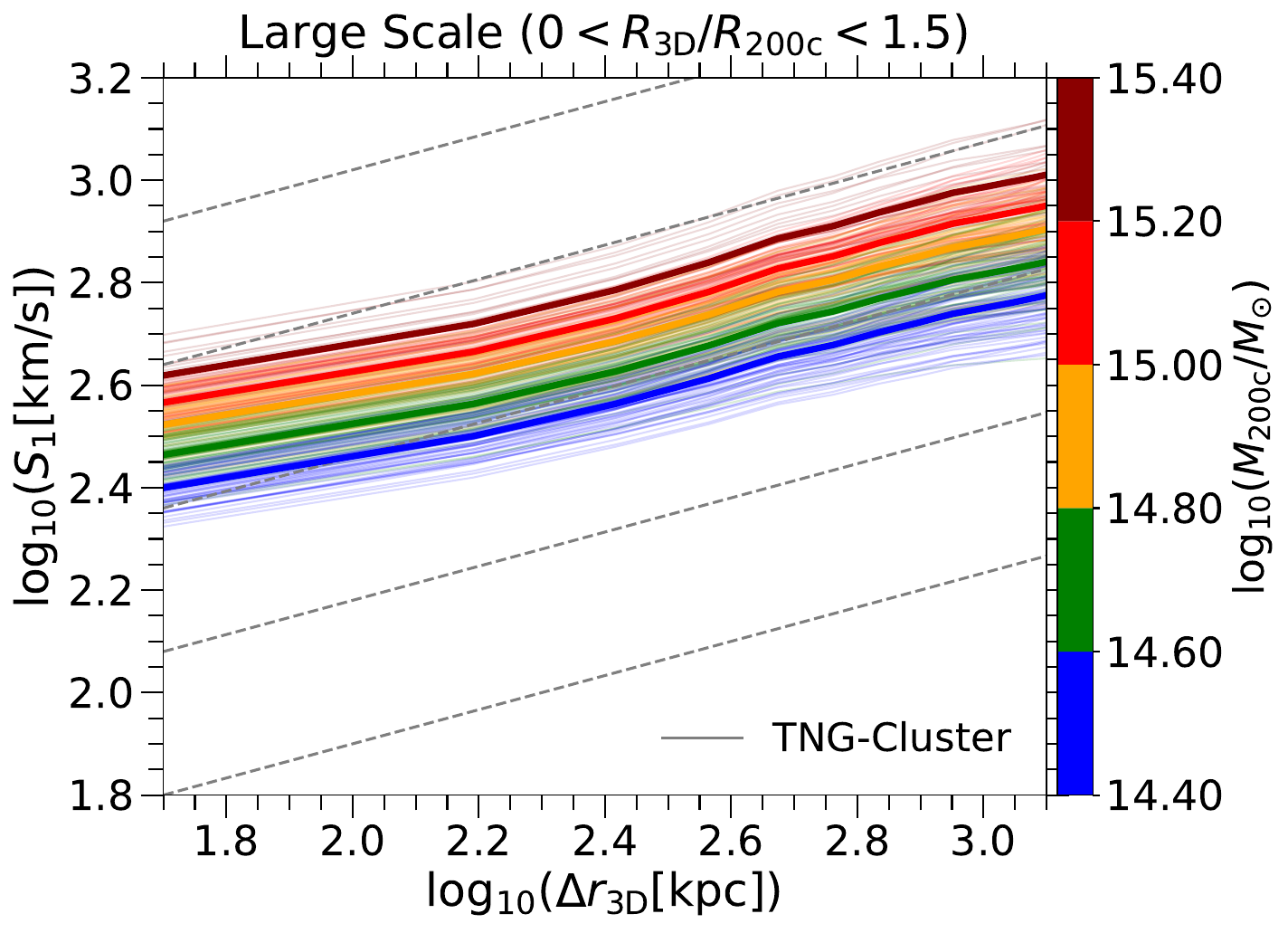}
    \includegraphics[width=0.33\textwidth]{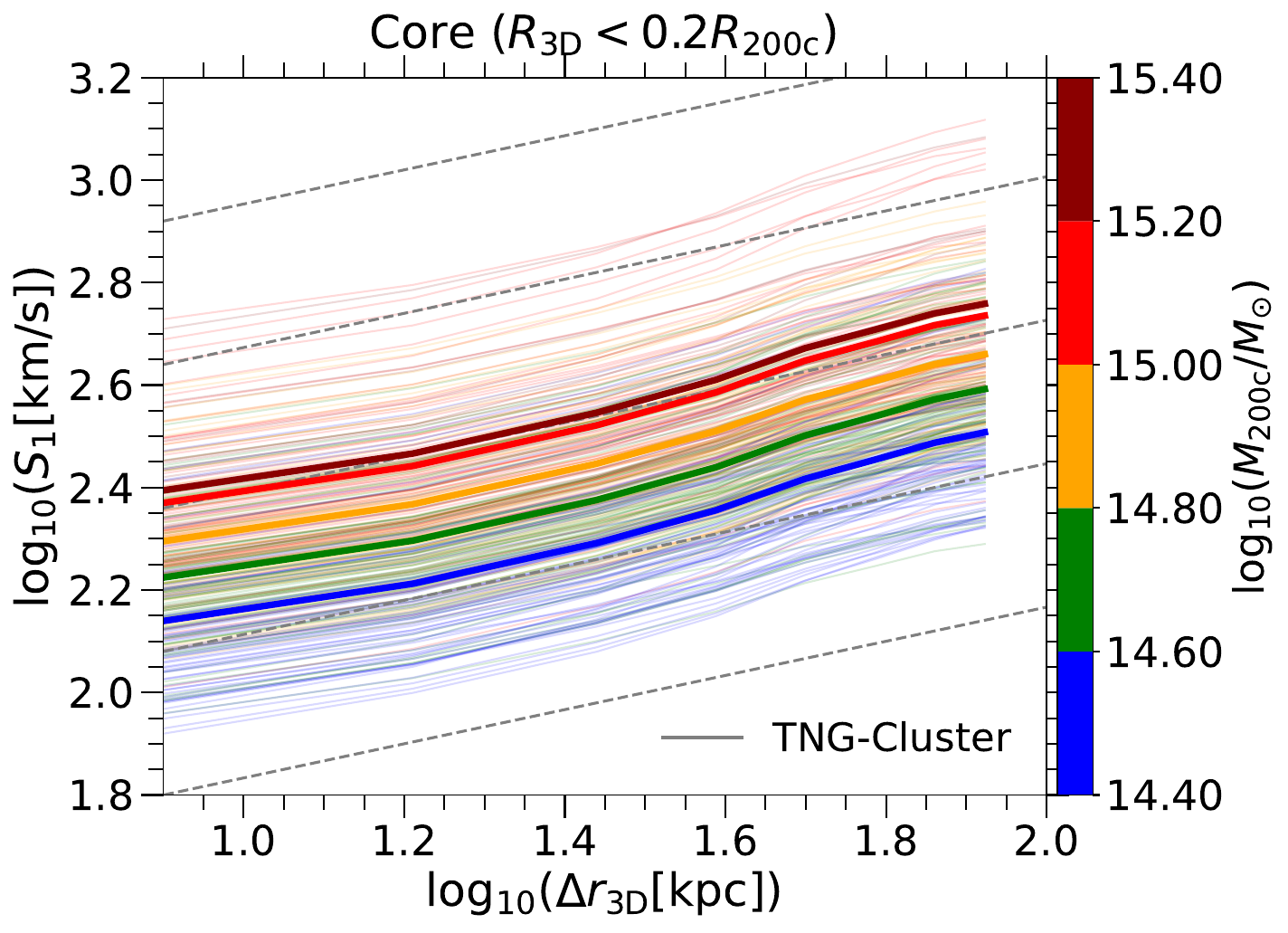}
    \includegraphics[width=0.33\textwidth]{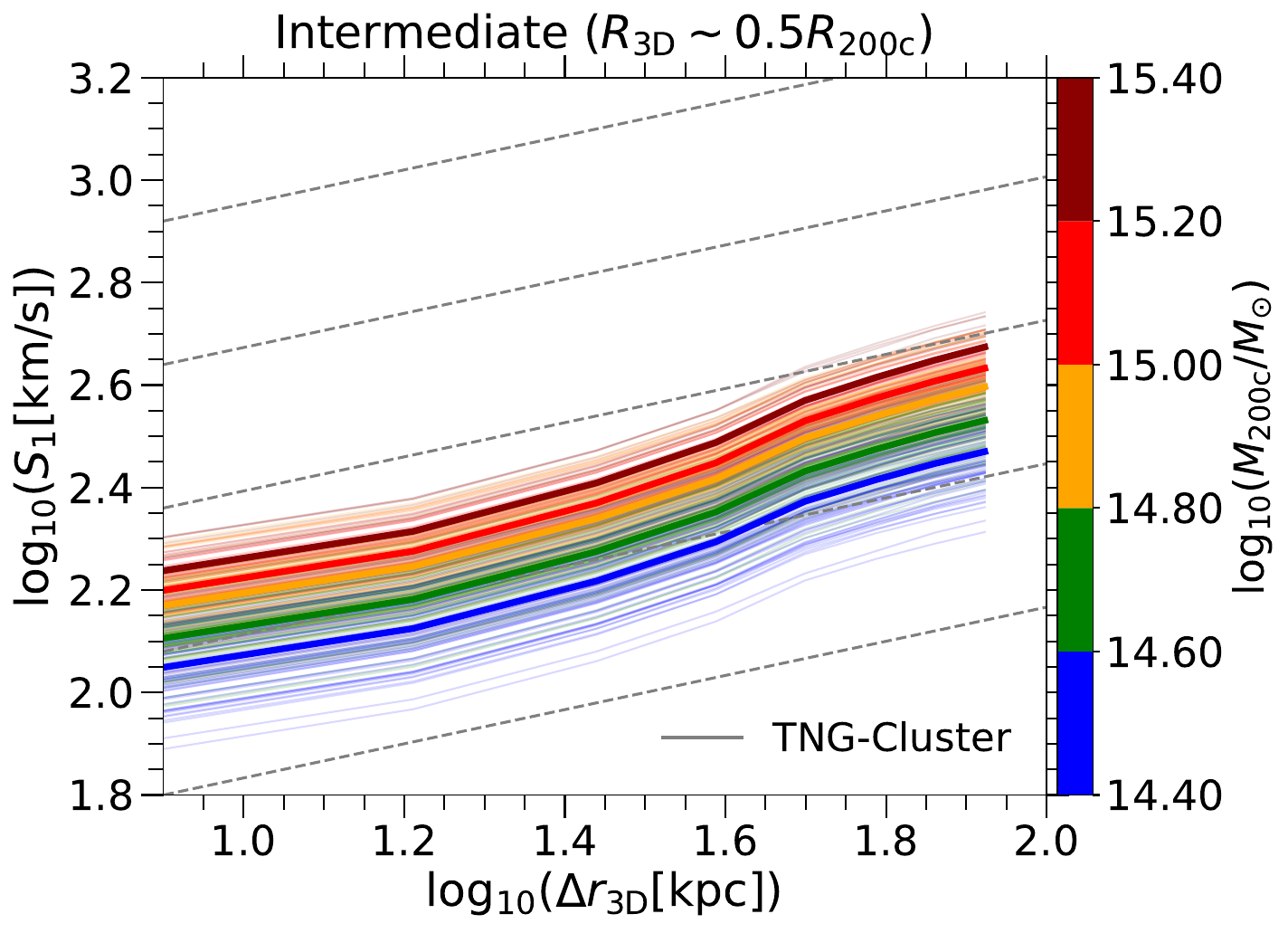}
    \includegraphics[width=0.33\textwidth]{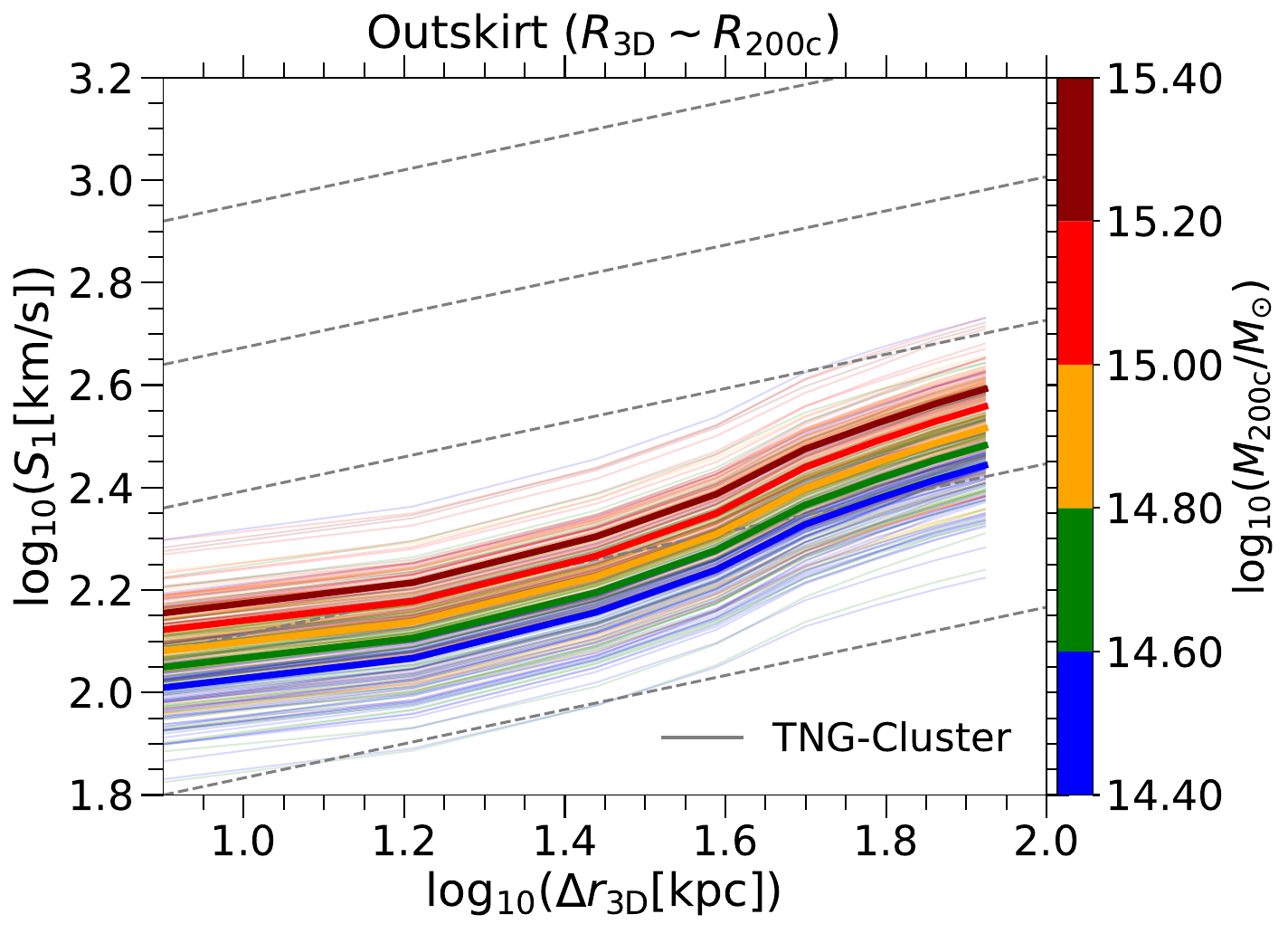}
    \caption{First-order velocity structure function (VSF) versus 3D separation distance for all TNG-Cluster halos at $z=0$. The top row shows the VSF for small separation scales ($\lesssim 10kpc$) near the cluster center (left), and larger separation scales ($100{\rm kpc}<\Delta r_{\rm 3D}<2{\rm Mpc}$) in the entire cluster and its outskirts. The bottom row shows the VSF at medium separation scales ($10{\rm kpc}<\Delta r_{\rm 3D}<100{\rm kpc}$) in the cluster core (bottom left), cluster intermediate regions (bottom center), and cluster outskirts (bottom right). Each color corresponds to a different halo mass bin. The thick solid lines depict the mean VSF, while the thin lines represent individual clusters. The diagonal lines correspond to the Kolmogorov slope of 1/3. Gas motion becomes less coherent with increasing separation distance, and is most coherent in the outskirts and least coherent in the cores of clusters, due to the combined effects of gravity and baryonic processes.}
\label{Fig: vsf1_3D}
\end{figure*}
In summary, our findings indicate that cool gas with high metallicity and hot gas with low metallicity are predominantly inflowing. On the other hand, it is challenging to discern a significant trend for outflowing gas based on either temperature or metallicity, although the outflowing gas is on average slightly warmer and has higher metallicity than the inflowing gas. This complexity underscores the intricate interplay of various factors governing gas kinematics in galaxy clusters.

\subsection{Turbulence and coherence of the ICM kinematics}
\label{subsec: velocity structure function}
To probe the scale dependence of gas kinematics in galaxy clusters, we turn our attention to the two-point statistical analysis of the velocity structure function (VSF). Fig. \ref{Fig: vsf1_3D} is an illustrative guide showing how the first-order VSF, $S_1$, captures the coherency of gas motion in TNG-Cluster at $z=0$. We now divide our halos into five zones or regions: core small scale ($R_{\rm 3D}<50\, \rm kpc$), cluster core ($R_{\rm 3D}<0.2 R_{\rm 200c}$), intermediate ($R_{\rm 3D}\sim 0.5R_{\rm 200c}$), outskirts ($R_{\rm 3D}\sim R_{\rm 200c}$), and large scale, including the entire halo and its outskirts ($R_{\rm 3D}<1.5 R_{\rm 200c}$).

For the intermediate and outskirts zones, we consider a shell thickness of 200 kpc around their respective radii. The VSF is computed for each halo (thin lines) and subsequently averaged across all halos within each mass bin (thick lines). The diagonal lines represent the \cite{Kolmogorov1941Local} slope of 1/3.\footnote{Note that the separation scales i.e. x-axis ranges differ across panels. The smallest separation scales are in the top-left panel, while the largest are in the top-right. The bottom three panels, representing the core, intermediate, and outskirts zones, maintain consistent separation scales.}

We find that across all regions of the halo, from scales of a few kpc (top left panel) to scales of a few Mpc (top right panel), the VSF increases with separation distance. This implies that the greater the distance between two parcels of gas, the less coherent their motion becomes. Additionally, a mass-dependent trend is evident. At a fixed separation distance (x-axis) and within a fixed halo zone (panels), the gas in more massive halos exhibits less coherent motion (larger $S_1$). Except for regions in close proximity to the center (top left panel), the VSF largely adheres to a power-law Kolmogorov relation with a slope of 1/3, which appears as a linear relationship in the log-log panels.

The VSF exhibits a pronounced dependence on the halo zone in which it is measured. As seen in the bottom row of panels, for a fixed separation scale and fixed halo mass (colors), the VSF is highest in the core and decreases towards the outskirts. This suggests that gas motion is more coherent in the outskirts and increasingly turbulent and chaotic towards the core. This behavior may reflect a transition from, and the interplay between, gravitational and baryonic feedback processes.

\begin{figure*}
    \centering
    \includegraphics[width=0.49\textwidth]{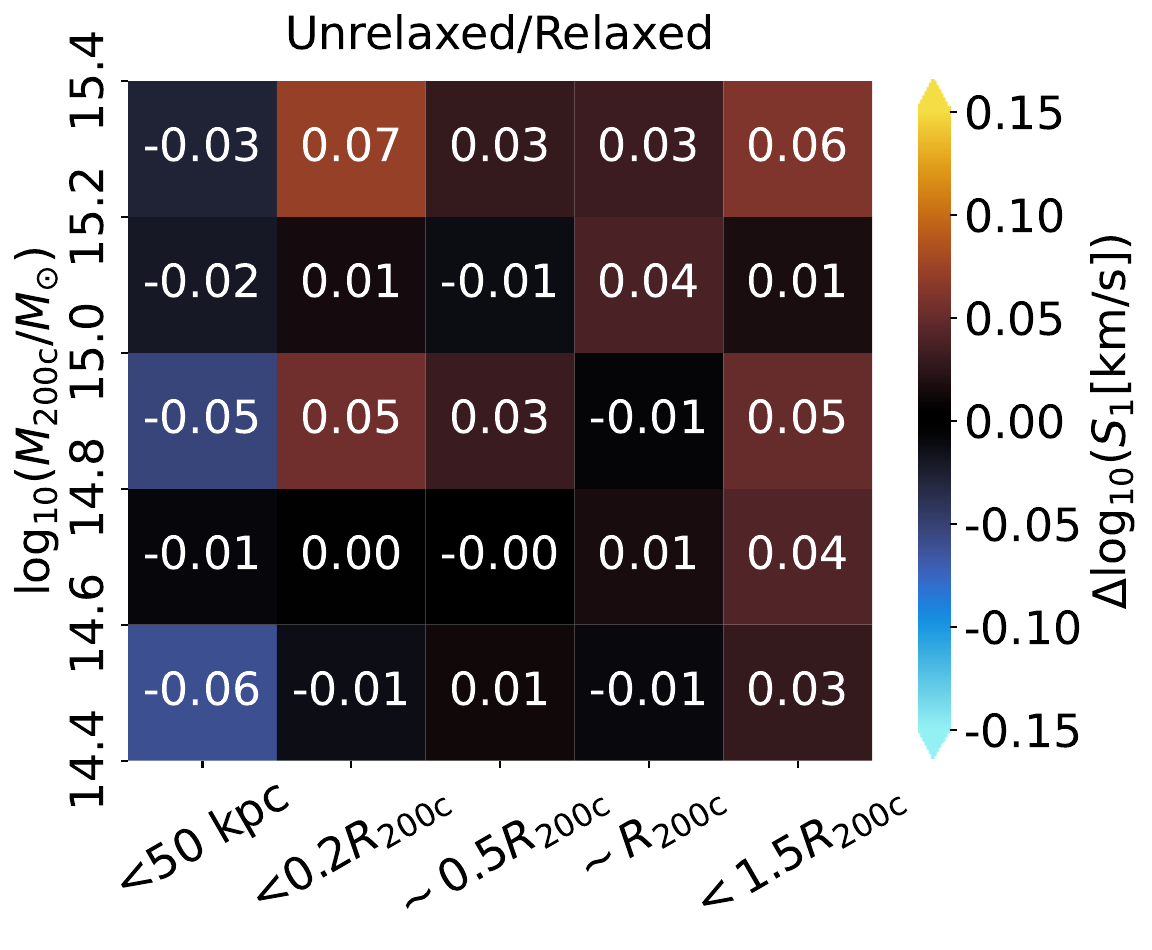}
    \includegraphics[width=0.49\textwidth]{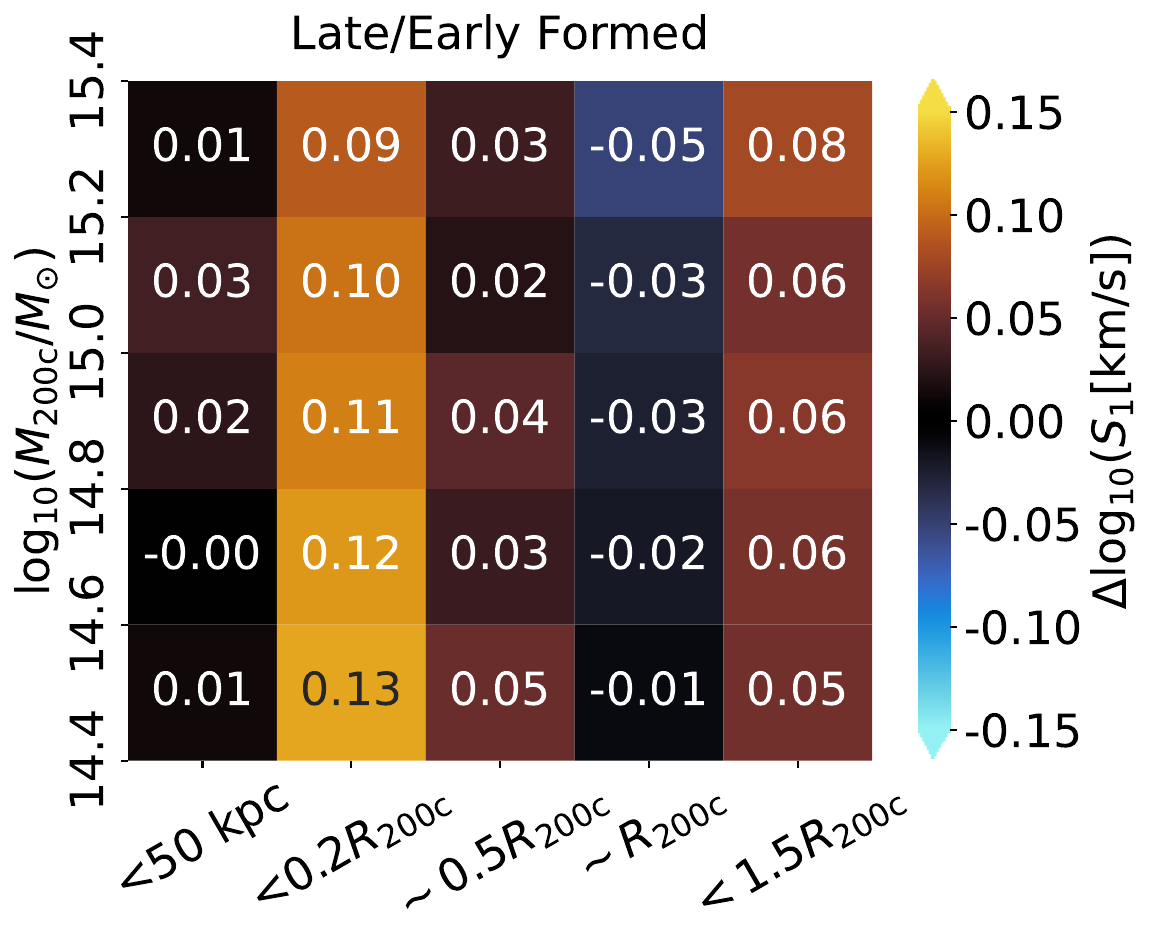}
    \includegraphics[width=0.49\textwidth]{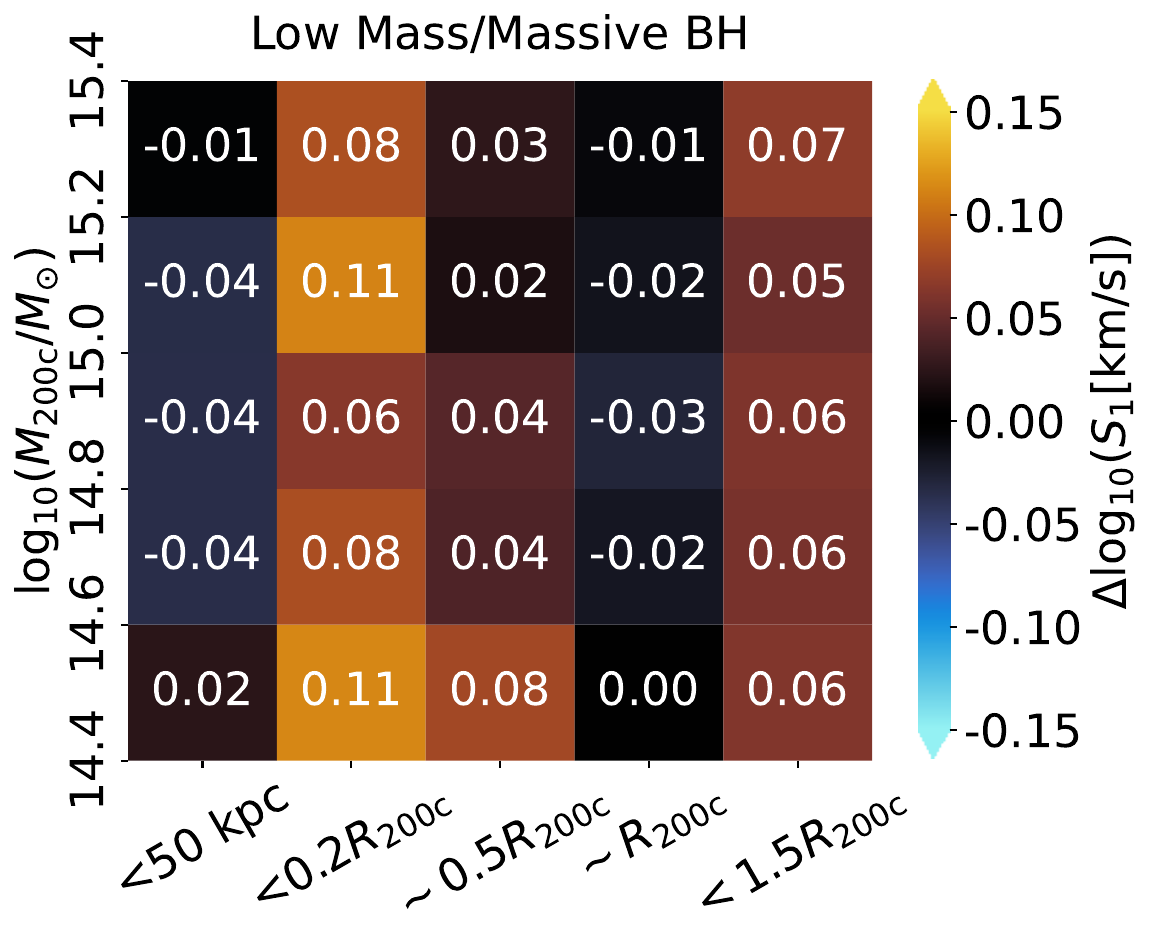}
    \includegraphics[width=0.49\textwidth]{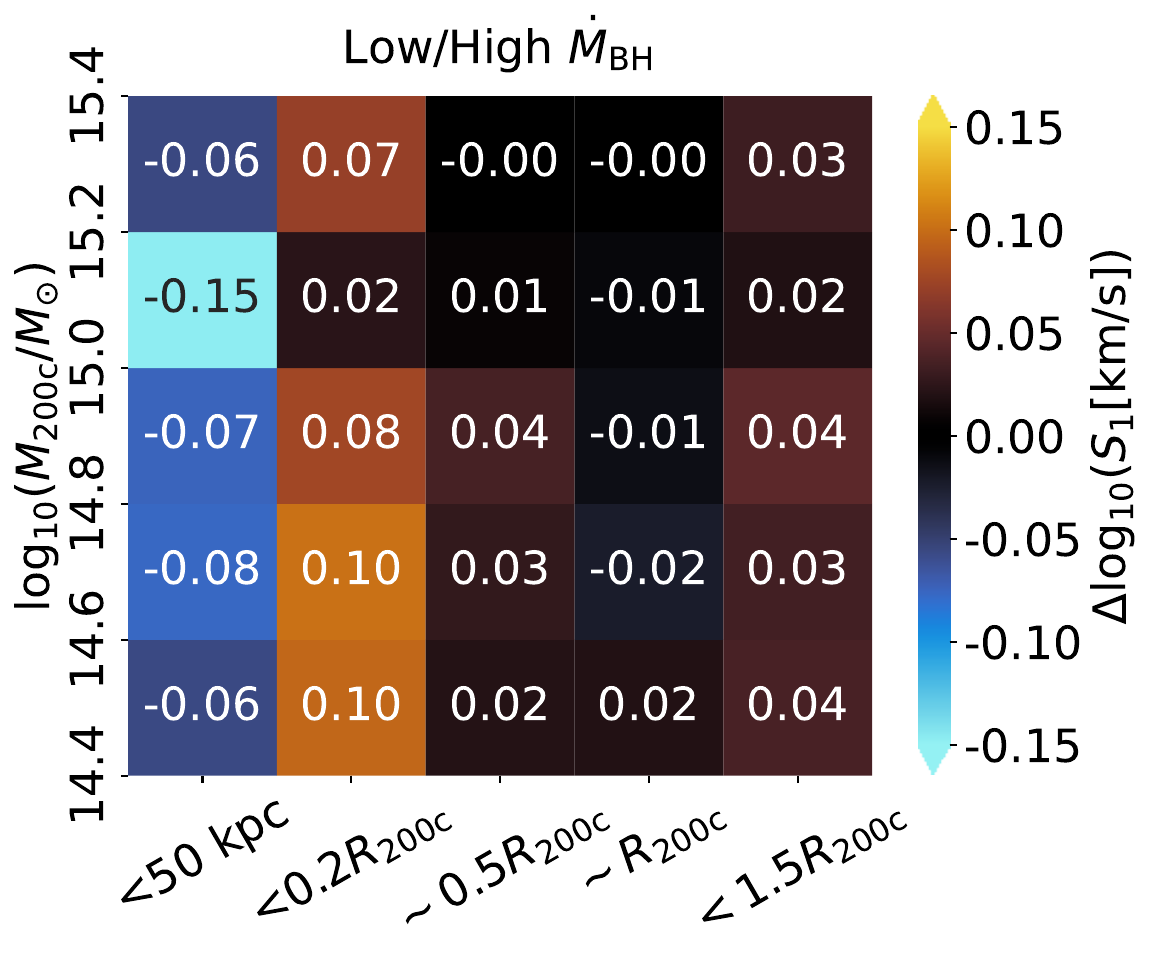}
    \caption{This figure illustrates the variations in the first-order Velocity Structure Function (VSF) across halos with distinct attributes. Each panel represents a specific property: relaxedness (top left), formation redshift (top right), SMBH mass (bottom left), and SMBH accretion rate (bottom right). The color scale quantifies the difference in VSF between relaxed and unrelaxed halos (top left), early and late-forming halos (top right), halos with high and low SMBH mass (bottom left), and halos with high and low SMBH accretion rates (bottom right). The y-axis denotes the halo mass, while the x-axis indicates the halo radial zone within which the VSF is measured. For each radial zone, the VSF is averaged over all separation scales as presented in Fig. \ref{Fig: vsf1_3D}. We note that averaging over all separation scales is justified, as we found the variation in $\Delta\log_{10}(S_{1})$ across these scales to be insignificant. The figure emphasizes that the VSF is sensitive to halo properties.}
\label{Fig: vsf1_3D_heatmap}
\end{figure*}

The velocity dispersion has the inverse trend with respect to halocentric distance (see Fig. \ref{Fig: vrad_disp}, Fig. \ref{Fig: profiles_main}). The apparent discrepancy is reconciled by the operational definitions of these two metrics. While the VSF quantifies the velocity difference between pairs of gas cells at a given relatively small separation scale, velocity dispersion indicates the deviation of velocities from the mean, computed for all gas within a specified halo zone or spherical shell. At a fixed small separation scale, the amplitude of the VSF is higher in cluster centers, which is also where the driving scale of turbulence is smaller, with AGN effects at smaller scales versus mergers at larger scales. The VSF in the outskirts, if it were to probe larger separation scales as in the upper right panel of Fig. \ref{Fig: vsf1_3D}, would show a greater amplitude on these large scales, which would align the trend with that seen in Fig. \ref{Fig: profiles_main}.

The VSF also exhibits significant halo-to-halo variability across different halo zones. Specifically, zones closer to the halo center (top left and bottom left panels) show a greater degree of halo-to-halo variation in the VSF. In contrast, in the outskirts, halos of comparable mass demonstrate more uniformity in the coherency of their gas motion. Furthermore, there are no abrupt transitions in the VSF between small and large scales, or between different regions of cluster halos. This is evident when comparing the VSF at a fixed separation scale (e.g., $\log_{10}\Delta r_{\rm 3D}\sim 1.8$) between the top right panel and the panels in the bottom row.

In Fig. \ref{Fig: vsf1_3D_heatmap}, we examine how the VSF varies based on halo properties: relaxedness, formation redshift, SMBH mass, and SMBH accretion rate, as in our previous analyses. Each panel shows a specific halo attribute. The color scale quantifies the VSF difference between two sub-populations of clusters, classified as: relaxed vs. unrelaxed halos (top left), early vs. late-forming halos (top right), high vs. low SMBH mass (bottom left), and high vs. low SMBH accretion rates (bottom right).

In each panel, the y-axis represents halo mass, and the x-axis specifies the spatial region where the VSF is calculated. For each region we average the VSF over all separation scales, adopting this choice because the variation in $\Delta\log_{10}(S_{1})$ across different scales is negligible. This allows us to summarize the difference between two VSFs by a single number, as given in each colored box. In the following discussion we focus on the most significant correlations present.

\begin{figure*}
    \centering
    \includegraphics[width=0.495\textwidth]{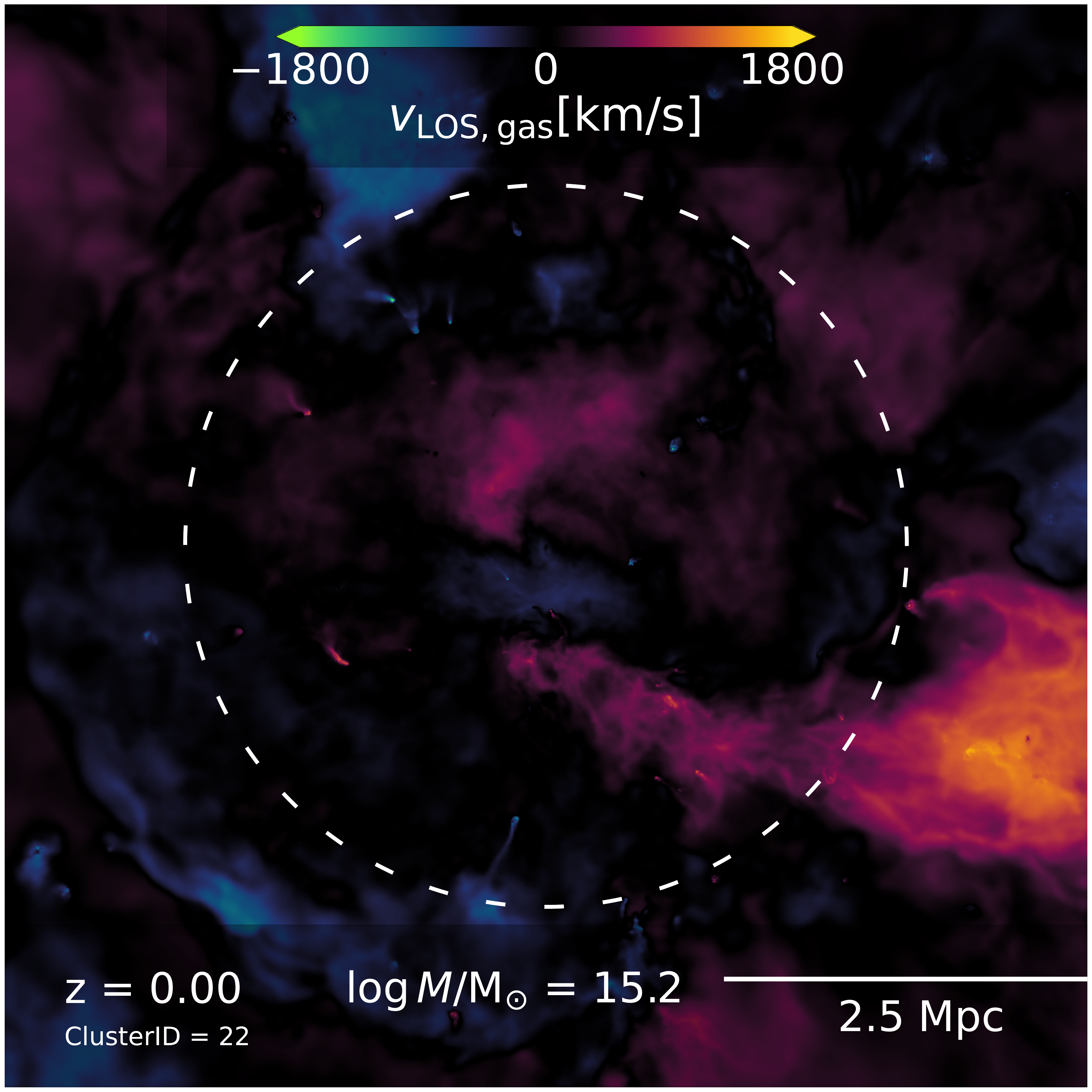}
    \includegraphics[width=0.495\textwidth]{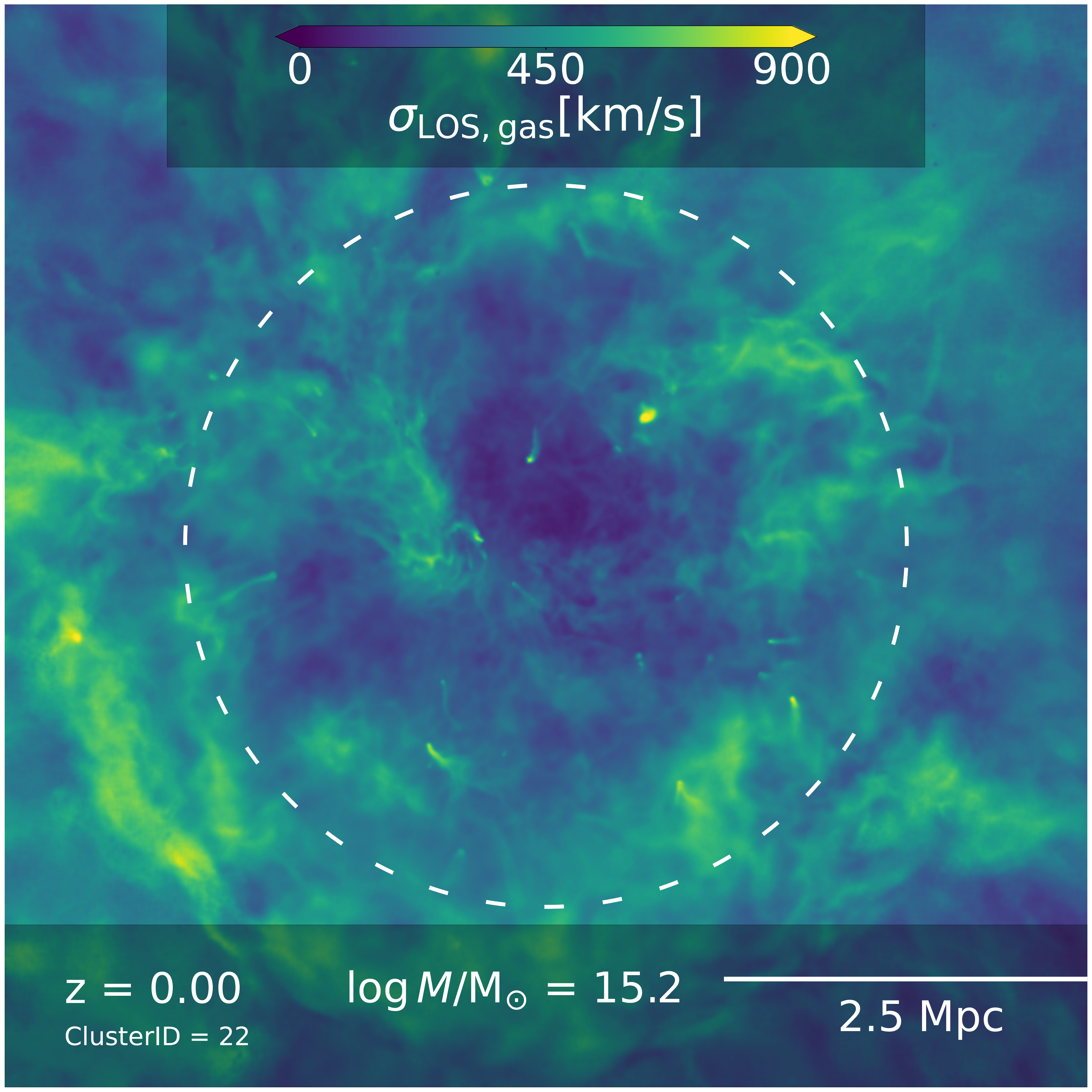}
    \caption{Maps of observable kinematics for the same halo as in Fig. \ref{Fig: vrad_box}. The left panel shows line of sight velocity, while the right panel shows line of sight velocity dispersion. Both are weighted by X-ray luminosity, such that they preferentially trace the flows of hot gas, as would be characterized by X-ray observations. The halo $R_{\rm 200c}$ is indicated by the white circles.}
\label{Fig: vlos_box}
\end{figure*}

\begin{figure*}
    \centering
    \includegraphics[width=0.24\textwidth]{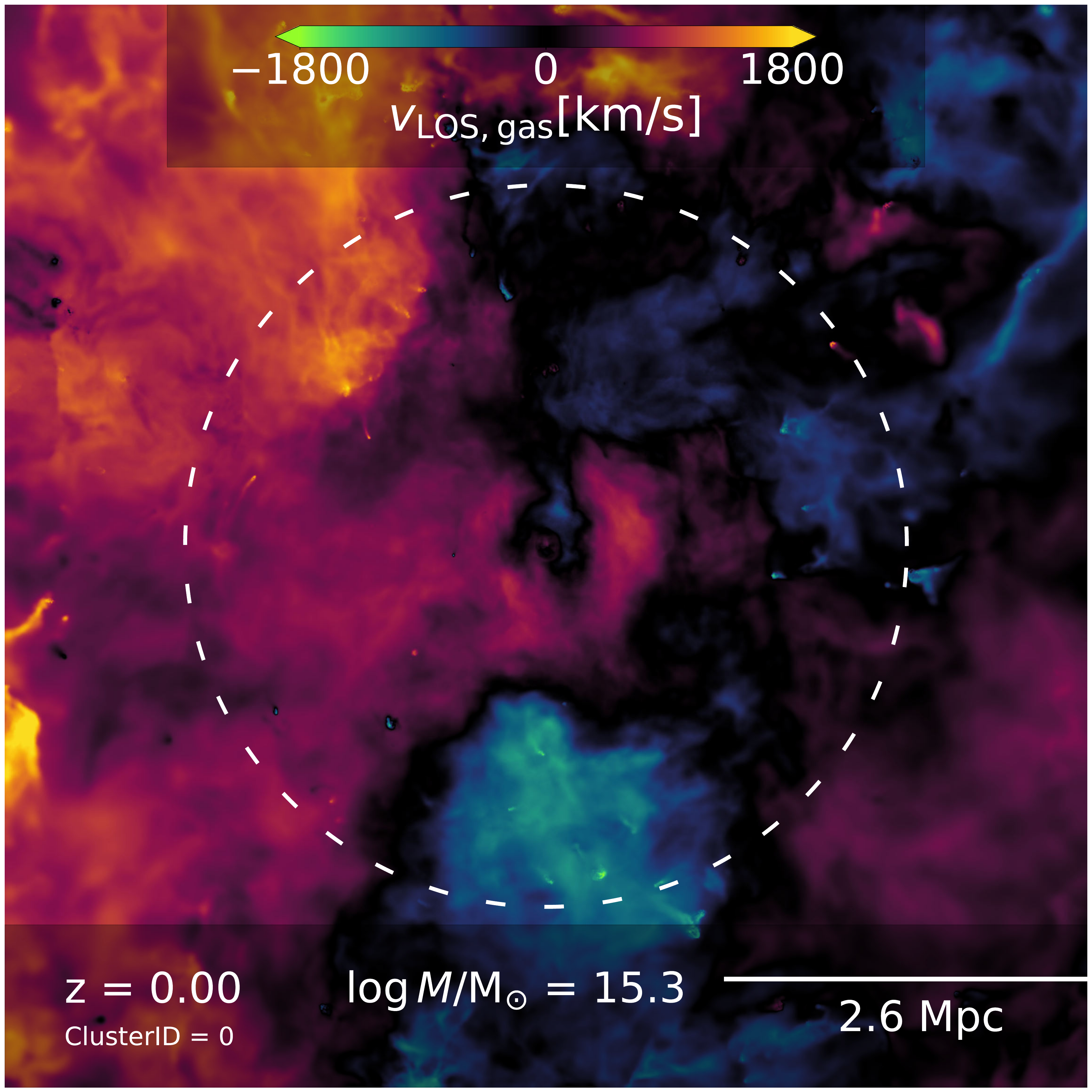}
    \includegraphics[width=0.24\textwidth]{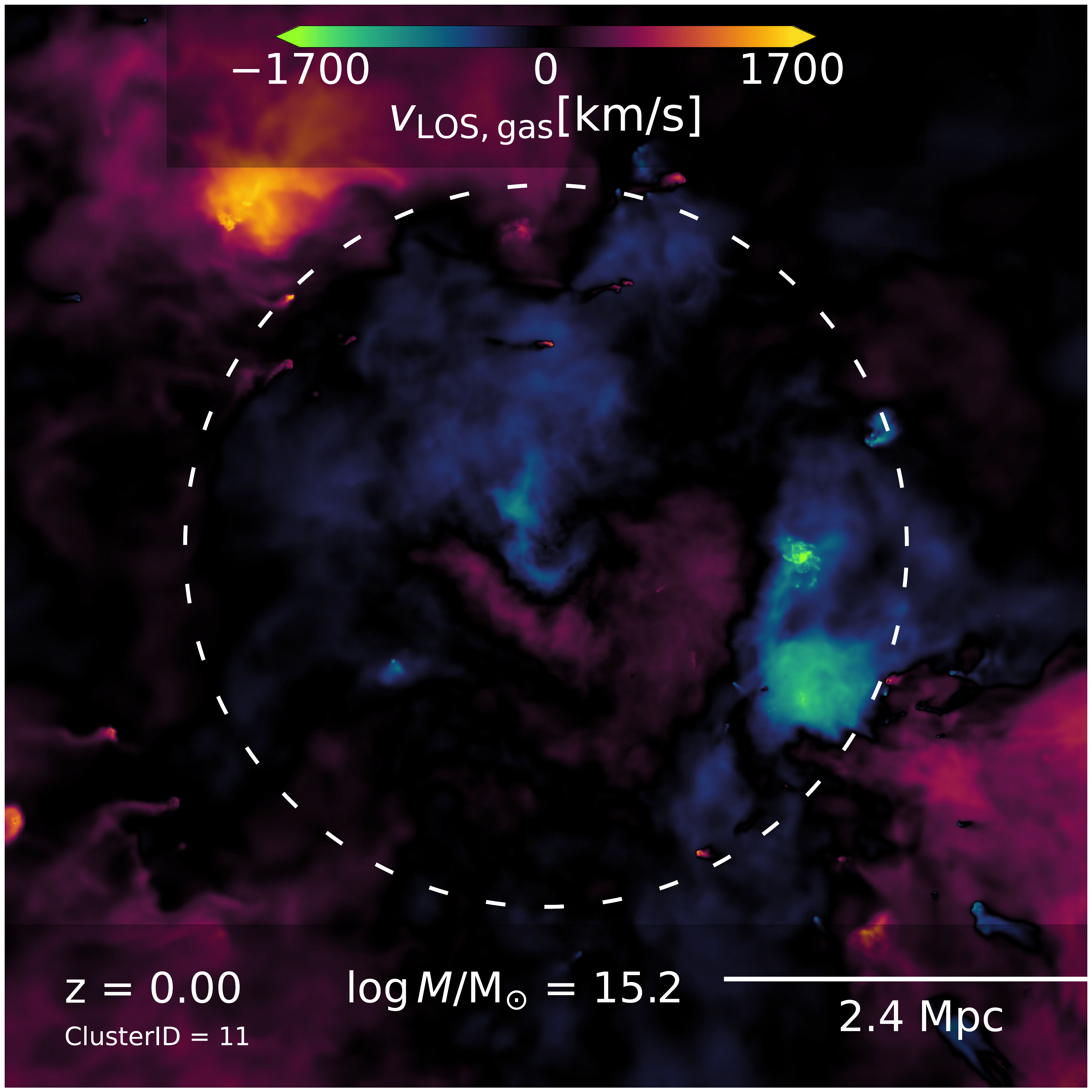}
    \includegraphics[width=0.24\textwidth]{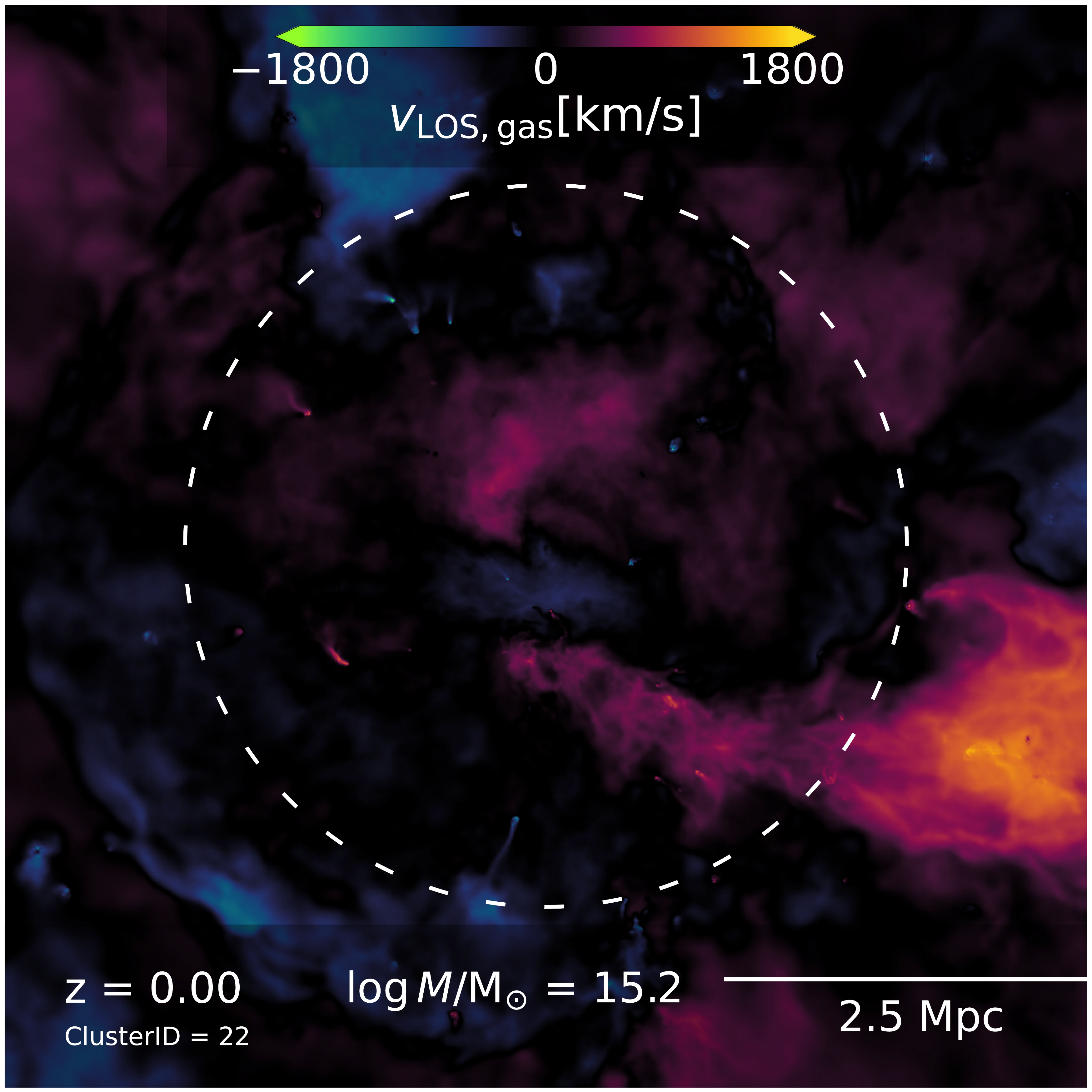}
    \includegraphics[width=0.24\textwidth]{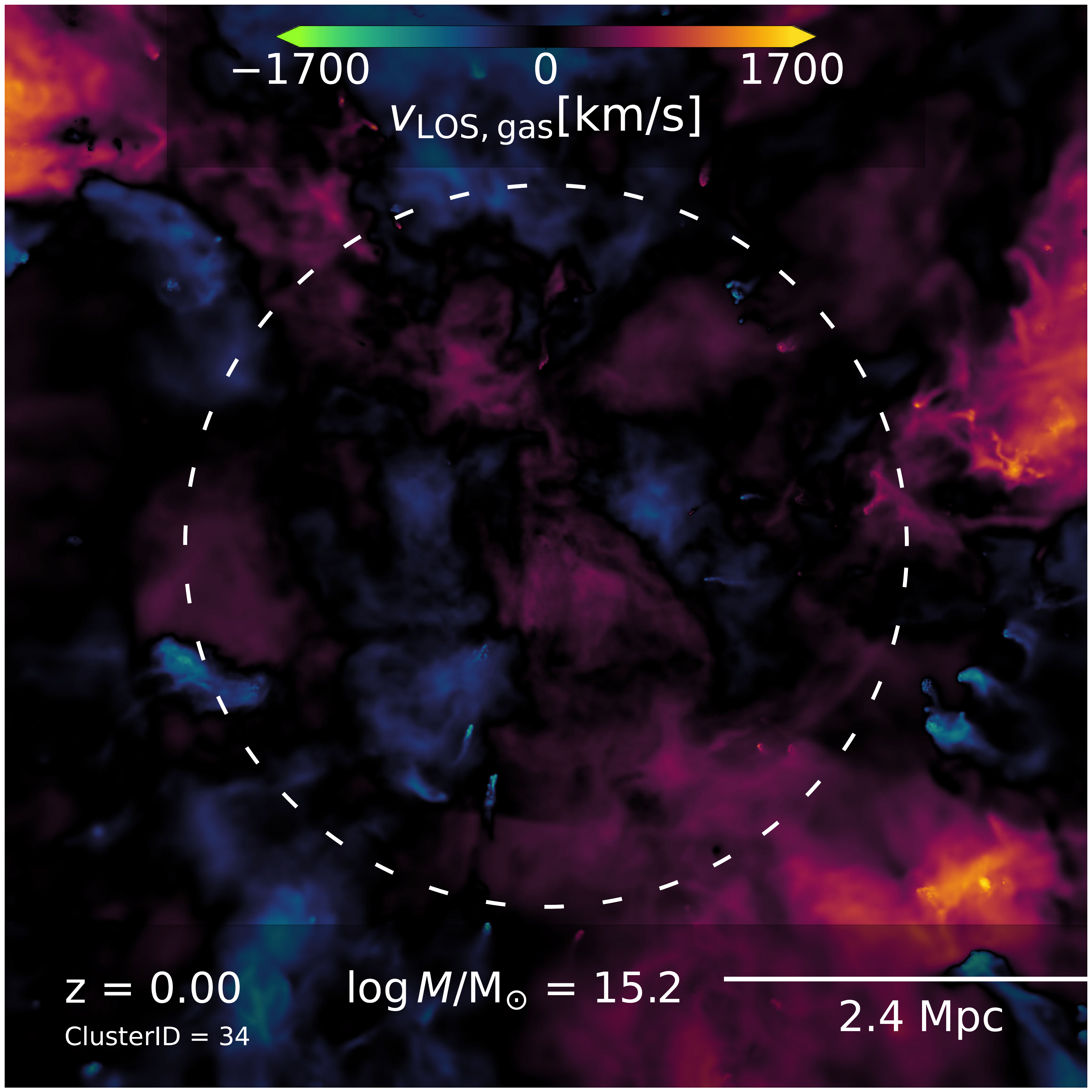}

    \includegraphics[width=0.24\textwidth]{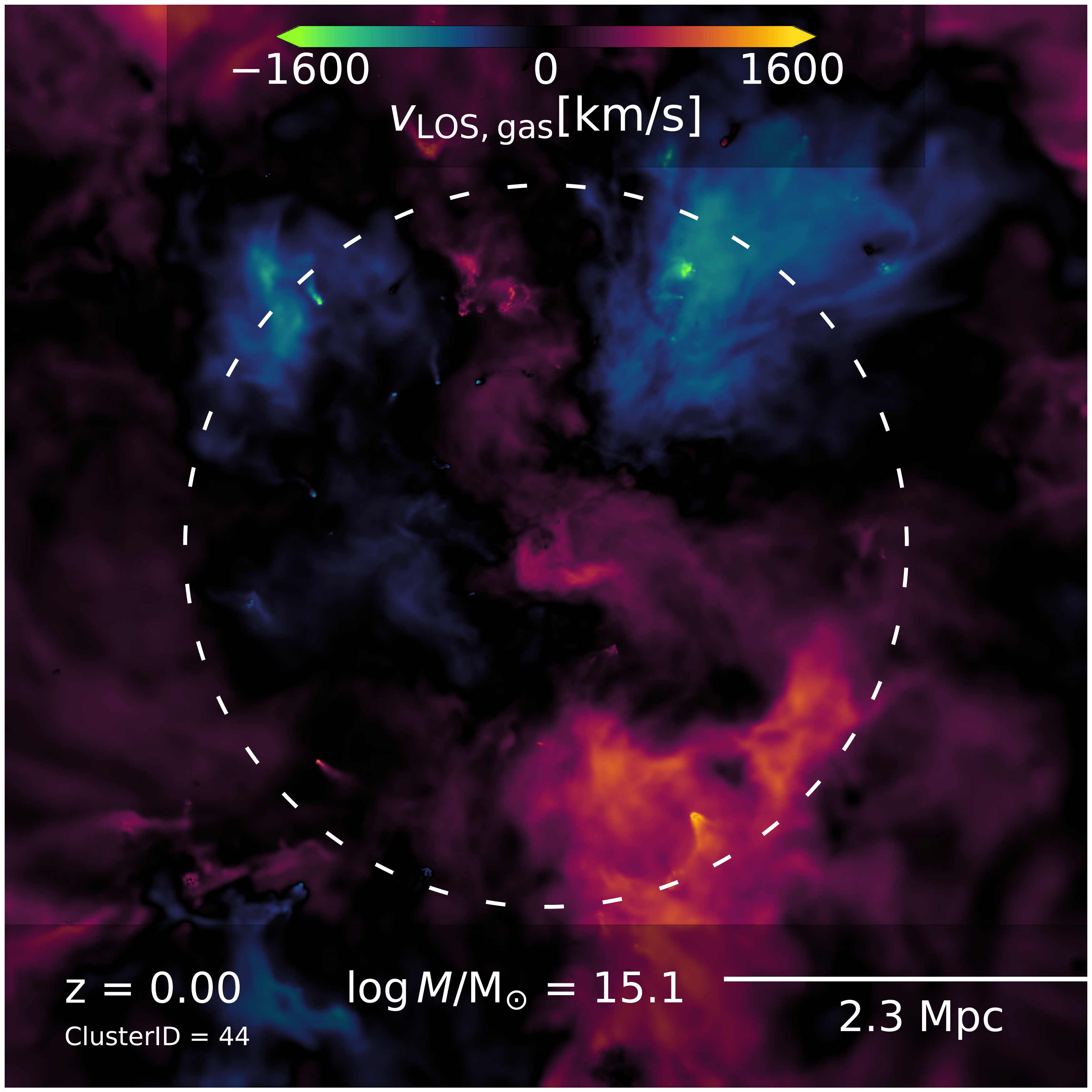}
    \includegraphics[width=0.24\textwidth]{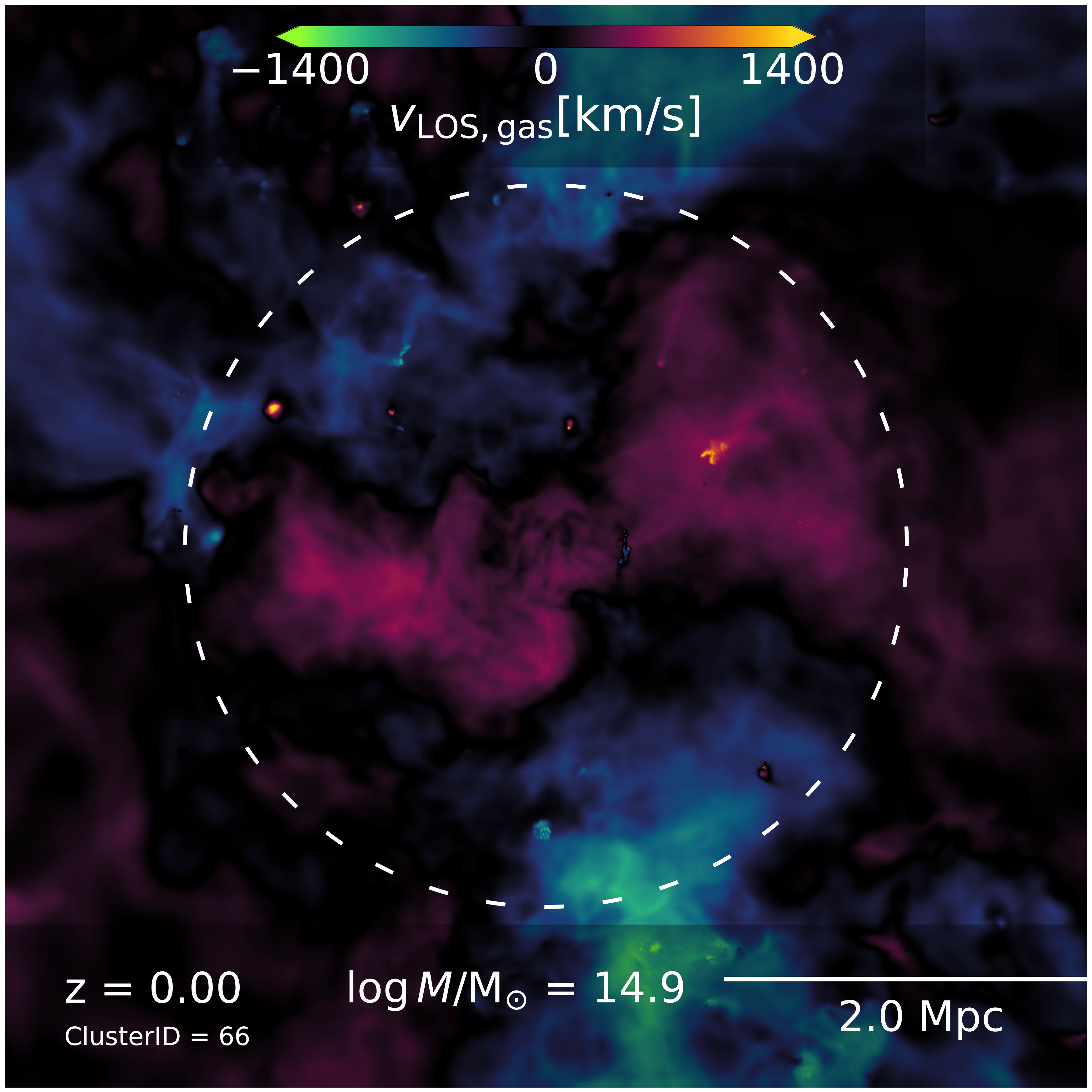}
    \includegraphics[width=0.24\textwidth]{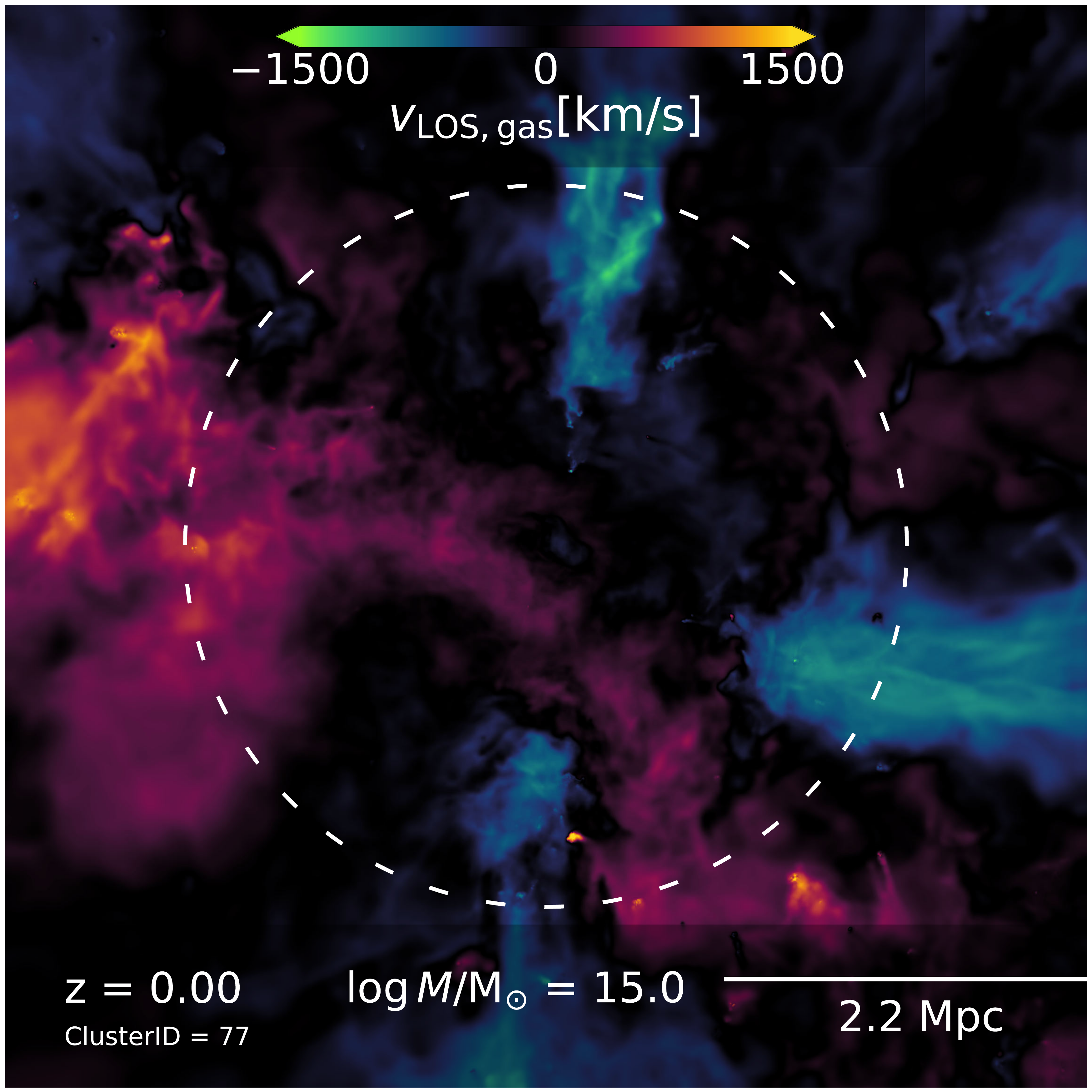}
    \includegraphics[width=0.24\textwidth]{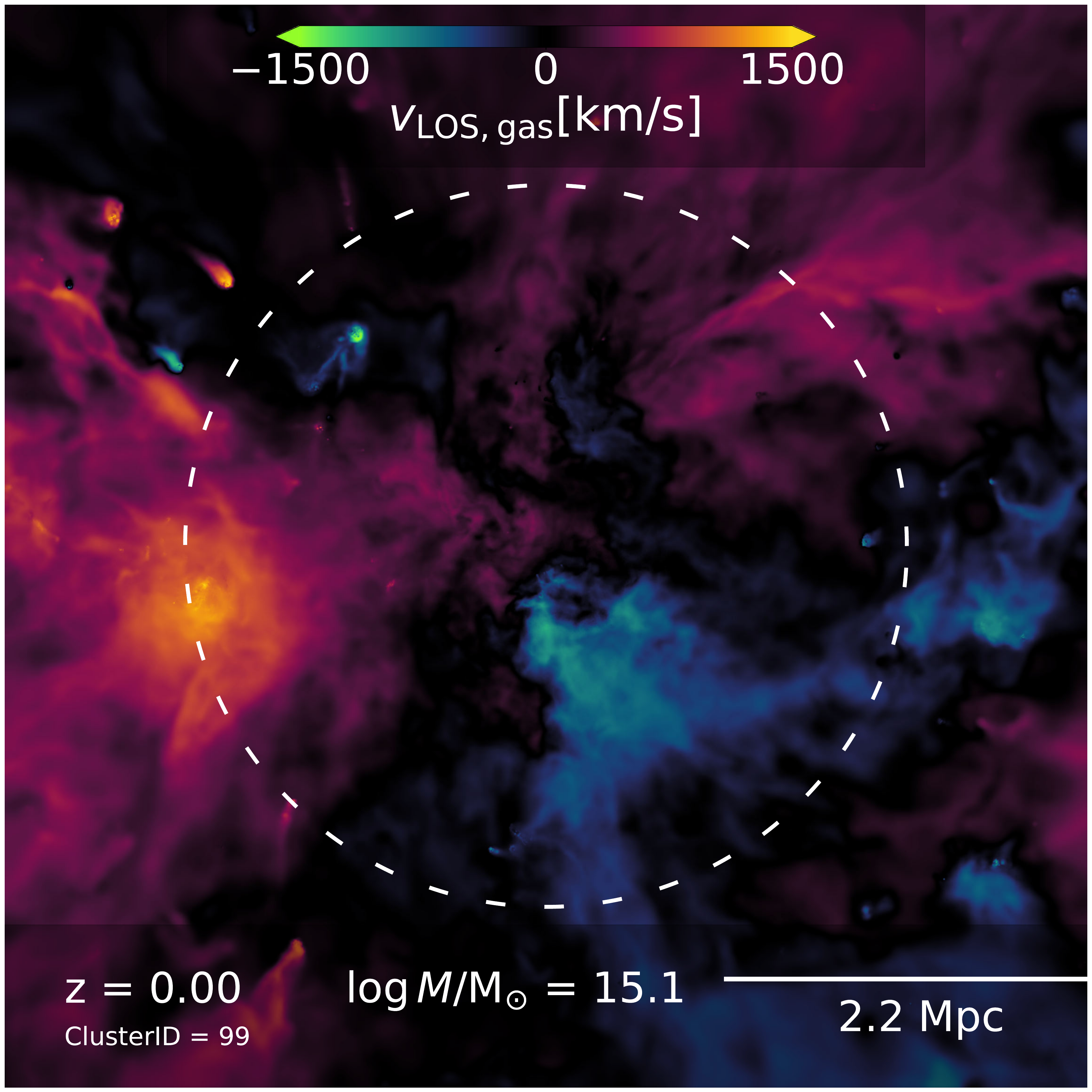}

    \includegraphics[width=0.24\textwidth]{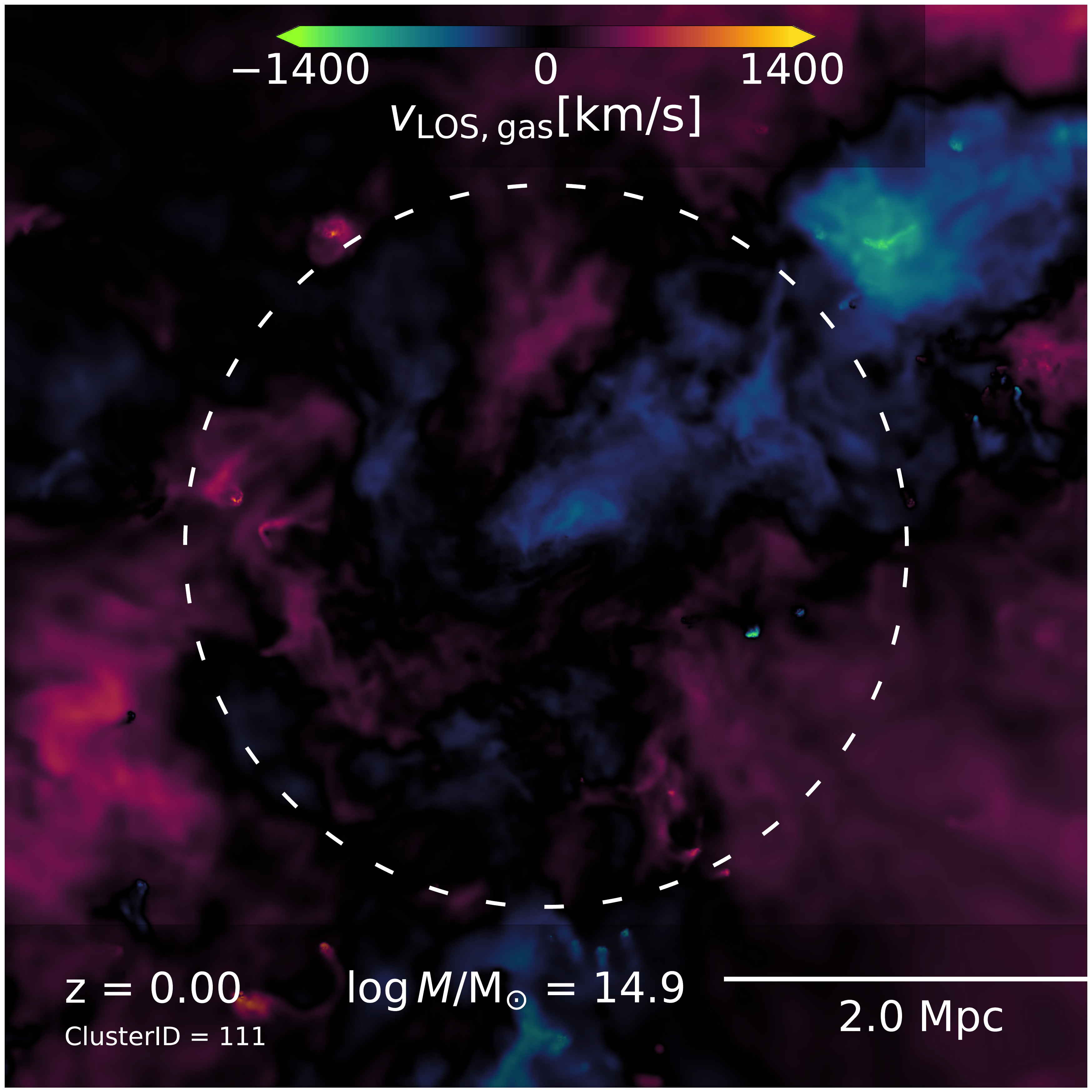}
    \includegraphics[width=0.24\textwidth]{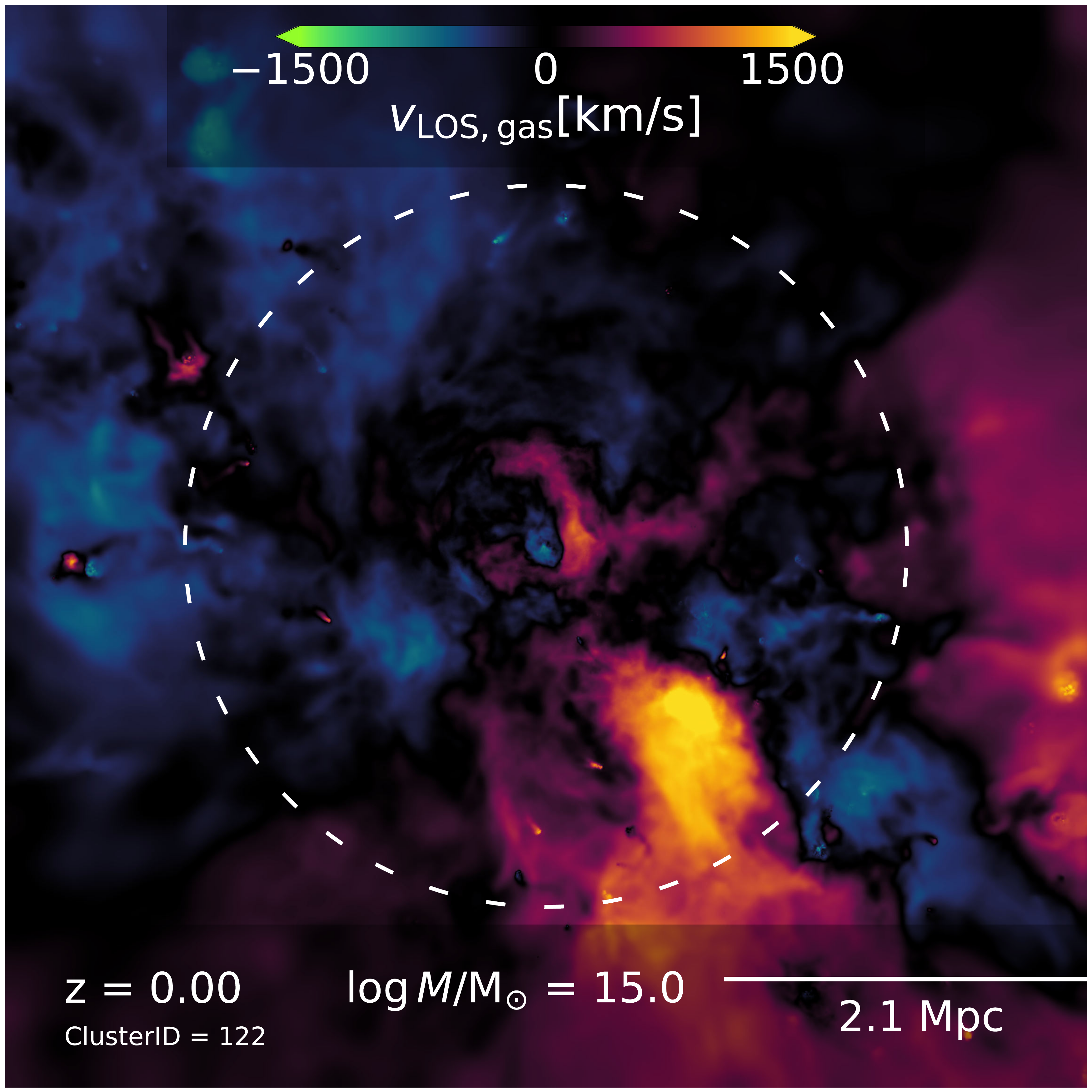}
    \includegraphics[width=0.24\textwidth]{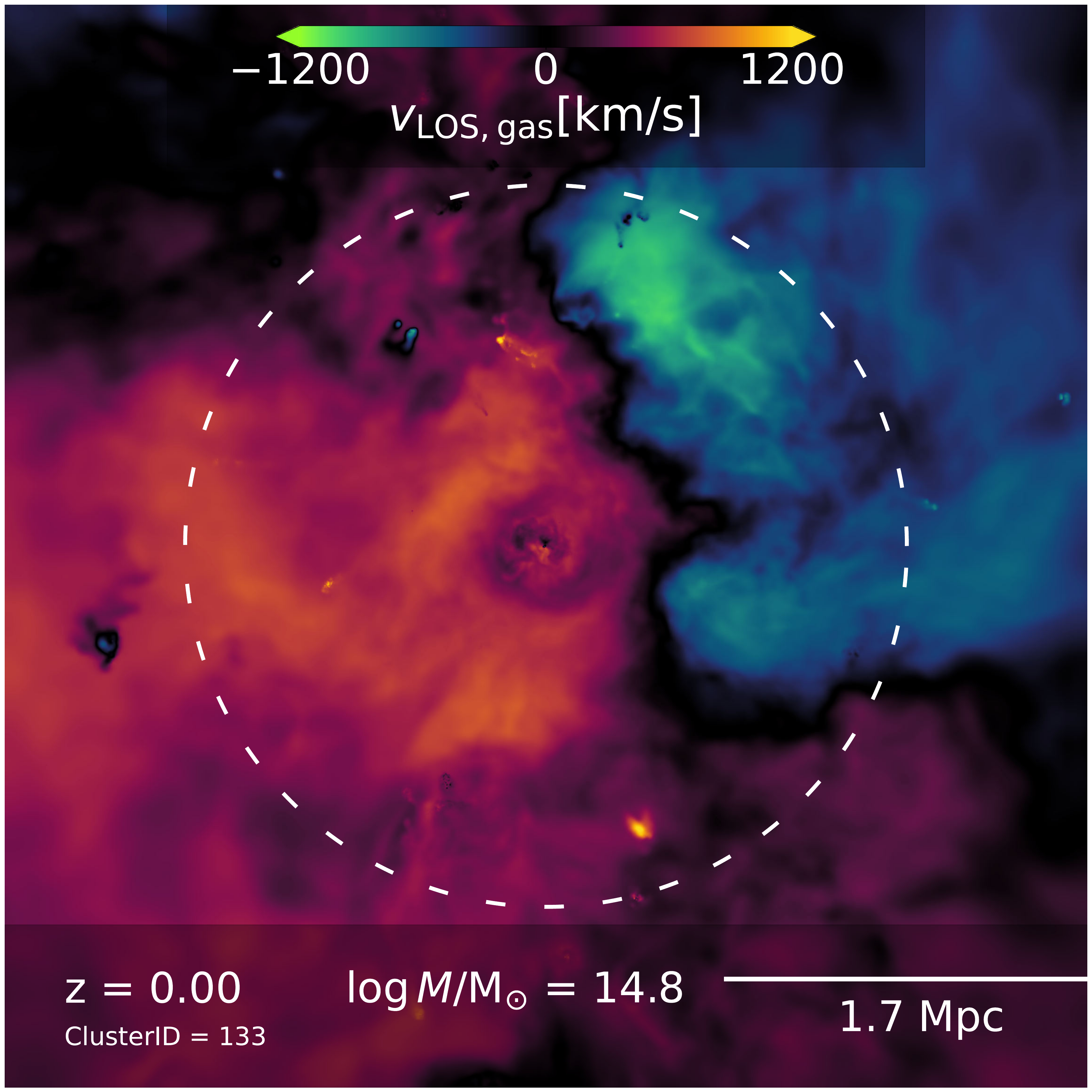}
    \includegraphics[width=0.24\textwidth]{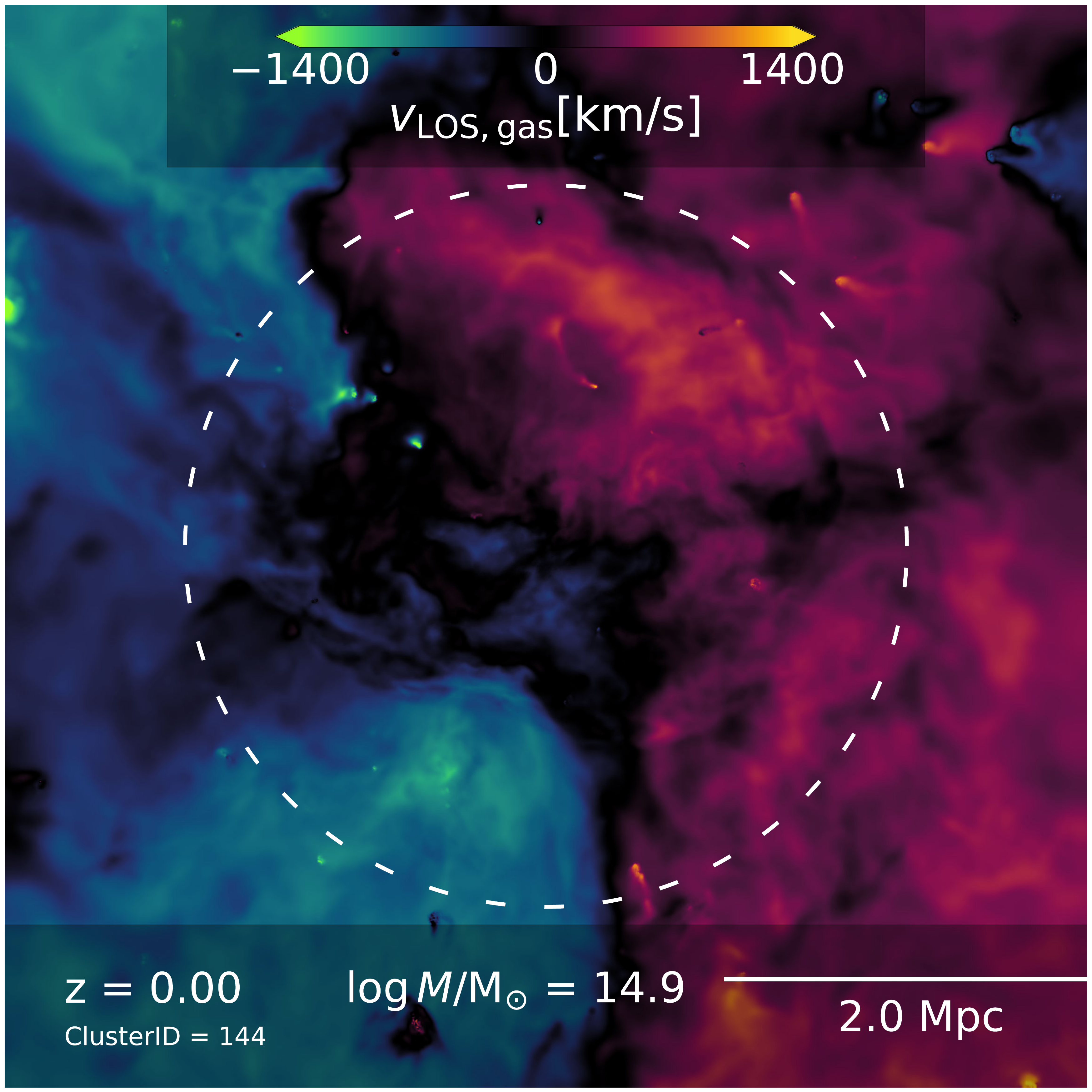}

    \includegraphics[width=0.24\textwidth]{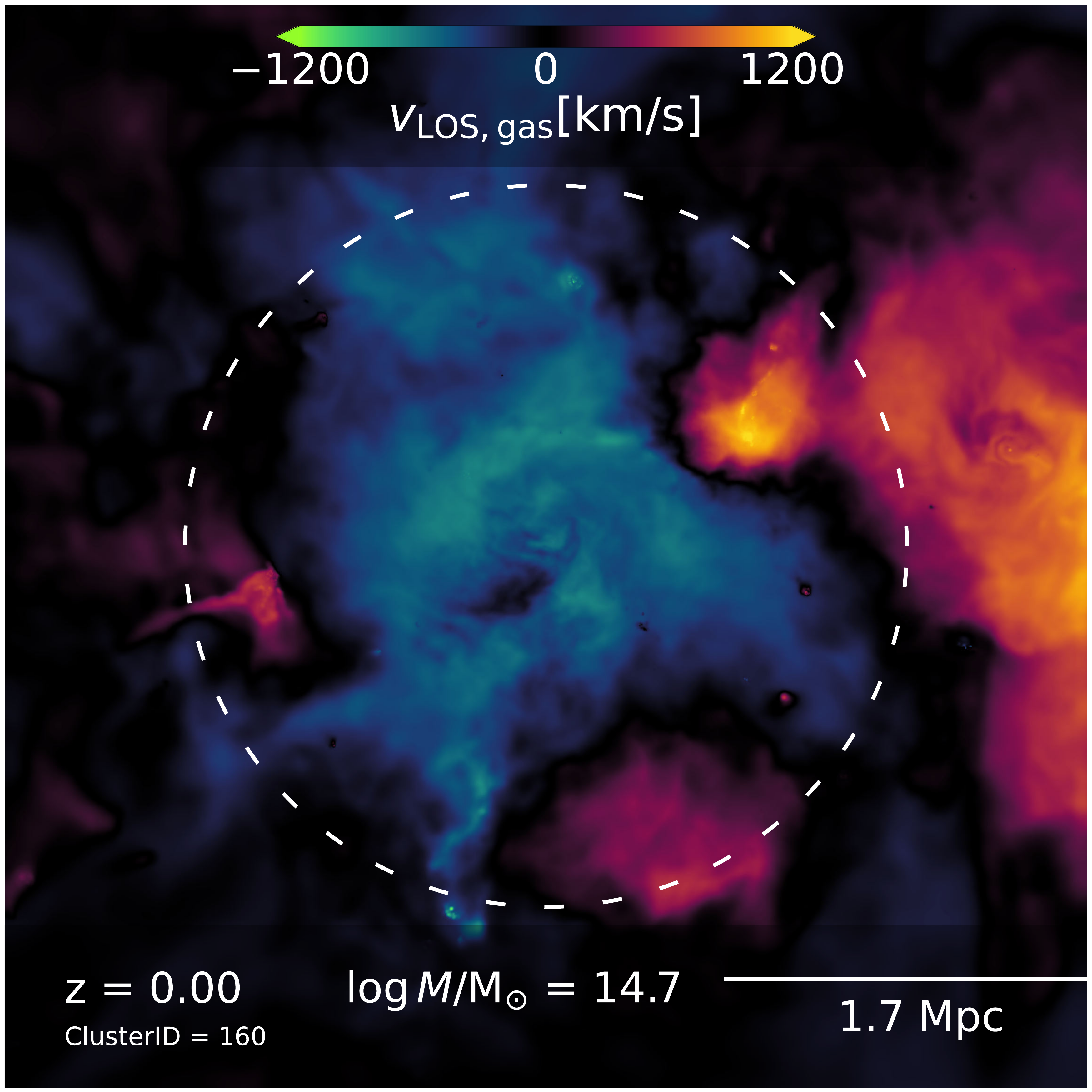}
    \includegraphics[width=0.24\textwidth]{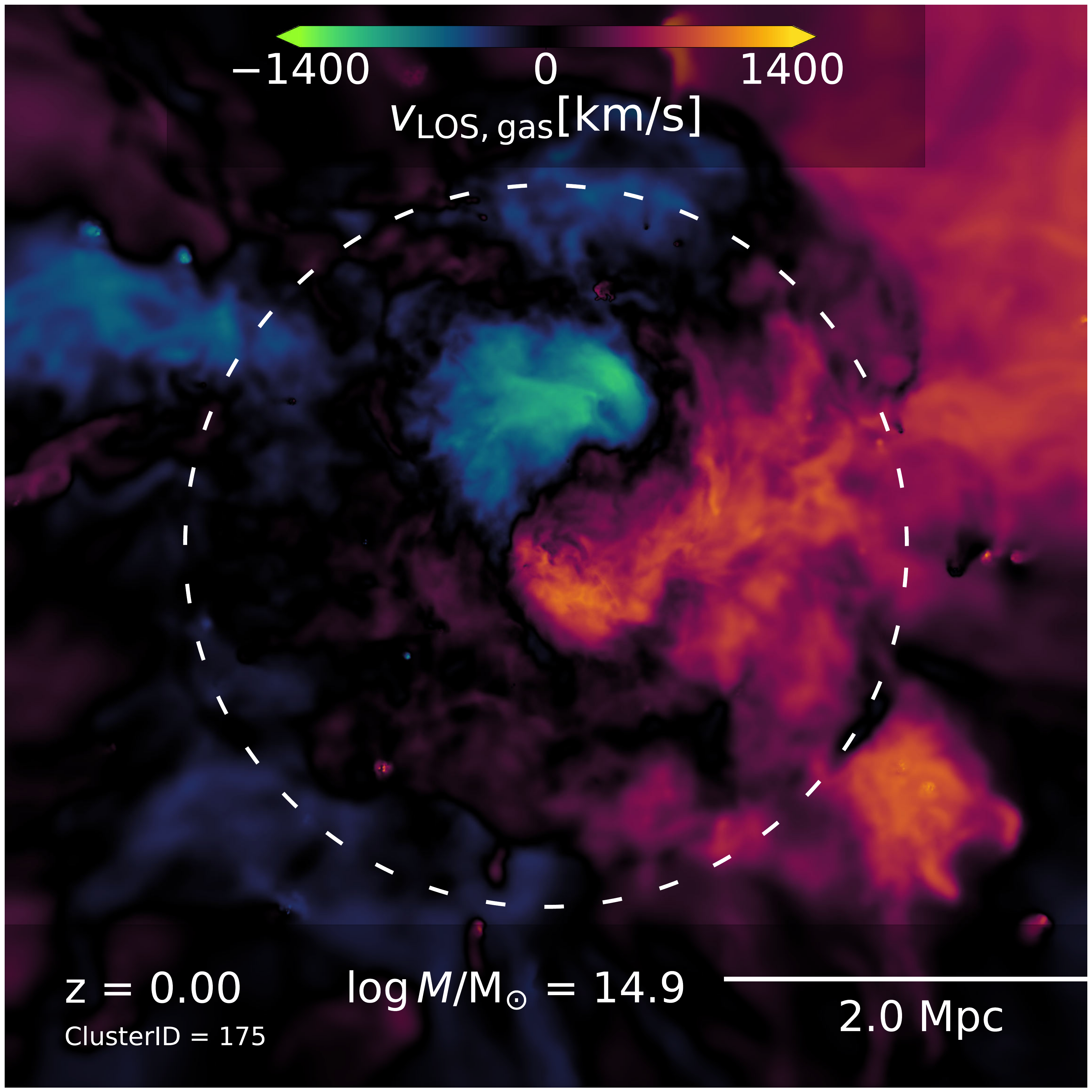}
    \includegraphics[width=0.24\textwidth]{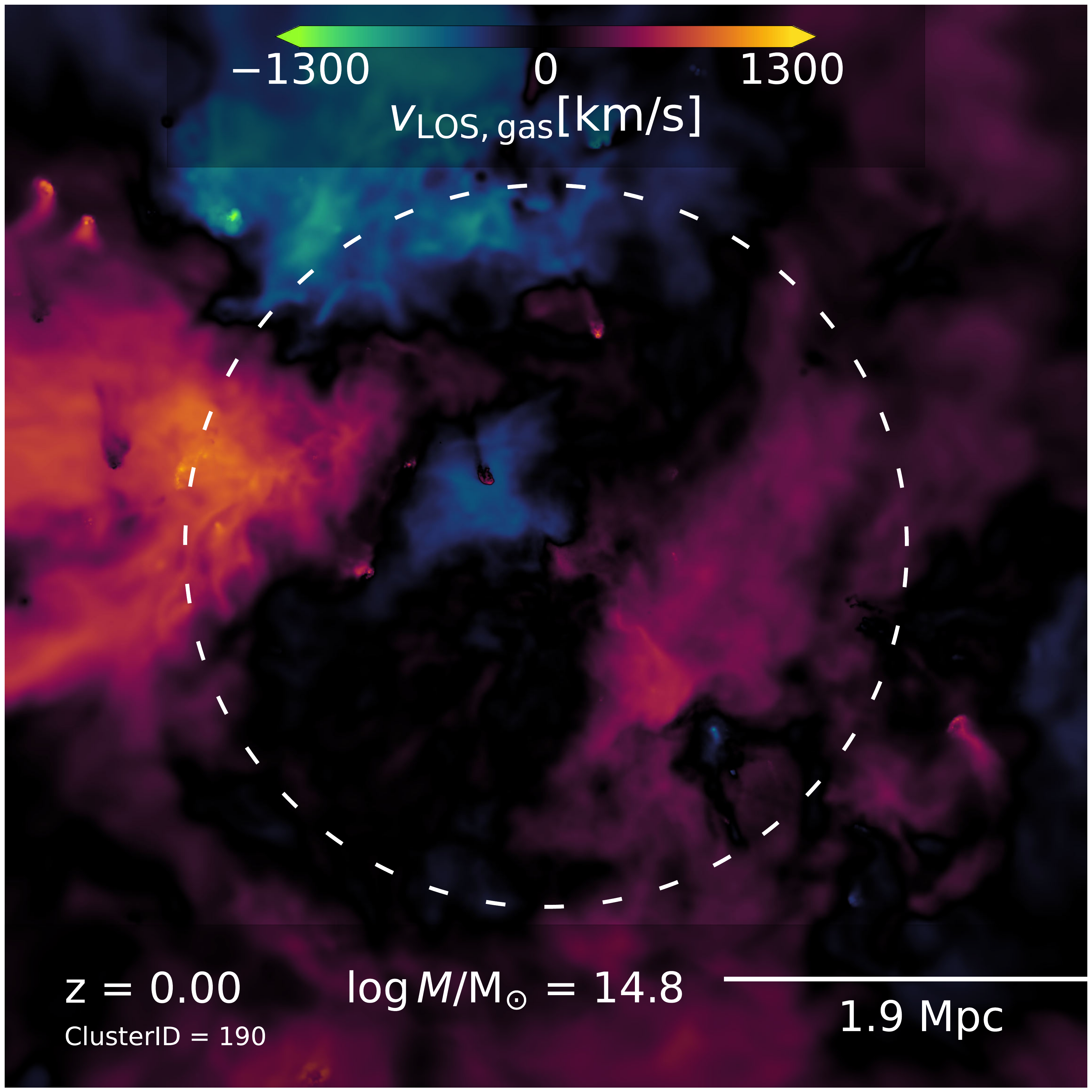}
    \includegraphics[width=0.24\textwidth]{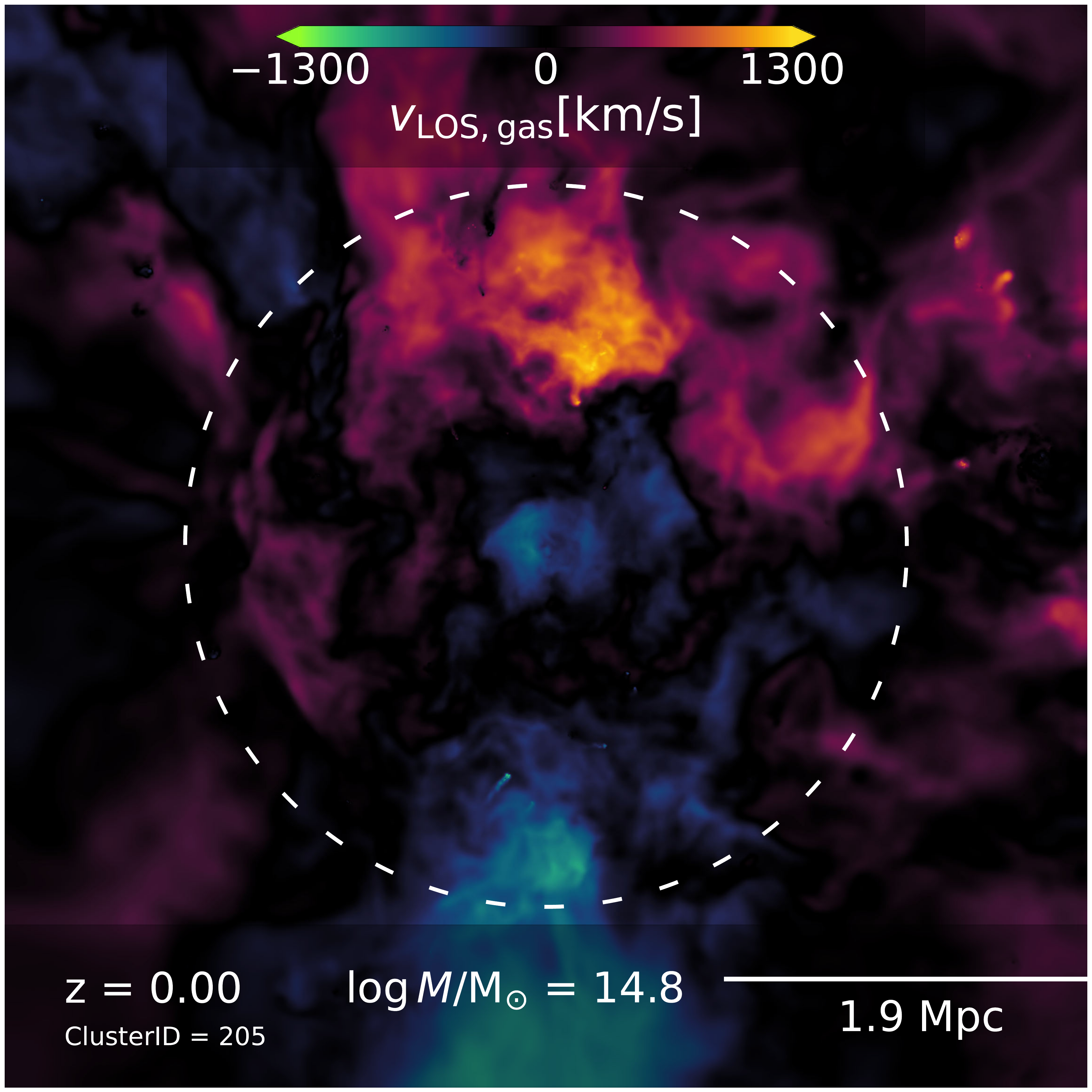}

    \includegraphics[width=0.24\textwidth]{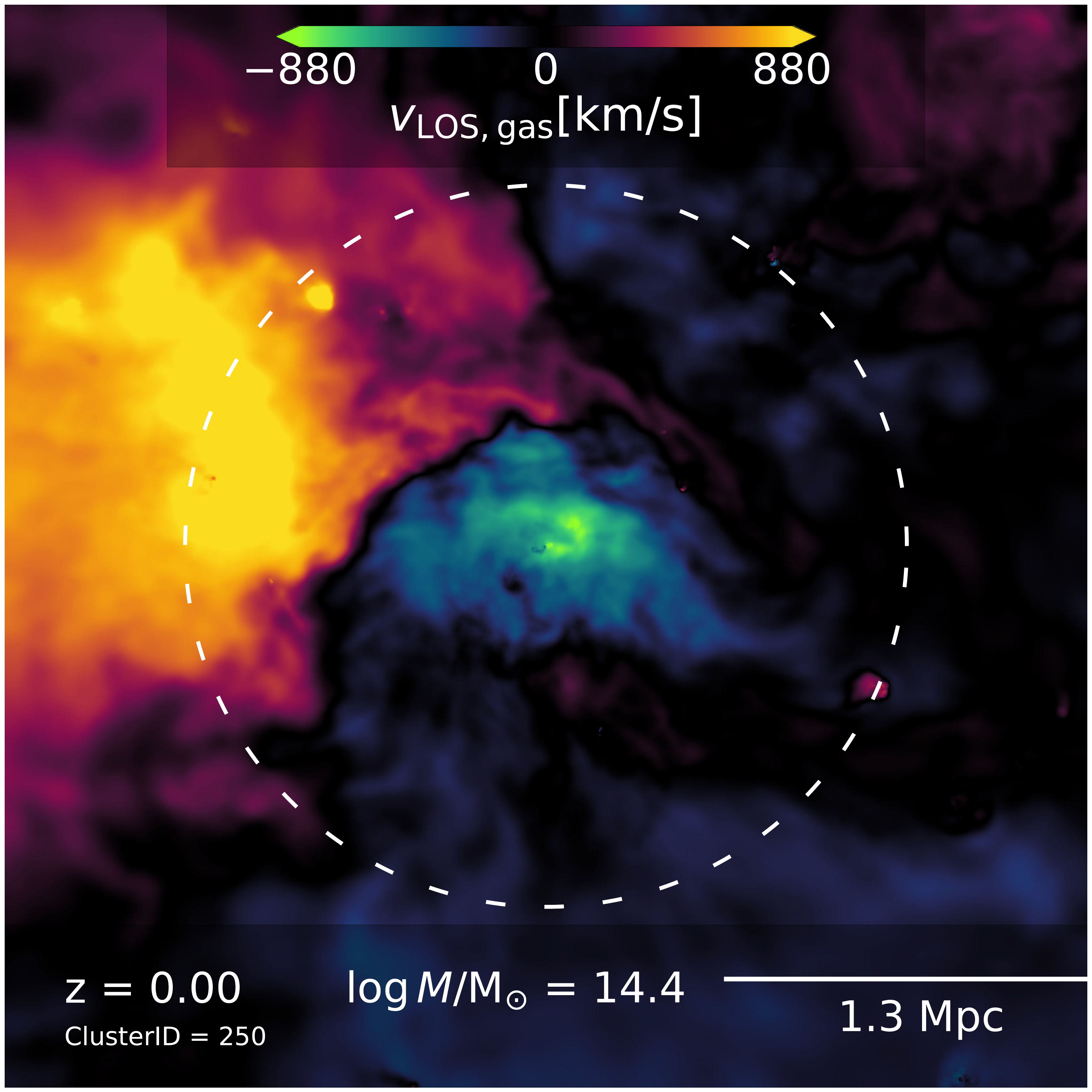}
    \includegraphics[width=0.24\textwidth]{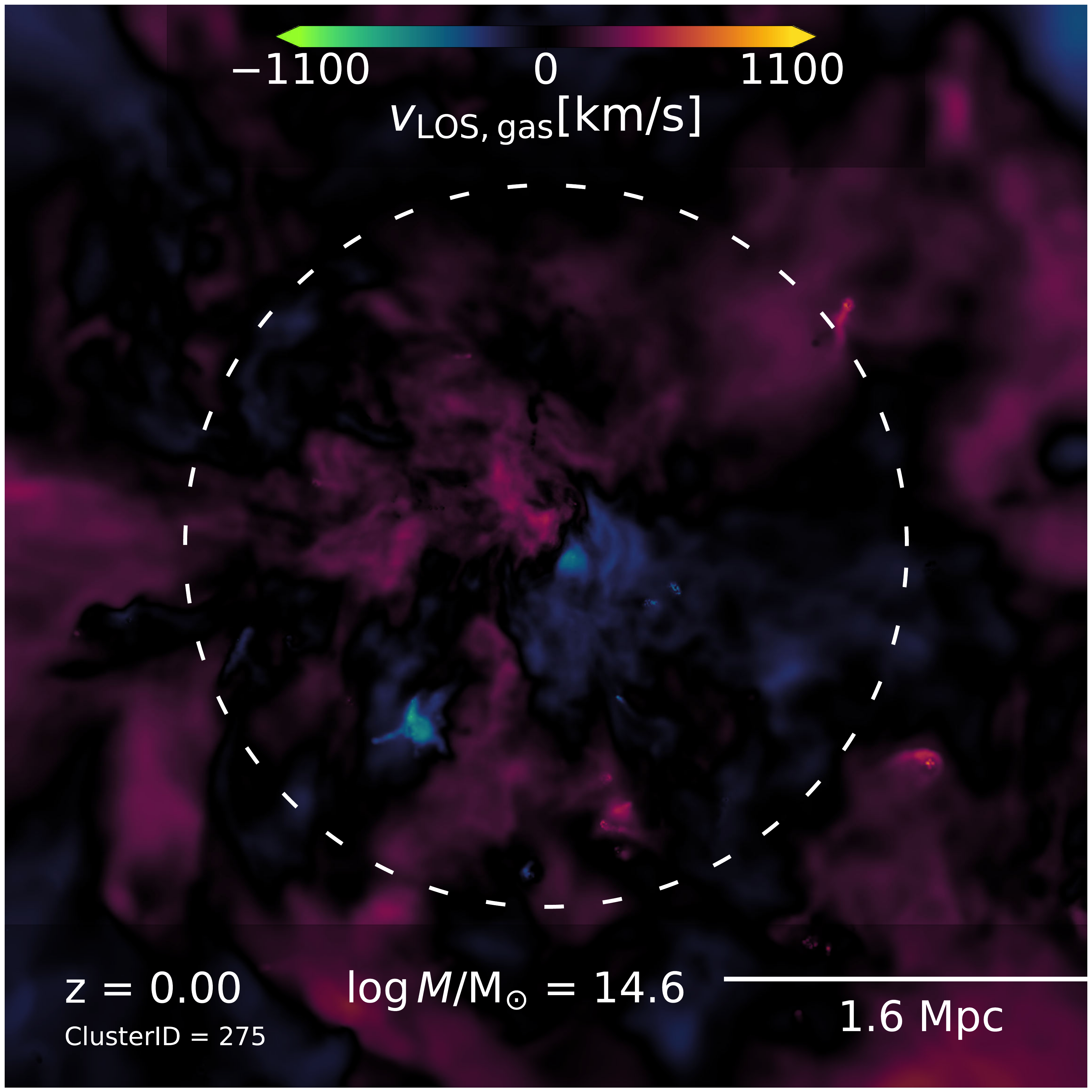}
    \includegraphics[width=0.24\textwidth]{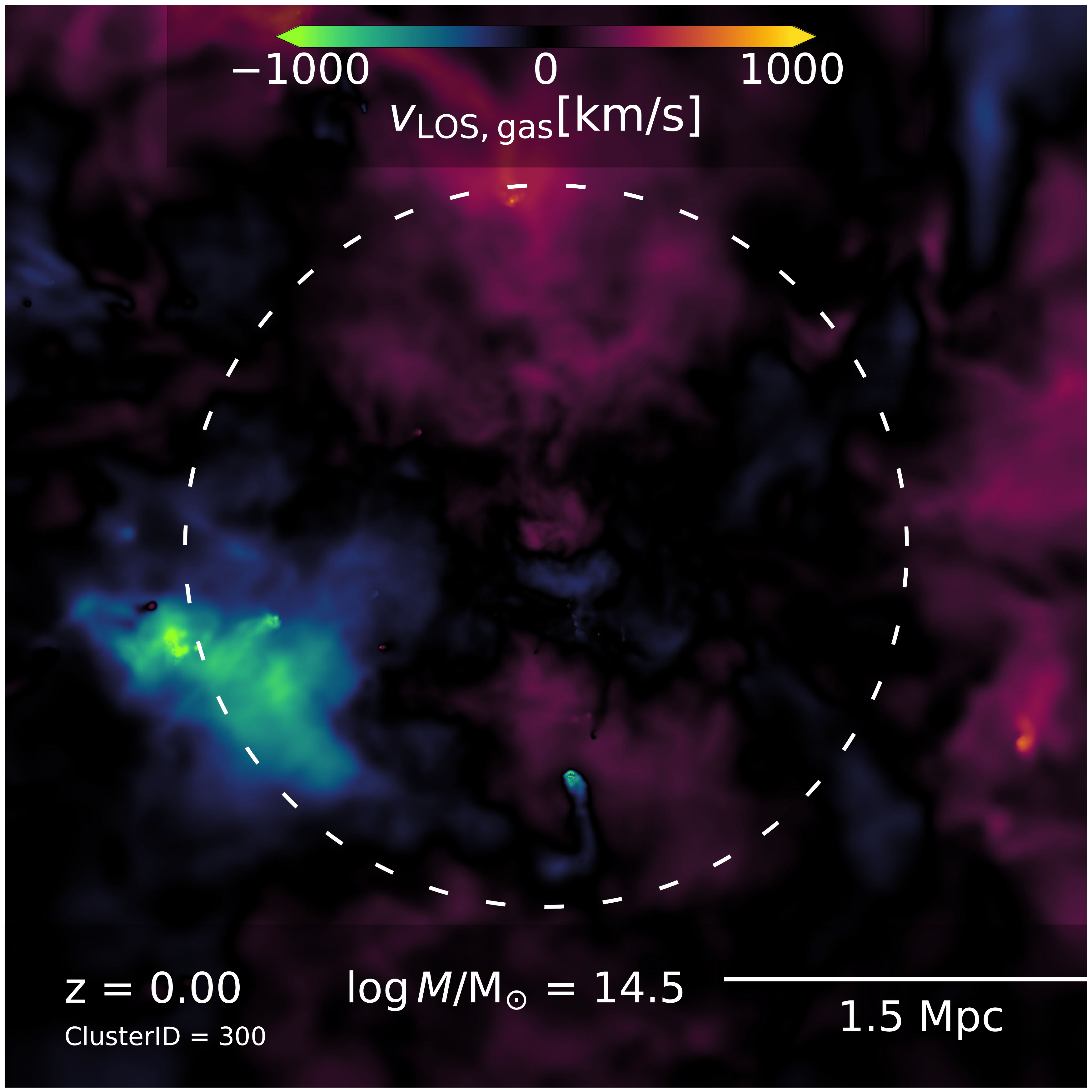}
    \includegraphics[width=0.24\textwidth]{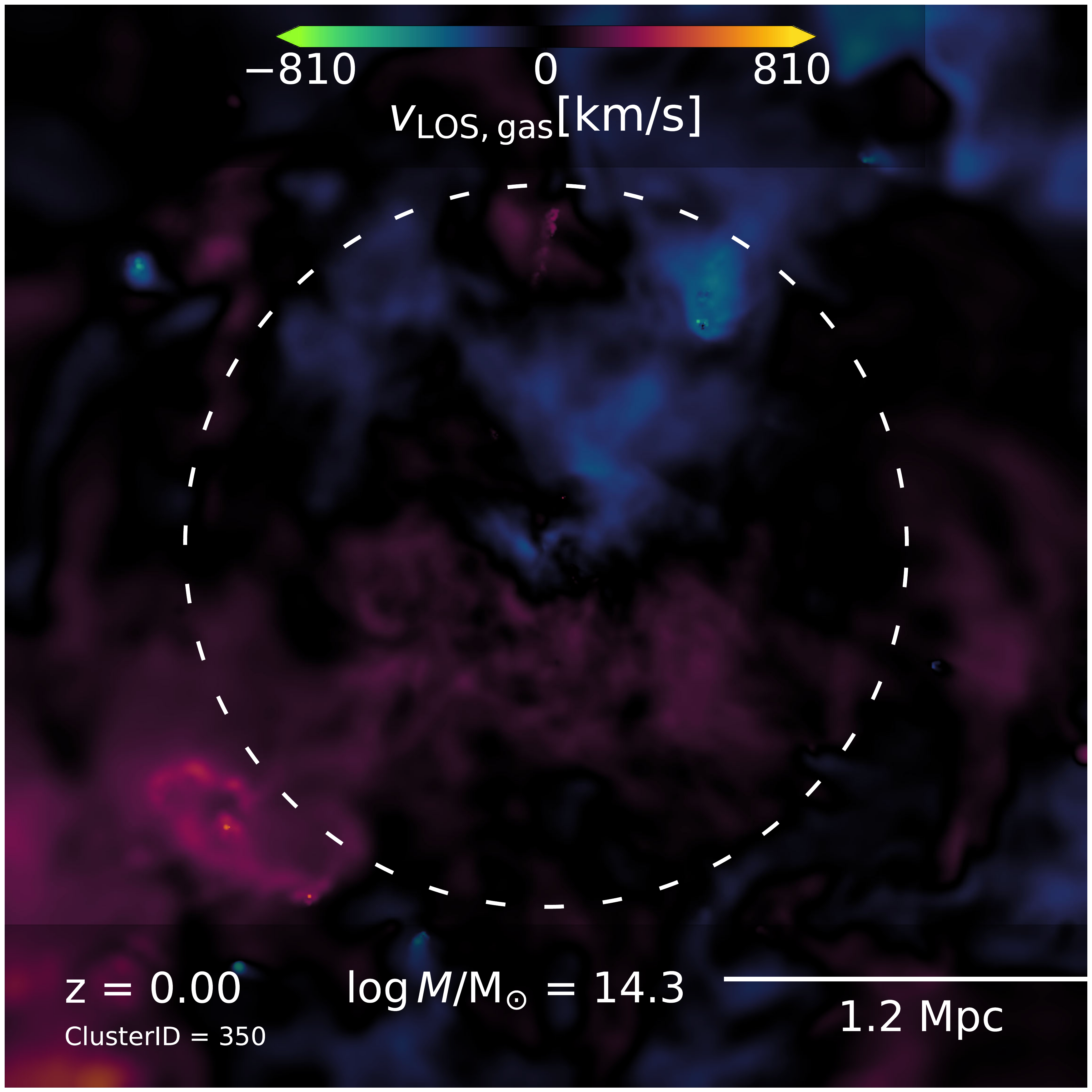}

    \caption{Gallery of projected line of sight velocities for a sub-sample of our simulated galaxy clusters. The figure highlights the diversity in the kinematics of TNG-Cluster halos at $z=0$, demonstrating their diverse dynamics and the underlying, hidden connections with galaxy and halo properties which are correlated with ICM motions. The scale written on the bottom left of each panel corresponds to the halo virial radius, $R_{\rm 200c}$. Positive (negative) velocities correspond to gas moving away from (towards) the observer.}
\label{Fig: vlos_box_sample}
\end{figure*}

In the vicinity of the halo center (leftmost column in all panels, $R_{\rm 3D}<50 \, \rm kpc$), the VSF exhibits a notable correlation with SMBH accretion rate (as seen in the bottom right panel). This correlation is absent for other halo properties. This suggests that AGN feedback significantly influences central gas kinematics, particularly in halos with higher accretion rates onto their central SMBHs. Variation of this trend with halo mass is non-monontonic and not particularly clear.

Within the halo core ($R_{\rm 3D}<0.2 R_{\rm 200c}$, second columns), the VSF shows varying degrees of correlation with all the halo properties shown. Overall, unrelaxed halos, late-forming halos, halos with less massive SMBHs, and halos with lower SMBH accretion rates have more dynamic gas motions and less coherency within their cores. This is suggestive that the kinematics of these regions are primarily driven by formation and assembly, rather than AGN-driven perturbations.

Moreover, the intermediate regions of clusters ($R_{\rm 3D}\sim 0.5 R_{\rm 200c}$, third columns) exhibit similar trends to those in the core, albeit with less pronounced impact on the VSF. In contrast, the outskirts of the halo (fourth column) are minimally influenced by any of the halo properties in a meaningful way. An exception is a correlation with formation time, that is primarily visible only in the most massive clusters of TNG-Cluster, where statistics become limited.

Finally, the VSF across the entire halo ($R_{\rm 3D}<1.5 R_{\rm 200c}$, rightmost column in all panels) correlates with all examined halo properties. The detected trend appears to be an average of the behaviors seen in different halo zones. Specifically, the VSF is higher (indicating less coherent gas motion) in unrelaxed halos, late-forming halos, halos with less massive SMBHs, and halos with lower accretion rates onto their SMBHs.


\section{Towards observable ICM kinematics in 2D projections}
\label{sec: obs_kinematics}
In this section, we transition from 3D intrinsic motion to observable 2D kinematics, focusing on line-of-sight (LOS) velocity and velocity dispersion as key observables for the ICM and cluster outskirts. Fig. \ref{Fig: vlos_box} presents the LOS velocity (left panel) and LOS velocity dispersion (right panel) for the same massive, relaxed galaxy cluster depicted in Fig. \ref{Fig: vrad_box}, with a mass of $M_{\rm 200c} \approx 1.6 \times 10^{15} \, \rm M_{\odot}$. Both the LOS velocity and velocity dispersion are weighted by X-ray luminosity, thereby emphasizing the hot gas flows that would be captured in X-ray observations.

The structure of LOS velocity is notably different from radial velocity, emphasizing the restricted information content available from observables in projection. The LOS velocity dispersion diverges from its radial counterpart as well. However, the difference is less than in the velocities themselves. The discrepancy between radial and line of sight velocities arises from the deeper projection along the LOS and the single i.e. observable LOS component. A gallery of LOS velocity maps for sixteen randomly selected clusters is displayed in Fig. \ref{Fig: vlos_box_sample}, highlighting the variation in gas kinematics across the $z=0$ population.

\subsection{Line-of-sight velocities and velocity dispersion}
\label{subsec: vLOS_disp}

\begin{figure*}
    \centering
    \includegraphics[width=1\columnwidth]{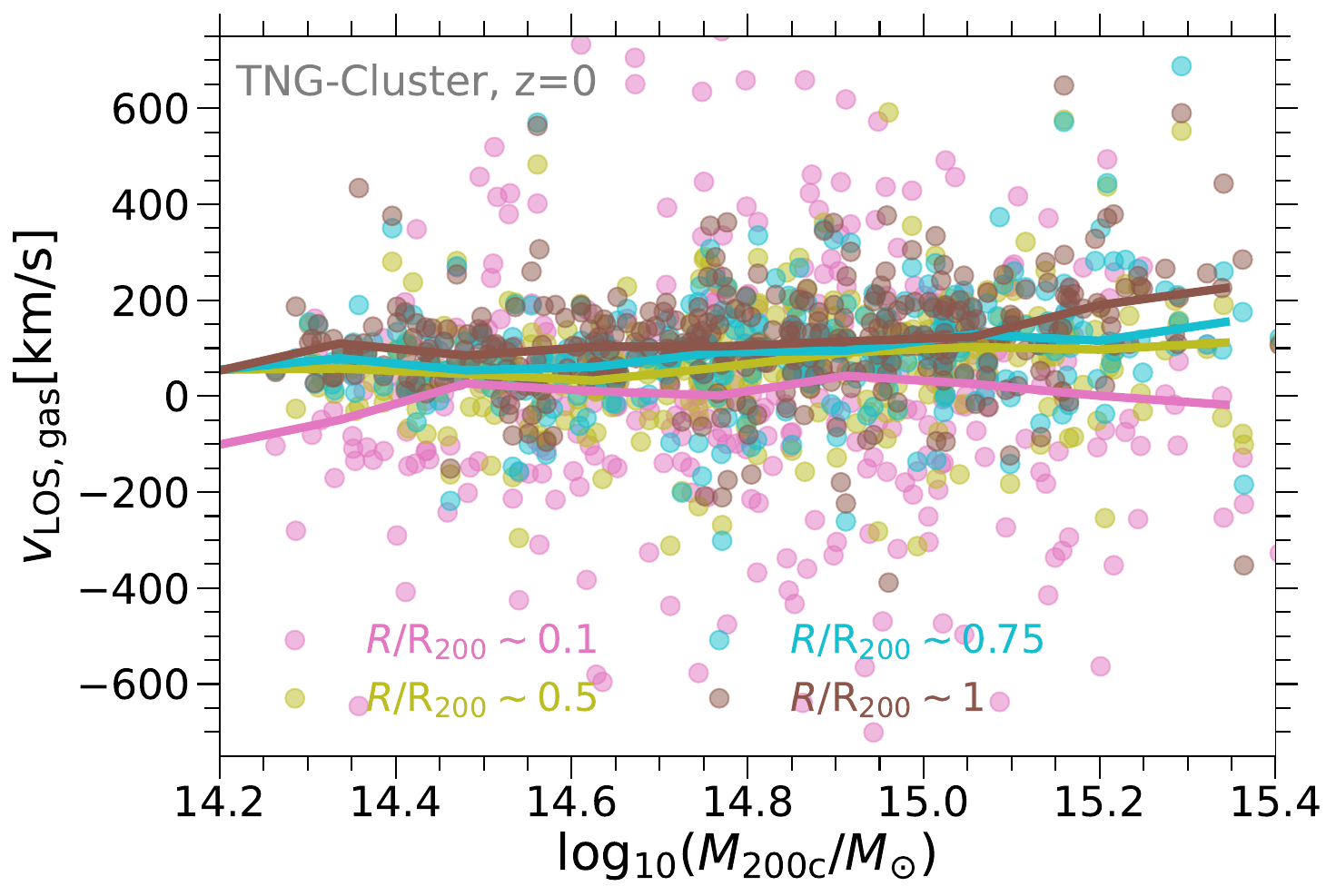}
    \includegraphics[width=1\columnwidth]{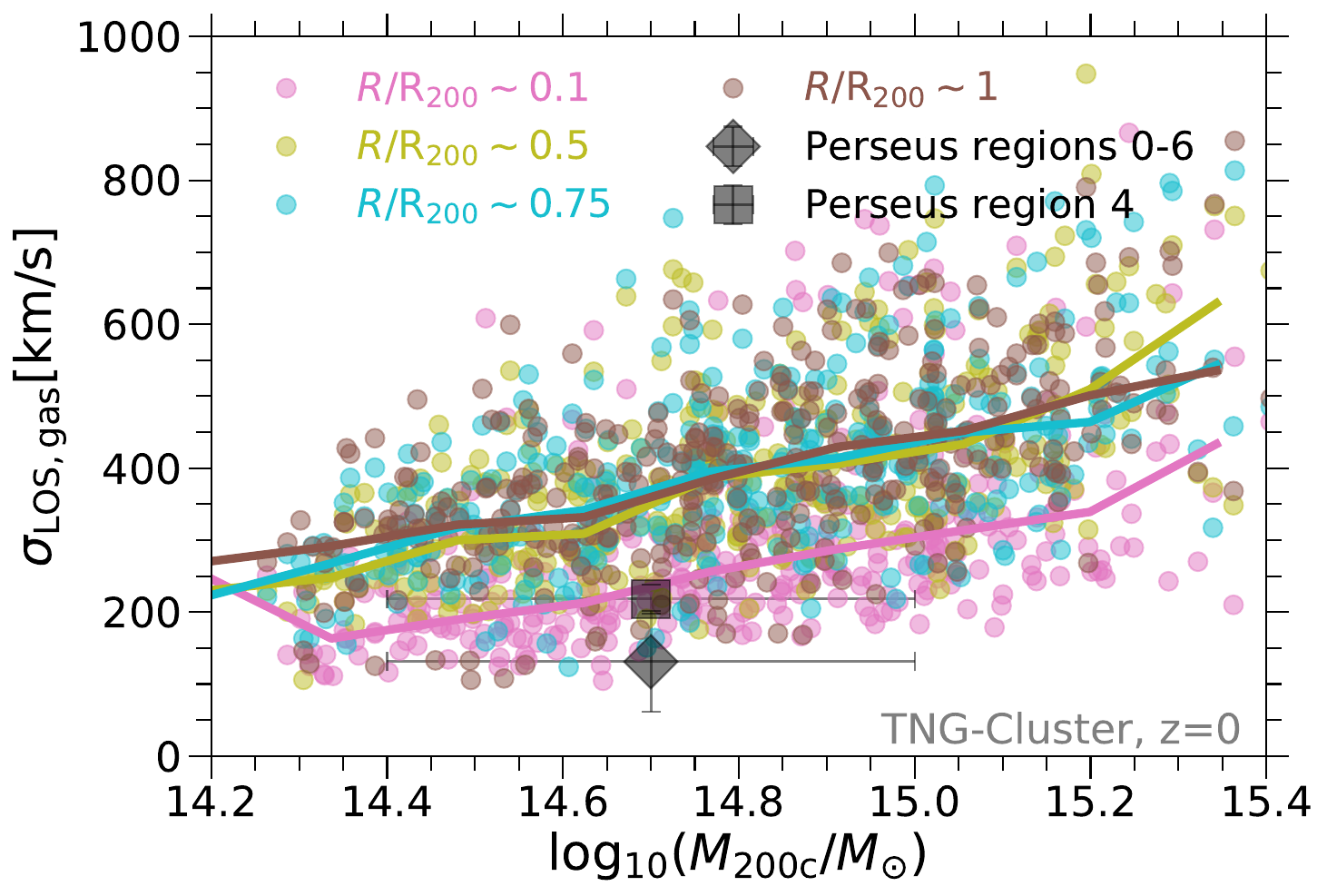}
    \caption{Observable kinematics of TNG-Cluster halos: line of sight velocity (top) and velocity dispersion (bottom) as a function of halo mass. Each dot corresponds to a halo, with the color representing the halocentric distance at which the velocity is measured. Lines show running medians. The big square and big diamond correspond to the Hitomi measurements of the velocity dispersion of the core of the Perseus cluster \citep{Hitomi2018Atmospheric}. Velocity dispersion increases rapidly for more massive halos, while the annular averaged line of sight velocities remain near zero on average, as expected. At the same time, there is significant diversity i.e. large scatter across the sample.}
\label{Fig: vlos_lines}
\end{figure*}

In Fig. \ref{Fig: vlos_lines}, we present the line-of-sight velocity (left panel) and velocity dispersion (right panel) as functions of halo mass, using all halos in the TNG-Cluster simulation at $z=0$. Each point represents a cluster, and the colors denote different regions within these clusters, ranging from the core to the outskirts. These values are X-ray luminosity weighted, as in Fig. \ref{Fig: vlos_box}. The measurements for each halo region are conducted within a 2D annulus with a width of 200 kpc. For the core measurements (pink dots), the analysis extends down to the cluster center.

The LOS velocity exhibits substantial halo-to-halo variation, reflecting the diverse dynamical states of these halos. No discernible trend with halo mass is detectable for LOS velocity, as the running median lines are roughly zero at all masses. While individual regions in individual clusters can have large line-of-sight velocities, our averages for each of the four regions, and across halos, indicate a tendency towards global ICM equilibrium.

In contrast, the LOS velocity dispersion is strongly correlated with halo mass across all regions of the halo, from the center (pink dots) to the outskirts (brown dots). In particular, $\sigma_{\rm LOS}$ increases with halo mass, by a factor of $\sim 2-4$ across our one dex range of cluster masses. A trend with halocentric radius is also evident: the velocity dispersion is higher in the outskirts versus core regions of clusters. This trend is most pronounced when transitioning from the halo center (pink) to intermediate and outer regions. However, the velocity dispersion remains relatively constant from $R/R_{\rm 200c} \sim 0.5$ to $R/R_{\rm 200c} \sim 1$.

We qualitatively compare our findings with the Hitomi measurements of the velocity dispersion in the core $R < 100 \, \text{kpc}$ of the Perseus cluster \citep{Hitomi2018Atmospheric}. Hitomi reports a peak velocity dispersion of approximately $\sim 200 \, \text{km/s}$ in regions influenced by AGN activity, versus $\sim 100 \, \text{km/s}$ in the other core areas of Perseus. When we compare our relevant measurements (cluster core, pink dots) with those from Hitomi, we see that at least some TNG-Cluster halos have central velocity dispersions compatible with Perseus. This broad agreement, intended here at face value only, validates TNG-Cluster for studies of gas kinematics in galaxy clusters. A detailed comparison with mock X-ray derived kinematics, quantitative comparison with the Hitomi result, and predictions for the X-ray spectrometer XRISM are the topic of a companion paper (\textcolor{blue}{Truong et al. submitted}).

\begin{figure}
    \centering
    \includegraphics[width=0.45\textwidth]{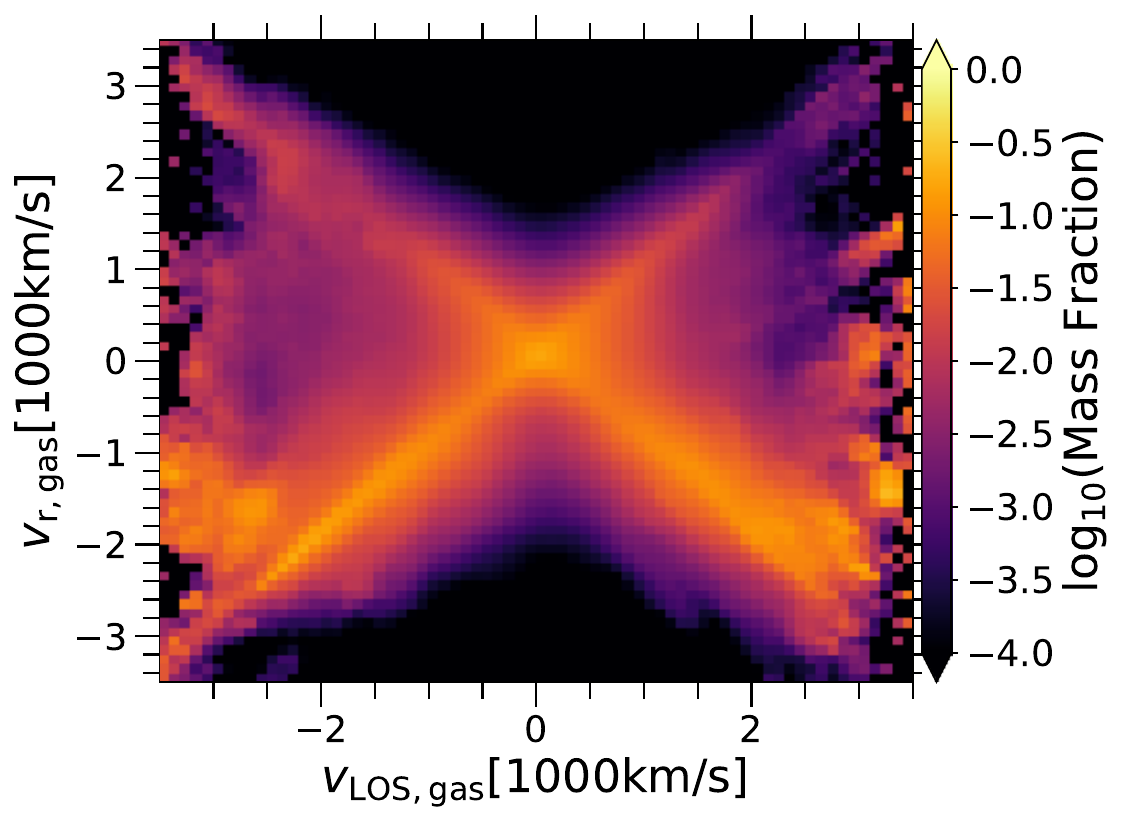}
    \includegraphics[width=0.45\textwidth]{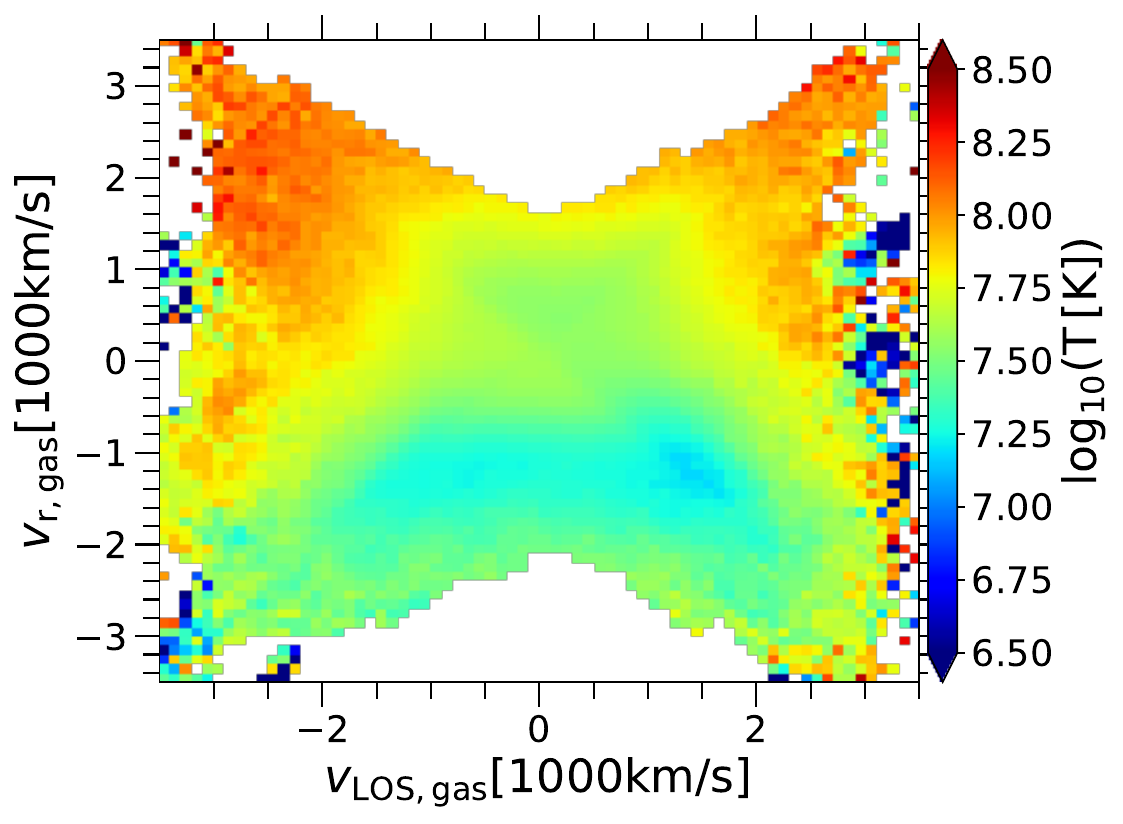}
    \includegraphics[width=0.45\textwidth]{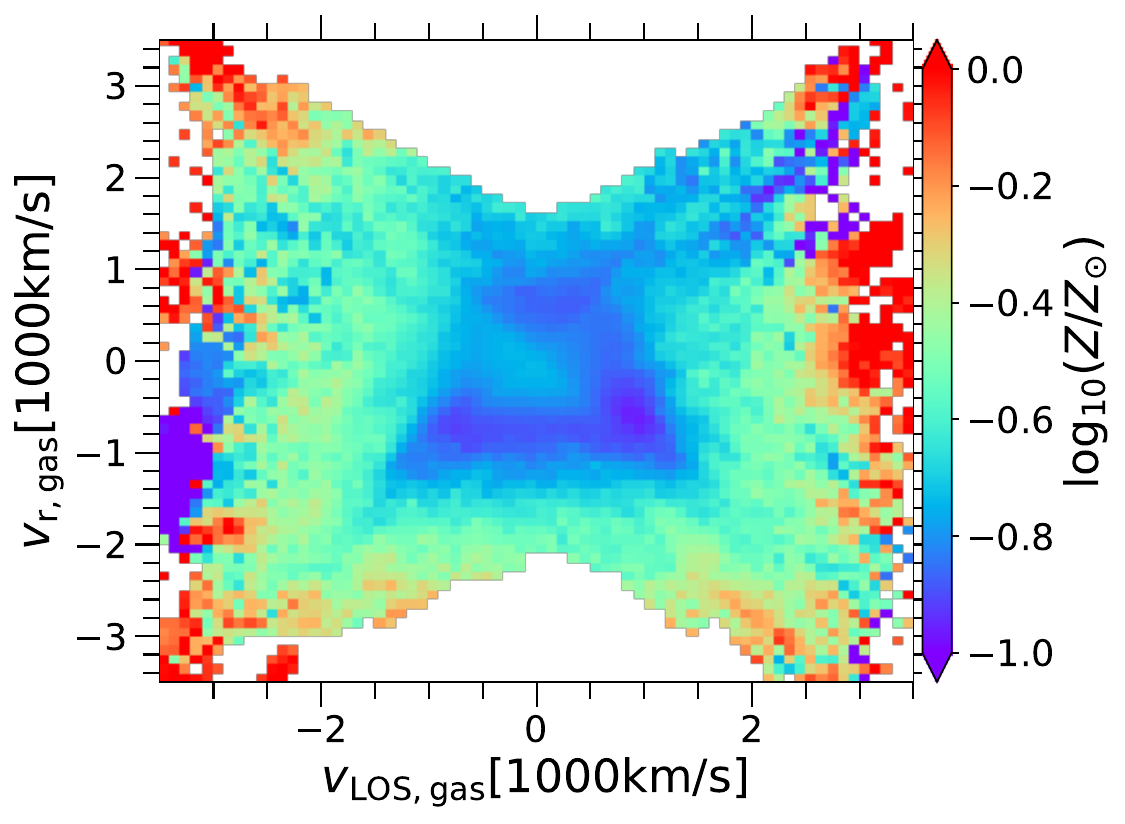}
    \caption{The relationship between true (3D) radial velocity and observable (projected) line of sight velocity. Colors represent mass fraction (top), mean temperature (middle), and mean metallicity (bottom). Here we stack all TNG-Cluster halos and use all gas cells together, giving equal weights to individual halos. In the top panel only, each column is normalized such that the sum equals unity. It is not possible to infer the inflowing (negative radial velocity) versus outflowing (positive radial velocity) gas solely from its line of sight velocity. However, strong correlations are evident between velocity and temperature which can help break the degeneracy.}
\label{Fig: vlos_vrad}
\end{figure}

The line-of-sight velocity, although a crucial observable for gas kinematics, may not be an accurate proxy for the true radial velocity, which reflects the dynamics of inflowing and outflowing gas. Fig. \ref{Fig: vlos_vrad} therefore shows the distribution of true radial velocities as a function of line of sight velocities. Colors denote mass fraction (top panel), mean temperature (middle panel), and mean metallicity (bottom panel).

We find a substantial scatter in LOS velocity for a given radial velocity. Specifically, for inflowing gas (negative radial velocity), the LOS velocity can be either negative or positive. As a result, it is not possible to deduce the directionality of gas flows -- whether inflowing or outflowing -- based solely on observed LOS velocities. Therefore, caution is warranted when interpreting LOS velocities as indicators of radial gas movements.
The temperature of the inflowing gas is, on average, lower than that of the outflowing gas. Therefore, at a fixed line-of-sight (LOS) velocity, whether positive or negative, temperature can help break the degeneracy between inflowing and outflowing gas, as shown in the middle panel of Fig. \ref{Fig: vlos_vrad}. The trend with metallicity is less clear, making it more challenging to use for identifying inflows and outflows (bottom panel).
Overall, LOS velocity alone is a limited diagnostic tool for gas kinematics. To obtain a more comprehensive understanding, it is imperative to integrate LOS velocity with other observables, such as temperature and metallicity.

In observational contexts, high LOS velocity values are often interpreted as either high-velocity outflows or inflows. However, we see that low LOS velocity values do not necessarily imply a static or slow-moving gas. This is because LOS velocity is a projection of the full 3D velocity, which may have significant components in the plane of the sky, orthogonal to the observed line of sight. 

\begin{figure*}
    \centering
    \includegraphics[width=0.7\textwidth]{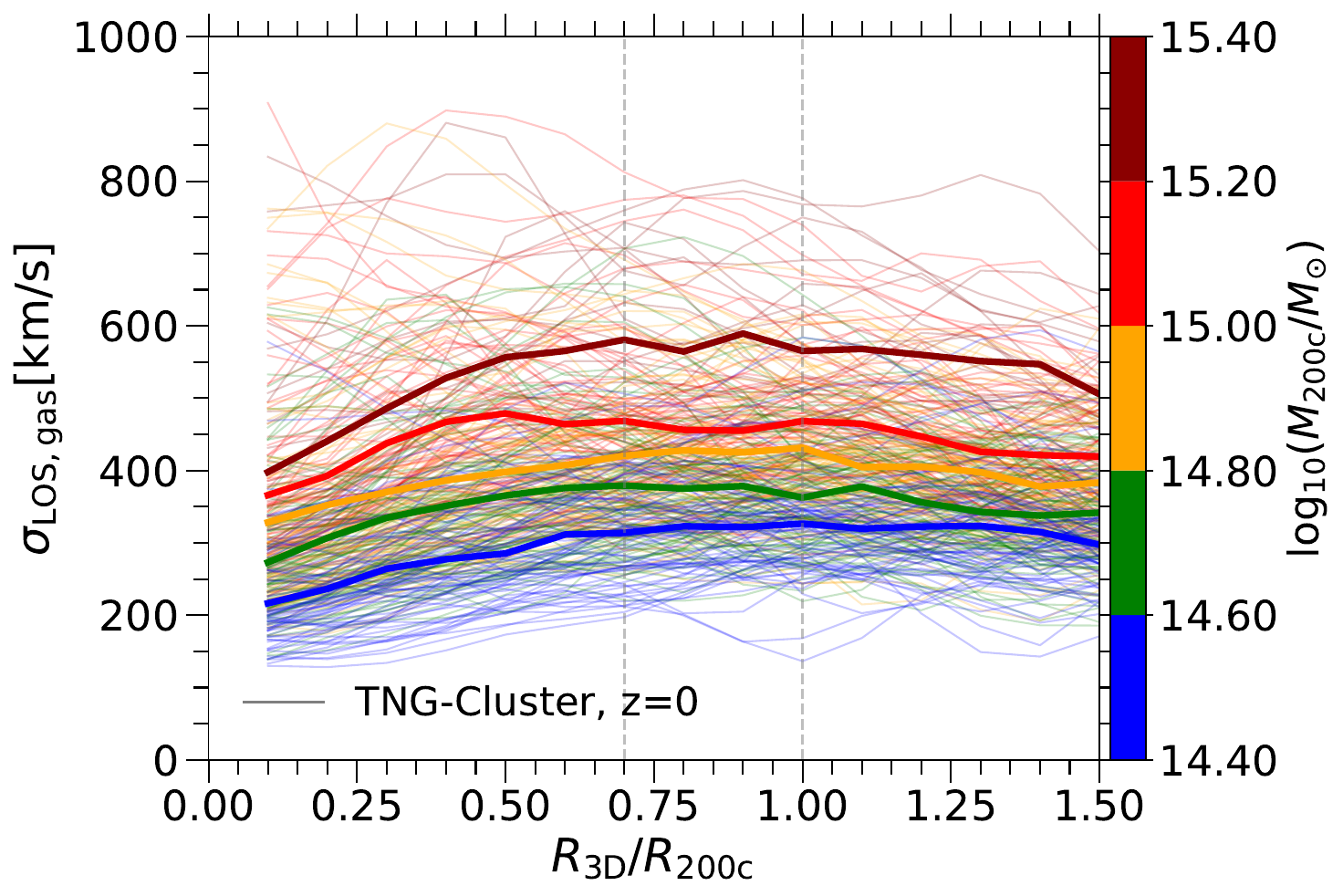}
    \includegraphics[width=0.46\textwidth]{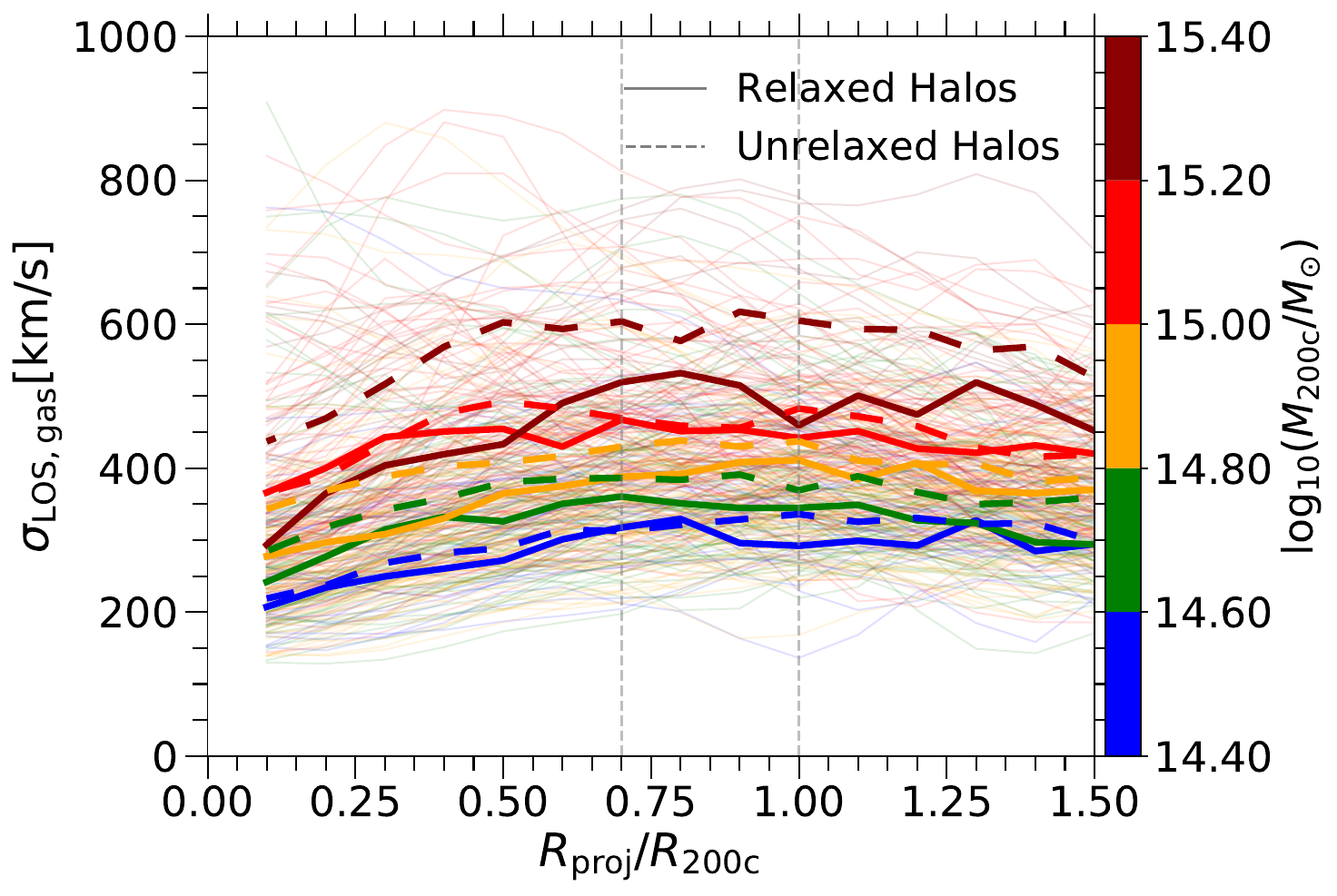}
    \includegraphics[width=0.46\textwidth]{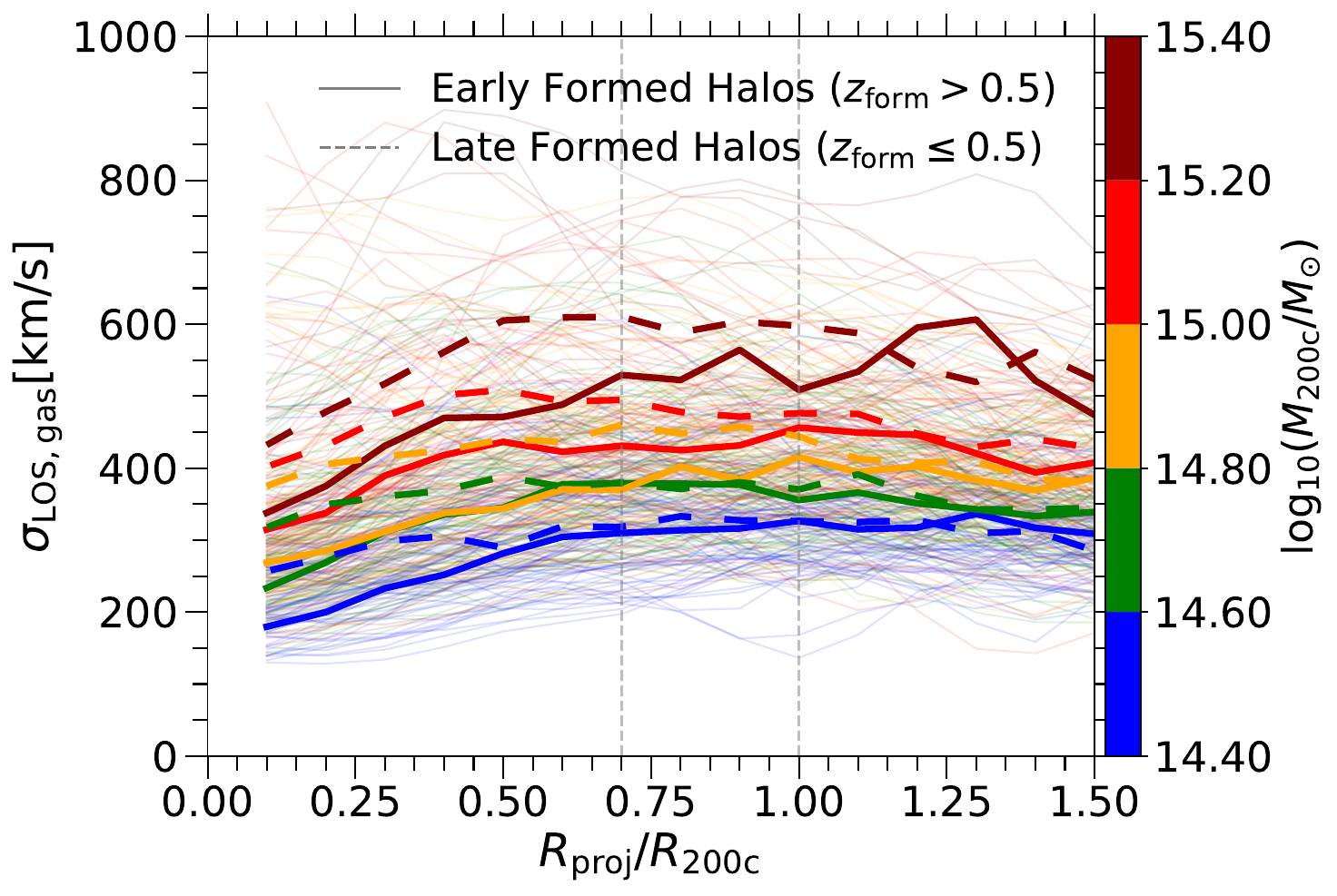}
    \includegraphics[width=0.46\textwidth]{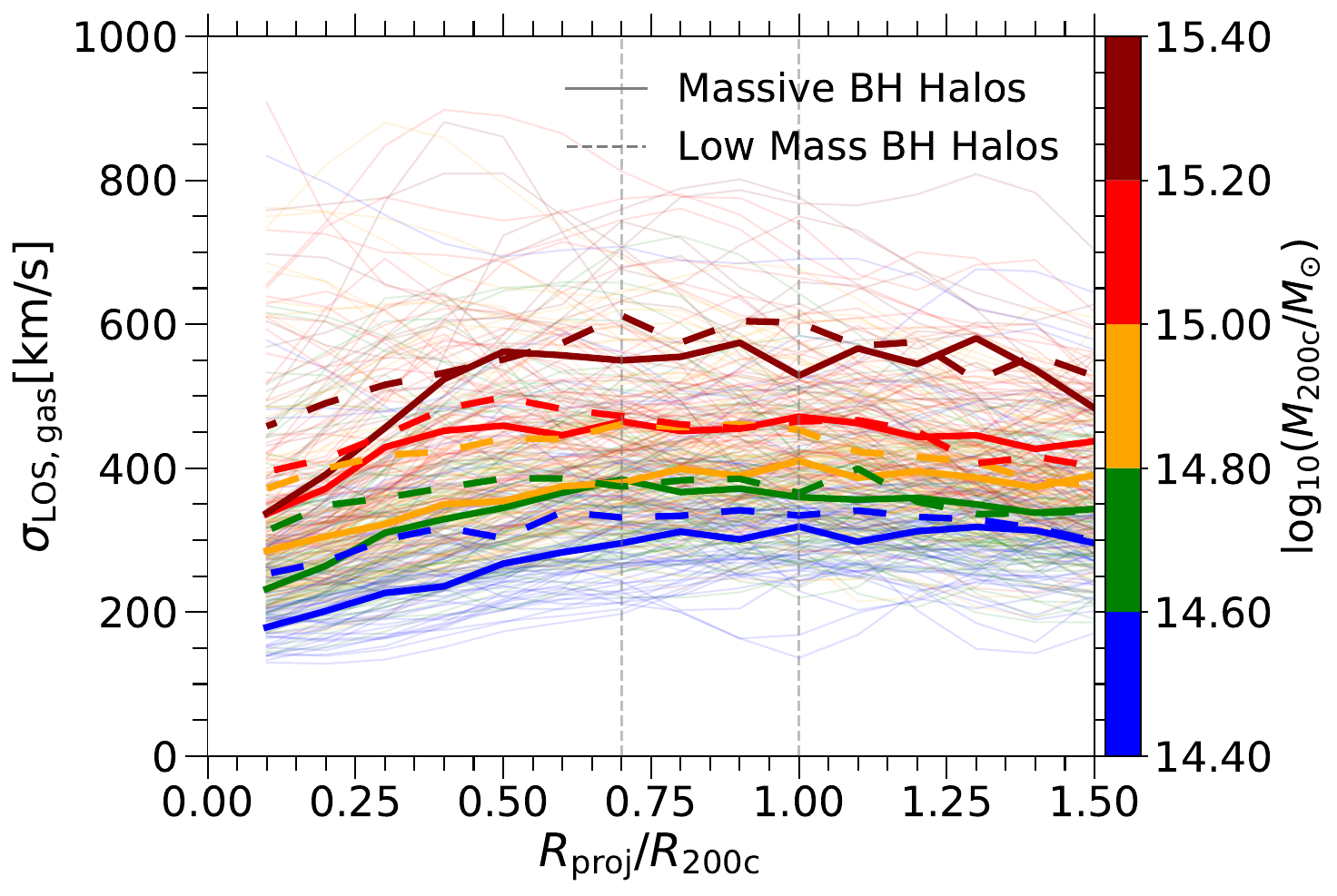}
    \includegraphics[width=0.46\textwidth]{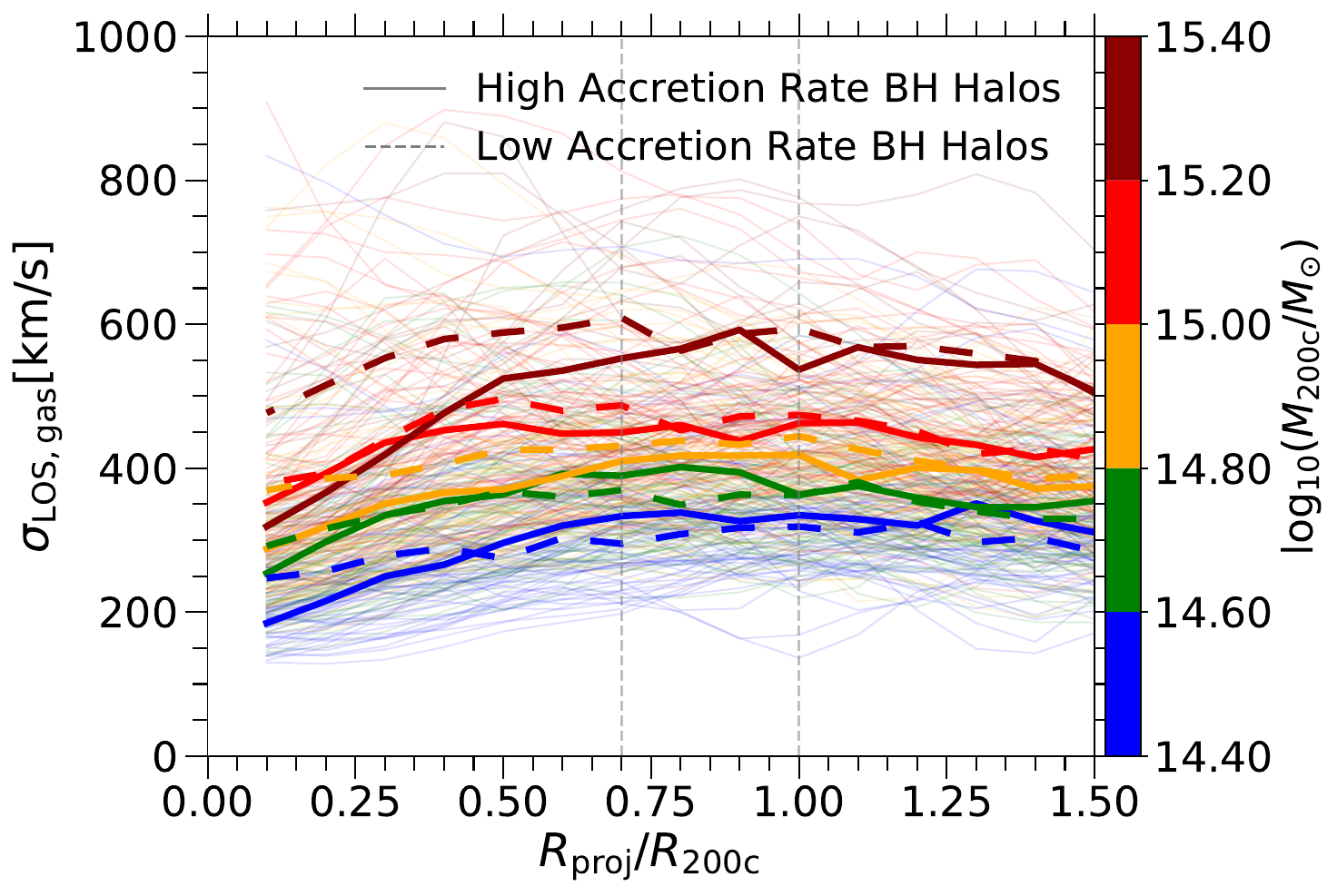}
    \caption{Radial profiles of the line of sight velocity dispersion for halos. The top panel shows this quantity for all halos, while the other panels show the same quantity for halos classified based on their relaxedness (middle left), and formation time (middle right), SMBH mass (bottom left), SMBH accretion rate (bottom right). Each color corresponds to a different halo mass bin. The thick solid lines depict the mean profile, while the thin lines represent individual clusters. The figure underscores the significant impact of halo properties on the line of sight velocity dispersion.}
\label{Fig: vLOS_disp}
\end{figure*}

We now examine the velocity dispersion of clusters as a function of halo properties, as shown in Fig. \ref{Fig: vLOS_disp}. The top panel presents the line-of-sight velocity dispersion for all clusters, highlighting the most important trends: with mass and radius. $\sigma_{\rm LOS}$ strongly depends on halo mass, with more massive clusters exhibiting higher velocity dispersion. Additionally, it also varies with halocentric distance. From the core to approximately $0.5 - 0.7 \, R_{\rm 200c}$, (or $R_{\rm 500c}$, indicated by the first vertical dashed line on the left), $\sigma_{\rm LOS}$ rises with increasing halo mass. Beyond this it stays constant until $R_{\rm 200c}$, and then decreases. This behavior contrasts with that of the radial velocity dispersion (see Fig. \ref{Fig: profiles_vrad_disp}), which generally increases monotonically with increasing halocentric distance. This difference is likely due to both projection effects and the inherent differences between radial and line-of-sight velocities, as previously discussed (Fig. \ref{Fig: vlos_vrad}).

The four lower subpanels study the influence of various halo properties on velocity dispersion: relaxedness (middle left), formation time (middle right), SMBH mass (lower left), and SMBH accretion rate (lower right). We find that unrelaxed halos generally exhibit higher velocity dispersions, as expected, although the difference is not statistically significant except for the most massive halos ($\log_{10}M_{200c}/{\rm M_{\odot}} > 15.2$, dark red lines). Late-forming halos display significantly higher velocity dispersions, particularly near the halo center, with the difference decreasing towards the outskirts. Within $0.2\, R_{\rm 200c}$, the average difference in $\sigma_{\rm LOS}$ between late- and early-forming halos can reach $\sim 200 \, \rm km/s$. This connection is consistent with the similar correlation for radial velocity dispersion (Fig. \ref{Fig: profiles_vrad_disp}). The considerable difference near the core suggests that early-forming halos have had ample time to relax and approach a quasi-equilibrium state, whereas late-forming halos are still dynamically evolving, leaving their cores in a non-equilibrium state.

The bottom row shows how velocity dispersion also correlates with SMBH properties. Overall, clusters with more massive SMBHs have lower velocity dispersion. This is consistent with more massive SMBHs residing in earlier formed halos (see Appendix \ref{app: halo correlations}).
There is also a significant trend with halocentric distance: this correlation is larger in the center and decreases with halocentric distance, vanishing completely at $R\sim 1.5\, R_{\rm 200c}$. Finally, SMBH accretion rate also correlates with velocity dispersion: higher accretion rate SMBHs have lower velocity dispersion. This is, however, less significant than the trend with the SMBH mass, because the instantaneous SMBH accretion rate is not a direct indicator of the evolution history of the SMBH, as opposed to its mass or the halo formation time.

\subsection{The velocity structure function versus observations}
\label{subsec: vsf_obs}

\begin{figure*}
    \centering
    \includegraphics[width=0.49\textwidth]{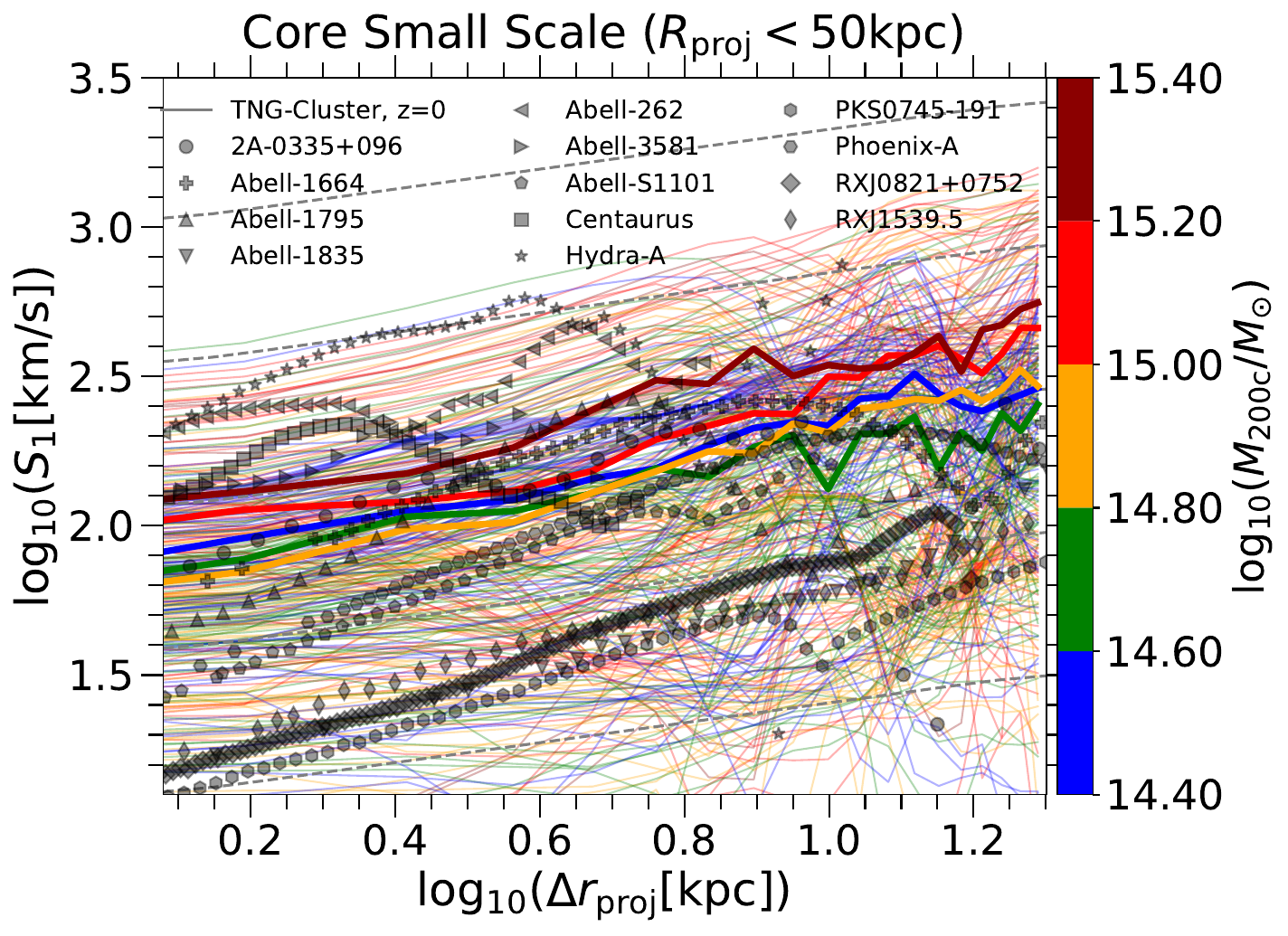}
    \includegraphics[width=0.49\textwidth]{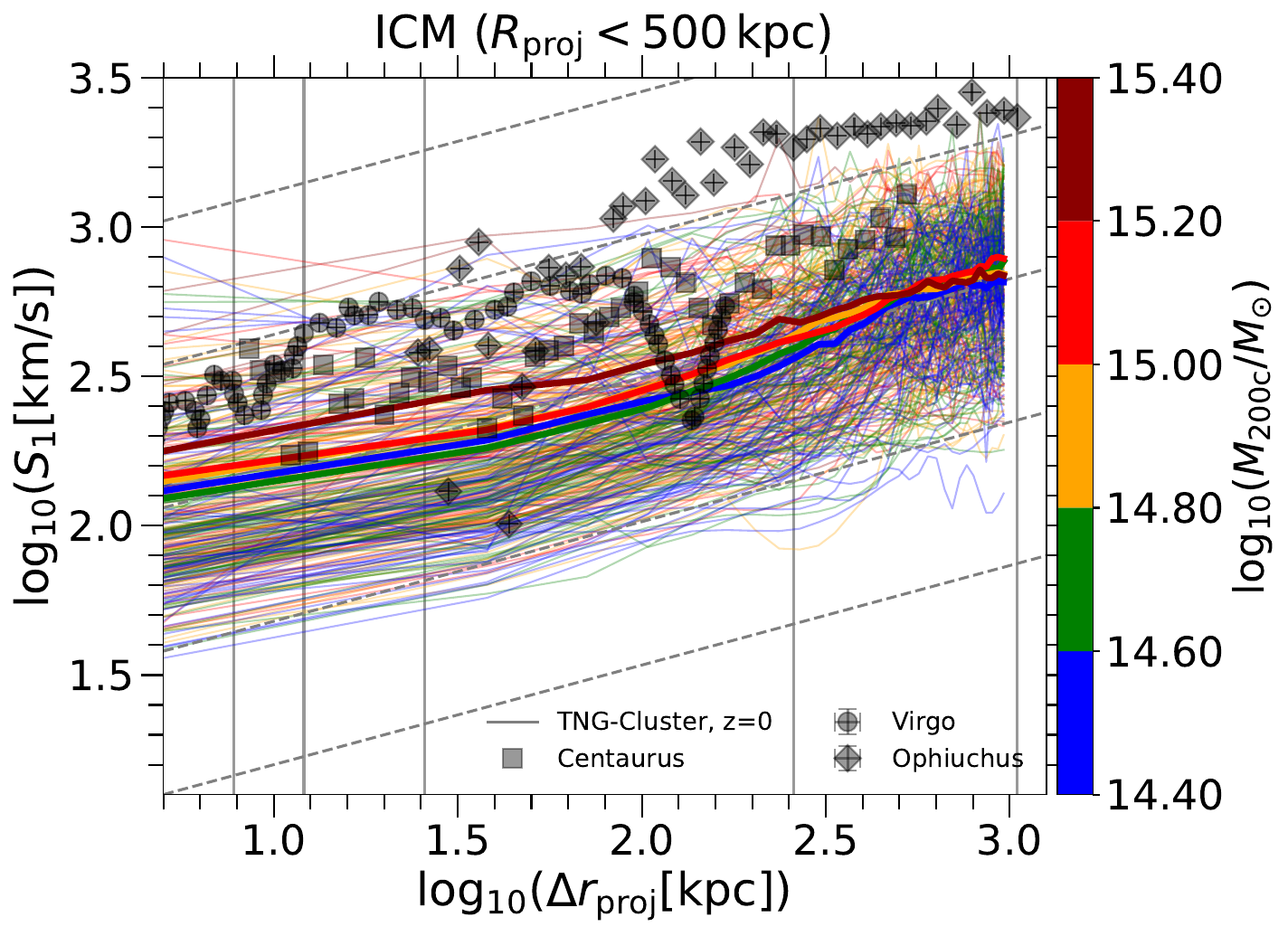}
    \caption{Comparison of the velocity structure function from the full sample of the TNG-Cluster simulation at $z=0$ with available observational data. The left panel compares simulated VSFs with observations of cold gas in the cores of galaxy clusters \citep{Li2020Direct,Ganguly2023nature}. The right panel contrasts the simulated VSF with inferences from X-ray observations of the hot ICM \citep{Gatuzz2023Measuring}. In both cases, the VSF is density weighted (see text). In the right panel, the vertical errorbars, shown for only a few data points as representative of the whole sample, are of order of $1000\rm\, km/s$. These correspond to large observational uncertainties, as discussed in the text. Overall, we find that there are, within the population of our simulated clusters, examples that broadly reproduce the behavior of the observed VSFs. The possible exception is the large scale measurements for Ophiuchus, where only a rough qualitative agreement exists given the observational uncertainties. At the same time, there is significant cluster-to-cluster variability of the VSF, indicating the complex physics governing the most massive objects in the Universe.}
\label{Fig: vsf1_obs}
\end{figure*}

We compare our measurements of the VSF with observations on both small and large separation scales, ranging from $\sim$\,kpc to $\sim$\,Mpc. The left panel of Fig. \ref{Fig: vsf1_obs} shows our results on small scales, which are specifically for the kinematics of cool gas. In our simulated galaxy clusters, we project along the line of sight and select gas with $\rm T \lesssim 3 \times 10^4 \, K$ within 50 kpc of the centers of our clusters to mimic the observational tracers of cold gas. The VSF measurements and binning are done as explained in Section \ref{subsec: methods_vsf} with bins weighted based on their densities. Individual colored lines show individual halos, while thick colored lines show the mean VSF in halo mass bins, as before. Gray symbols show recent measurements of the VSF for cold gas from \citet{Li2020Direct,Ganguly2023nature}, taken in the cores of 13 galaxy clusters \citep[see also][for VSF characterization of cool-phase CGM turbulence at lower mass scales]{Chen2023Empirical,Chen2023Ensemble}. We find that the VSF of cool gas in TNG-Cluster halos is broadly compatible with these observational inferences, i.e. among the simulated sample of 352 high-mass clusters, several systems have VSFs similar to those observed, ranging from $\sim 1 \, \rm kpc$ to a few tens of kpc.

The right panel of Fig. \ref{Fig: vsf1_obs} shows the VSF of the hot ICM on larger separation scales. We compare our VSF measurements with those from \cite{Gatuzz2023Measuring}, who measure the VSF of the ICM for three observed galaxy clusters out to scales comparable to the halo virial radius. Among these, Ophiuchus (diamonds) is in a relaxed state, while Centaurus (squares) and Virgo (circles) are influenced by AGN feedback and plasma sloshing. In the latter two cases, there are unambiguous cases of simulated clusters with similar VSFs, from kpc to Mpc separation scales. The case of Ophiuchus is less clear: observed values of $S_1$ reach $\gtrsim 3000$\,km/s which are exceedingly rare across the entire TNG-Cluster population.\footnote{We have checked, and the VSF predicted by TNG-Cluster is in very good agreement with the amplitude of one-component velocity of gas motions within a few hundred kpc separation scales in several other observed galaxy clusters measured by \cite{Zhuravleva2018Gas}.} 

We note that observational measurements are not necessarily directly comparable to our simulations. Specifically, when calculating the large-scale VSF (right panel) from simulations, we are not subject to the same limitations as in observational data. Most notably, the use of adaptive spatial bins—i.e., constant signal-to-noise ratio bins—can lead to measurements of velocities dominated by different regions of a cluster. This common technique in observational data naturally gives higher weight to core regions due to the radial dependence of the bin size and the rapid drop of X-ray surface brightness away from cluster centers. Our VSF in the cluster core (left panel), on the other hand, does not change significantly with different binning or weighting schemes. This consistency is primarily because all measurements are confined within the cluster core, avoiding the mixing of different cluster zones that occurs with VSF measurements on large scales.

For the VSF on large scales (right panel of Fig. \ref{Fig: vsf1_obs}), we examined various binning and weighting methods, including constant bin size, adaptive bin size, and weighting based on X-ray luminosity, mass, density, and volume. We found that the large-scale VSF is very sensitive to these choices. For instance, the choice of spatial binning impacts the VSF, predominantly due to its strong radial dependence (see Fig. \ref{Fig: vsf1_3D}). It exhibits a higher VSF in the core than in the outskirts. To address these complexities, a detailed mock analysis is required, starting with the generation of X-ray images from simulation data and then applying all the methodology, uncertainties, and biases inherent in the observational analysis pipeline. This is beyond the scope of this paper, and we defer it to future work. Instead, here we employed a density weighted VSF (see Section \ref{subsec: methods_vsf}) to increase the impact of the core regions, allowing a qualitative comparison with observations. The impact of weighting can indeed be significant, and is discussed in Appendix \ref{app: vsf biases}.

An additional complexity arises due to the large uncertainties in the observed VSF, which are of the order of 1000 km/s across almost all separation scales. Moreover, the substantial halo-to-halo variation in both simulated and observed VSFs of galaxy clusters is evident at both small (left panel) and ICM (right panel) scales. This variation is showcased in the figure and underscores the complexity and diversity of galaxy cluster dynamics \citep[see also][at lower mass scales]{Ramesh2023Zooming}. As expected, this diversity is highly correlated with the same properties of the host halo that we have previously discussed. This opens the way for future studies to use TNG-Cluster in order to disentangle the physics which contribute to the observable VSF of the ICM of galaxy clusters as a function of mass, region, scale, and phase.


\section{Summary and Conclusions}
\label{sec: summary}

This paper is part of a series introducing first results from the new TNG-Cluster simulation, a suite of 352 high-mass galaxy clusters ($14.3<\log_{10} M_{\rm 200c}/{\rm M_{\odot}}<15.4$ at $z=0$) simulated using the IllustrisTNG model. In this work, we present a comprehensive atlas of gas motions within these galaxy clusters at $z=0$, ranging from the core to the outskirts. The gas dynamics predicted by TNG-Cluster and quantified in this paper are the result of magneto-hydrodynamics, gravity, mergers, satellites and smooth accretion, feedback from star formation and from SMBHs, and all the phenomena captured within the full cosmological framework of IllustrisTNG. With respect to intrinsic kinematics in 3D space, our key findings are:

\begin{itemize}
    \item Gas motions in all cluster regions exhibit a very complex dynamics. In cluster cores and intermediate regions the gas motions are largely balanced between inflows and outflows. Radial velocities are Gaussian distributed about zero velocity. In the outskirts, the distributions become asymmetric and a second peak, reflecting cosmic accretion, emerges (Fig. \ref{Fig: vel_hist1d}).
    \item The velocity dispersion of gas is highest in the outskirts and decreases towards the core, reflecting the complex dynamics of cluster outskirts. More massive clusters have higher velocity dispersion (Figs. \ref{Fig: vrad_disp}, \ref{Fig: profiles_main}).
    \item Gas inflow rates are low in cluster cores and increase towards the outskirts, flattening beyond $\sim 0.5 R_{\rm 200c}$. More massive halos have higher inflow rates. Outflow rates, on the other hand, peak at intermediate distances from cluster centers and decline towards the core as well as the outskirts (Fig. \ref{Fig: profiles_main}).
\end{itemize}

Exploring the correlation between gas motions and halo properties, we find:

\begin{itemize}
    \item Unrelaxed clusters, typically undergoing mergers, exhibit faster inflows and outflows compared to relaxed halos.
    \item Late-forming clusters are more dynamically evolving compared to their early-forming counterparts, with more time to relax and reach quasi-equilibrium.
    \item Clusters with less massive central SMBHs show faster inflows and outflows compared to those with more massive SMBHs, at a fixed halo mass. This is partially because more massive SMBHs reside in earlier formed and relaxed halos.
    \item At fixed halo mass, clusters with lower SMBH accretion rates are more dynamically evolving, especially near halo centers, suggesting that SMBH feedback may impact cluster gas motions (Fig. \ref{Fig: profiles_vrad}).
    \item Gas metallicity and temperature can help distinguish between inflows and outflows. Cool, high metallicity gas as well as hot, low metallicity gas are both predominantly inflowing (Fig \ref{Fig: 2dhists}).
\end{itemize}

We then extend our study to the line of sight motion of the gas in 2D observational space. We find:

\begin{itemize}
    \item Line of sight velocity cannot uniquely identify inflows versus outflows in galaxy clusters, as it does not necessarily correlate with (true) radial velocity. Gas temperature can help break this degeneracy (Fig. \ref{Fig: vlos_vrad}).
    \item On the other hand, as for the 3D intrinsic counterpart, LOS velocity dispersion is strongly correlated with halo mass. It also increases by a factor of $\sim 2-4$ from the cores to the outskirts of clusters (Fig. \ref{Fig: vlos_lines}).
    \item The LOS velocity dispersion also retains the secondary dependencies seen for the 3D quantity: late-forming and unrelaxed halos have higher velocity dispersions near the core. Additionally, at fixed halo mass, clusters with either more massive SMBHs, or SMBHs with higher accretion rates, have lower velocity dispersion than otherwise (Fig. \ref{Fig: vLOS_disp}).
\end{itemize}

Finally, we investigate turbulence and the coherency of gas motion using the velocity structure function (VSF). Our main results are as follow:

\begin{itemize}
    \item Gas motion becomes less coherent with increasing separation distance, and with increasing halo mass. The intrinsic VSF roughly follows a Kolmogorov power-law relation with a slope of 1/3, except near the cores (Fig. \ref{Fig: vsf1_3D}).
    \item The VSF increases with distance from halo centers. This indicates more coherent gas motion in the outskirts and increasing turbulence towards the cores (Fig. \ref{Fig: vsf1_3D}).
    \item Halo-to-halo variability in the VSF is more pronounced closer to the halo center, while the outskirts are remarkably consistent across clusters (Fig. \ref{Fig: vsf1_3D}).
    \item The amplitude of the VSF in cluster cores ($< 50$\,kpc) correlates with SMBH accretion rate, suggesting AGN feedback impacts central ICM dynamics (Fig. \ref{Fig: vsf1_3D_heatmap}).
    \item In the halo core, unrelaxed, late-forming, and clusters with less massive SMBHs have more dynamic gas motion and less coherency. In the outskirts, minimal correlations with halo properties are present, except for a weak connection with formation time in the most massive clusters (Fig. \ref{Fig: vsf1_3D_heatmap}).
    \item The VSF predicted by TNG-Cluster compares favorably to observations, once observational effects are accounted for: this is the case on small and large separation scales, and for both cool gas in the core and the hot, extended ICM (Fig. \ref{Fig: vsf1_obs}).
\end{itemize}

Overall, our findings lay the groundwork for future studies that aim to delve deeper into the complex interplay between gas dynamics, halo properties, mergers, and supermassive black hole activity in galaxy clusters. The TNG-Cluster simulation provides a rich dataset that can be further exploited to understand the role of various physical processes in shaping the ICM and influencing cluster evolution. Specifically, investigating protoclusters—the progenitors of our current clusters at higher redshifts—could provide valuable insights into the temporal evolution of galaxy clusters. Additionally, the VSF could be a powerful tool for comparing simulations with upcoming high-resolution X-ray observations such as LEM \citep{Kraft2022Line,Mernier2023Exploring} and Athena \citep{Barret2016Athena}, thereby providing a more comprehensive picture of the physics of the ICM.

\section*{acknowledgements}
MA thanks Guinevere Kauffmann, Simon White, Philipp Girichidis, Ralf Klessen, Eugene Churazov, Jeremy Sanders, and Efrain Gatuzz for fruitful discussions. AP thanks Luca Sorriso-Valvo for useful conversations.
MA and DN acknowledge funding from the Deutsche Forschungsgemeinschaft (DFG) through an Emmy Noether Research Group (grant number NE 2441/1-1). YL acknowledges financial support from NSF grants AST-2107735 and AST-2219686, NASA grant 80NSSC22K0668, and Chandra X-ray Observatory grant TM3-24005X. KL acknowledges funding from the Hector Fellow Academy through a Research Career Development Award. KL is a fellow of the International Max Planck Research School for Astronomy and Cosmic Physics at the University of Heidelberg (IMPRS-HD).
This work is co-funded by the European Union (ERC, COSMIC-KEY, 101087822, PI: Pillepich).
The TNG-Cluster simulation has been run with compute time (i) awarded under the TNG-Cluster project on the HoreKa supercomputer, funded by the Ministry of Science, Research and the Arts Baden-Württemberg and by the Federal Ministry of Education and Research, (ii) on the bwForCluster Helix supercomputer, supported by the state of Baden-Württemberg through bwHPC and the German Research Foundation (DFG) through grant INST 35/1597-1 FUGG, (iii) with the Vera cluster of the Max Planck Institute for Astronomy (MPIA), as well as the Cobra and Raven clusters, all three operated by the Max Planck Computational Data Facility (MPCDF), and (iv) the BinAC cluster, supported by the High Performance and Cloud Computing Group at the Zentrum für Datenverarbeitung of the University of Tübingen, the state of Baden-Württemberg through bwHPC and the German Research Foundation (DFG) through grant no INST 37/935-1 FUGG.

\bibliographystyle{aa}
\bibliography{refbibtex}

\begin{thebibliography}{71}
\expandafter\ifx\csname natexlab\endcsname\relax\def\natexlab#1{#1}\fi

\bibitem[{{Arnaud} {et~al.}(2010){Arnaud}, {Pratt}, {Piffaretti}, {B{\"o}hringer}, {Croston}, \& {Pointecouteau}}]{Arnaud2010Universal}
{Arnaud}, M., {Pratt}, G.~W., {Piffaretti}, R., {et~al.} 2010, \aap, 517, A92

\bibitem[{{Ayromlou} {et~al.}(2023){Ayromlou}, {Nelson}, \& {Pillepich}}]{Ayromlou2023Feedback}
{Ayromlou}, M., {Nelson}, D., \& {Pillepich}, A. 2023, \mnras, 524, 5391

\bibitem[{{Barret} {et~al.}(2016){Barret}, {Lam Trong}, {den Herder}, {Piro}, {Barcons}, {Huovelin}, {Kelley}, {Mas-Hesse}, {Mitsuda}, {Paltani}, {Rauw}, {Ro{\.Z}anska}, {Wilms}, {Barbera}, {Bozzo}, {Ceballos}, {Charles}, {Decourchelle}, {den Hartog}, {Duval}, {Fiore}, {Gatti}, {Goldwurm}, {Jackson}, {Jonker}, {Kilbourne}, {Macculi}, {Mendez}, {Molendi}, {Orleanski}, {Pajot}, {Pointecouteau}, {Porter}, {Pratt}, {Pr{\^e}le}, {Ravera}, {Renotte}, {Schaye}, {Shinozaki}, {Valenziano}, {Vink}, {Webb}, {Yamasaki}, {Delcelier-Douchin}, {Le Du}, {Mesnager}, {Pradines}, {Branduardi-Raymont}, {Dadina}, {Finoguenov}, {Fukazawa}, {Janiuk}, {Miller}, {Naz{\'e}}, {Nicastro}, {Sciortino}, {Torrejon}, {Geoffray}, {Hernandez}, {Luno}, {Peille}, {Andr{\'e}}, {Daniel}, {Etcheverry}, {Gloaguen}, {Hassin}, {Hervet}, {Maussang}, {Moueza}, {Paillet}, {Vella}, {Campos Garrido}, {Damery}, {Panem}, {Panh}, {Bandler}, {Biffi}, {Boyce}, {Cl{\'e}net}, {DiPirro}, {Jamotton}, {Lotti}, {Schwander}, {Smith}, {van Leeuwen}, {van Weers},
  {Brand}, {Cobo}, {Dauser}, {de Plaa}, \& {Cucchetti}}]{Barret2016Athena}
{Barret}, D., {Lam Trong}, T., {den Herder}, J.-W., {et~al.} 2016, in Society of Photo-Optical Instrumentation Engineers (SPIE) Conference Series, Vol. 9905, Space Telescopes and Instrumentation 2016: Ultraviolet to Gamma Ray, ed. J.-W.~A. {den Herder}, T.~{Takahashi}, \& M.~{Bautz}, 99052F

\bibitem[{{Beckmann} {et~al.}(2022){Beckmann}, {Dubois}, {Pellissier}, {Polles}, \& {Olivares}}]{Beckmann2022AGN}
{Beckmann}, R.~S., {Dubois}, Y., {Pellissier}, A., {Polles}, F.~L., \& {Olivares}, V. 2022, \aap, 666, A71

\bibitem[{{Biffi} {et~al.}(2022){Biffi}, {ZuHone}, {Mroczkowski}, {Bulbul}, \& {Forman}}]{Biffi2022Velocity}
{Biffi}, V., {ZuHone}, J.~A., {Mroczkowski}, T., {Bulbul}, E., \& {Forman}, W. 2022, \aap, 663, A76

\bibitem[{{Bondi}(1952)}]{Bondi1952spherically}
{Bondi}, H. 1952, \mnras, 112, 195

\bibitem[{{Bondi} \& {Hoyle}(1944)}]{Bondi1944mechanism}
{Bondi}, H. \& {Hoyle}, F. 1944, \mnras, 104, 273

\bibitem[{{Chen} {et~al.}(2023{\natexlab{a}}){Chen}, {Chen}, {Rauch}, {Qu}, {Johnson}, {Li}, {Schaye}, {Rudie}, {Zahedy}, {Boettcher}, {Cooksey}, \& {Cantalupo}}]{Chen2023Empirical}
{Chen}, M.~C., {Chen}, H.-W., {Rauch}, M., {et~al.} 2023{\natexlab{a}}, \mnras, 518, 2354

\bibitem[{{Chen} {et~al.}(2023{\natexlab{b}}){Chen}, {Chen}, {Rauch}, {Qu}, {Johnson}, {Li}, {Schaye}, {Rudie}, {Zahedy}, {Boettcher}, {Cooksey}, \& {Cantalupo}}]{Chen2023Ensemble}
{Chen}, M.~C., {Chen}, H.-W., {Rauch}, M., {et~al.} 2023{\natexlab{b}} [\eprint[arXiv]{2310.18406}]

\bibitem[{{Chen} {et~al.}(2019){Chen}, {Heinz}, \& {En{\ss}lin}}]{Chen2019Jets}
{Chen}, Y.-H., {Heinz}, S., \& {En{\ss}lin}, T.~A. 2019, \mnras, 489, 1939

\bibitem[{Chiu {et~al.}(2018)Chiu, Mohr, McDonald, Bocquet, Desai, Klein, Israel, Ashby, Stanford, Benson, {et~al.}}]{chiu2018baryon}
Chiu, I., Mohr, J., McDonald, M., {et~al.} 2018, Monthly Notices of the Royal Astronomical Society, 478, 3072

\bibitem[{{Cicone} {et~al.}(2014){Cicone}, {Maiolino}, {Sturm}, {Graci{\'a}-Carpio}, {Feruglio}, {Neri}, {Aalto}, {Davies}, {Fiore}, {Fischer}, {Garc{\'\i}a-Burillo}, {Gonz{\'a}lez-Alfonso}, {Hailey-Dunsheath}, {Piconcelli}, \& {Veilleux}}]{Cicone2014Massive}
{Cicone}, C., {Maiolino}, R., {Sturm}, E., {et~al.} 2014, \aap, 562, A21

\bibitem[{{Croton} {et~al.}(2006){Croton}, {Springel}, {White}, {De Lucia}, {Frenk}, {Gao}, {Jenkins}, {Kauffmann}, {Navarro}, \& {Yoshida}}]{croton2006many}
{Croton}, D.~J., {Springel}, V., {White}, S. D.~M., {et~al.} 2006, \mnras, 365, 11

\bibitem[{{Cui} {et~al.}(2018){Cui}, {Knebe}, {Yepes}, {Pearce}, {Power}, {Dave}, {Arth}, {Borgani}, {Dolag}, {Elahi}, {Mostoghiu}, {Murante}, {Rasia}, {Stoppacher}, {Vega-Ferrero}, {Wang}, {Yang}, {Benson}, {Cora}, {Croton}, {Sinha}, {Stevens}, {Vega-Mart{\'\i}nez}, {Arthur}, {Baldi}, {Ca{\~n}as}, {Cialone}, {Cunnama}, {De Petris}, {Durando}, {Ettori}, {Gottl{\"o}ber}, {Nuza}, {Old}, {Pilipenko}, {Sorce}, \& {Welker}}]{Cui2018Three}
{Cui}, W., {Knebe}, A., {Yepes}, G., {et~al.} 2018, \mnras, 480, 2898

\bibitem[{{Davis} {et~al.}(1985){Davis}, {Efstathiou}, {Frenk}, \& {White}}]{Davis1985TheEvolution}
{Davis}, M., {Efstathiou}, G., {Frenk}, C.~S., \& {White}, S.~D.~M. 1985, \apj, 292, 371

\bibitem[{{Eckert} {et~al.}(2021){Eckert}, {Gaspari}, {Gastaldello}, {Le Brun}, \& {O'Sullivan}}]{Eckert2021Feedback}
{Eckert}, D., {Gaspari}, M., {Gastaldello}, F., {Le Brun}, A. M.~C., \& {O'Sullivan}, E. 2021, Universe, 7, 142

\bibitem[{{Ehlert} {et~al.}(2021){Ehlert}, {Weinberger}, {Pfrommer}, \& {Springel}}]{Ehlert2021Connecting}
{Ehlert}, K., {Weinberger}, R., {Pfrommer}, C., \& {Springel}, V. 2021, \mnras, 503, 1327

\bibitem[{{Fabian}(1994)}]{Fabian1994Cooling}
{Fabian}, A.~C. 1994, \araa, 32, 277

\bibitem[{{Fielding} {et~al.}(2020){Fielding}, {Tonnesen}, {DeFelippis}, {Li}, {Su}, {Bryan}, {Kim}, {Forbes}, {Somerville}, {Battaglia}, {Schneider}, {Li}, {Choi}, {Hayward}, \& {Hernquist}}]{Fielding2020First}
{Fielding}, D.~B., {Tonnesen}, S., {DeFelippis}, D., {et~al.} 2020, \apj, 903, 32

\bibitem[{{Ganguly} {et~al.}(2023){Ganguly}, {Li}, {Olivares}, {Su}, {Combes}, {Prakash}, {Hamer}, {Guillard}, \& {Ha}}]{Ganguly2023nature}
{Ganguly}, S., {Li}, Y., {Olivares}, V., {et~al.} 2023, Frontiers in Astronomy and Space Sciences, 10, 1138613

\bibitem[{{Gatuzz} {et~al.}(2023){Gatuzz}, {Mohapatra}, {Federrath}, {Sanders}, {Liu}, {Walker}, \& {Pinto}}]{Gatuzz2023Measuring}
{Gatuzz}, E., {Mohapatra}, R., {Federrath}, C., {et~al.} 2023, arXiv e-prints, arXiv:2307.02576

\bibitem[{{Ghirardini} {et~al.}(2019){Ghirardini}, {Eckert}, {Ettori}, {Pointecouteau}, {Molendi}, {Gaspari}, {Rossetti}, {De Grandi}, {Roncarelli}, {Bourdin}, {Mazzotta}, {Rasia}, \& {Vazza}}]{Ghirardini2019Universal}
{Ghirardini}, V., {Eckert}, D., {Ettori}, S., {et~al.} 2019, \aap, 621, A41

\bibitem[{Giodini {et~al.}(2009)Giodini, Pierini, Finoguenov, Pratt, Boehringer, Leauthaud, Guzzo, Aussel, Bolzonella, Capak, {et~al.}}]{giodini2009stellar}
Giodini, S., Pierini, D., Finoguenov, A., {et~al.} 2009, The Astrophysical Journal, 703, 982

\bibitem[{{Hitomi Collaboration} {et~al.}(2016){Hitomi Collaboration}, {Aharonian}, {Akamatsu}, {Akimoto}, {Allen}, {Anabuki}, {Angelini}, {Arnaud}, {Audard}, {Awaki}, {Axelsson}, {Bamba}, {Bautz}, {Blandford}, {Brenneman}, {Brown}, {Bulbul}, {Cackett}, {Chernyakova}, {Chiao}, {Coppi}, {Costantini}, {de Plaa}, {den Herder}, {Done}, {Dotani}, {Ebisawa}, {Eckart}, {Enoto}, {Ezoe}, {Fabian}, {Ferrigno}, {Foster}, {Fujimoto}, {Fukazawa}, {Furuzawa}, {Galeazzi}, {Gallo}, {Gandhi}, {Giustini}, {Goldwurm}, {Gu}, {Guainazzi}, {Haba}, {Hagino}, {Hamaguchi}, {Harrus}, {Hatsukade}, {Hayashi}, {Hayashi}, {Hayashida}, {Hiraga}, {Hornschemeier}, {Hoshino}, {Hughes}, {Iizuka}, {Inoue}, {Inoue}, {Ishibashi}, {Ishida}, {Ishikawa}, {Ishisaki}, {Itoh}, {Iyomoto}, {Kaastra}, {Kallman}, {Kamae}, {Kara}, {Kataoka}, {Katsuda}, {Katsuta}, {Kawaharada}, {Kawai}, {Kelley}, {Khangulyan}, {Kilbourne}, {King}, {Kitaguchi}, {Kitamoto}, {Kitayama}, {Kohmura}, {Kokubun}, {Koyama}, {Koyama}, {Kretschmar}, {Krimm}, {Kubota}, {Kunieda},
  {Laurent}, {Lebrun}, {Lee}, {Leutenegger}, {Limousin}, {Loewenstein}, {Long}, {Lumb}, {Madejski}, {Maeda}, {Maier}, {Makishima}, {Markevitch}, {Matsumoto}, {Matsushita}, {McCammon}, {McNamara}, {Mehdipour}, {Miller}, {Miller}, {Mineshige}, {Mitsuda}, {Mitsuishi}, {Miyazawa}, {Mizuno}, {Mori}, {Mori}, {Moseley}, {Mukai}, {Murakami}, {Murakami}, {Mushotzky}, {Nagino}, {Nakagawa}, {Nakajima}, {Nakamori}, {Nakano}, {Nakashima}, {Nakazawa}, {Nobukawa}, {Noda}, {Nomachi}, {O'Dell}, {Odaka}, {Ohashi}, {Ohno}, {Okajima}, {Ota}, {Ozaki}, {Paerels}, {Paltani}, {Parmar}, {Petre}, {Pinto}, {Pohl}, {Porter}, {Pottschmidt}, {Ramsey}, {Reynolds}, {Russell}, {Safi-Harb}, {Saito}, {Sakai}, {Sameshima}, {Sato}, {Sato}, {Sato}, {Sawada}, {Schartel}, {Serlemitsos}, {Seta}, {Shidatsu}, {Simionescu}, {Smith}, {Soong}, {Stawarz}, {Sugawara}, {Sugita}, {Szymkowiak}, {Tajima}, {Takahashi}, {Takahashi}, {Takeda}, {Takei}, {Tamagawa}, {Tamura}, {Tamura}, {Tanaka}, {Tanaka}, {Tanaka}, {Tashiro}, {Tawara}, {Terada}, {Terashima},
  {Tombesi}, {Tomida}, {Tsuboi}, {Tsujimoto}, {Tsunemi}, {Tsuru}, {Uchida}, {Uchiyama}, {Uchiyama}, {Ueda}, {Ueda}, {Ueno}, {Uno}, {Urry}, {Ursino}, {de Vries}, {Watanabe}, {Werner}, {Wik}, {Wilkins}, {Williams}, {Yamada}, {Yamaguchi}, {Yamaoka}, {Yamasaki}, {Yamauchi}, {Yamauchi}, {Yaqoob}, {Yatsu}, {Yonetoku}, {Yoshida}, {Yuasa}, {Zhuravleva}, \& {Zoghbi}}]{Hitomi2016quiescent}
{Hitomi Collaboration}, {Aharonian}, F., {Akamatsu}, H., {et~al.} 2016, \nat, 535, 117

\bibitem[{{Hitomi Collaboration} {et~al.}(2018){Hitomi Collaboration}, {Aharonian}, {Akamatsu}, {Akimoto}, {Allen}, {Angelini}, {Audard}, {Awaki}, {Axelsson}, {Bamba}, {Bautz}, {Blandford}, {Brenneman}, {Brown}, {Bulbul}, {Cackett}, {Canning}, {Chernyakova}, {Chiao}, {Coppi}, {Costantini}, {de Plaa}, {de Vries}, {den Herder}, {Done}, {Dotani}, {Ebisawa}, {Eckart}, {Enoto}, {Ezoe}, {Fabian}, {Ferrigno}, {Foster}, {Fujimoto}, {Fukazawa}, {Furuzawa}, {Galeazzi}, {Gallo}, {Gandhi}, {Giustini}, {Goldwurm}, {Gu}, {Guainazzi}, {Haba}, {Hagino}, {Hamaguchi}, {Harrus}, {Hatsukade}, {Hayashi}, {Hayashi}, {Hayashi}, {Hayashida}, {Hiraga}, {Hornschemeier}, {Hoshino}, {Hughes}, {Ichinohe}, {Iizuka}, {Inoue}, {Inoue}, {Inoue}, {Ishida}, {Ishikawa}, {Ishisaki}, {Iwai}, {Kaastra}, {Kallman}, {Kamae}, {Kataoka}, {Katsuda}, {Kawai}, {Kelley}, {Kilbourne}, {Kitaguchi}, {Kitamoto}, {Kitayama}, {Kohmura}, {Kokubun}, {Koyama}, {Koyama}, {Kretschmar}, {Krimm}, {Kubota}, {Kunieda}, {Laurent}, {Lee}, {Leutenegger}, {Limousin},
  {Loewenstein}, {Long}, {Lumb}, {Madejski}, {Maeda}, {Maier}, {Makishima}, {Markevitch}, {Matsumoto}, {Matsushita}, {McCammon}, {McNamara}, {Mehdipour}, {Miller}, {Miller}, {Mineshige}, {Mitsuda}, {Mitsuishi}, {Miyazawa}, {Mizuno}, {Mori}, {Mori}, {Mukai}, {Murakami}, {Mushotzky}, {Nakagawa}, {Nakajima}, {Nakamori}, {Nakashima}, {Nakazawa}, {Nobukawa}, {Nobukawa}, {Noda}, {Odaka}, {Ohashi}, {Ohno}, {Okajima}, {Ota}, {Ozaki}, {Paerels}, {Paltani}, {Petre}, {Pinto}, {Porter}, {Pottschmidt}, {Reynolds}, {Safi-Harb}, {Saito}, {Sakai}, {Sasaki}, {Sato}, {Sato}, {Sato}, {Sawada}, {Schartel}, {Serlemtsos}, {Seta}, {Shidatsu}, {Simionescu}, {Smith}, {Soong}, {Stawarz}, {Sugawara}, {Sugita}, {Szymkowiak}, {Tajima}, {Takahashi}, {Takahashi}, {Takeda}, {Takei}, {Tamagawa}, {Tamura}, {Tanaka}, {Tanaka}, {Tanaka}, {Tanaka}, {Tashiro}, {Tawara}, {Terada}, {Terashima}, {Tombesi}, {Tomida}, {Tsuboi}, {Tsujimoto}, {Tsunemi}, {Tsuru}, {Uchida}, {Uchiyama}, {Uchiyama}, {Ueda}, {Ueda}, {Uno}, {Urry}, {Ursino}, {Wang},
  {Watanabe}, {Werner}, {Wilkins}, {Williams}, {Yamada}, {Yamaguchi}, {Yamaoka}, {Yamasaki}, {Yamauchi}, {Yamauchi}, {Yaqoob}, {Yatsu}, {Yonetoku}, {Zhuravleva}, \& {Zoghbi}}]{Hitomi2018Atmospheric}
{Hitomi Collaboration}, {Aharonian}, F., {Akamatsu}, H., {et~al.} 2018, \pasj, 70, 9

\bibitem[{{Kauffmann} {et~al.}(2003){Kauffmann}, {Heckman}, {Tremonti}, {Brinchmann}, {Charlot}, {White}, {Ridgway}, {Brinkmann}, {Fukugita}, {Hall}, {Ivezi{\'c}}, {Richards}, \& {Schneider}}]{Kauffmann2003Host}
{Kauffmann}, G., {Heckman}, T.~M., {Tremonti}, C., {et~al.} 2003, \mnras, 346, 1055

\bibitem[{{Kolmogorov}(1941)}]{Kolmogorov1941Local}
{Kolmogorov}, A. 1941, Akademiia Nauk SSSR Doklady, 30, 301

\bibitem[{{Komatsu} {et~al.}(2011){Komatsu}, {Smith}, {Dunkley}, {Bennett}, {Gold}, {Hinshaw}, {Jarosik}, {Larson}, {Nolta}, {Page}, {Spergel}, {Halpern}, {Hill}, {Kogut}, {Limon}, {Meyer}, {Odegard}, {Tucker}, {Weiland}, {Wollack}, \& {Wright}}]{komatsu2011seven}
{Komatsu}, E., {Smith}, K.~M., {Dunkley}, J., {et~al.} 2011, \apjs, 192, 18

\bibitem[{{Kraft} {et~al.}(2022){Kraft}, {Markevitch}, {Kilbourne}, {Adams}, {Akamatsu}, {Ayromlou}, {Bandler}, {Bennett}, {Bhardwaj}, {Biffi}, {Bodewits}, {Bogdan}, {Bonamente}, {Borgani}, {Branduardi-Raymont}, {Bregman}, {Burchett}, {Cann}, {Carter}, {Chakraborty}, {Churazov}, {Crain}, {Cumbee}, {Dave}, {DiPirro}, {Dolag}, {Bertrand Doriese}, {Drake}, {Dunn}, {Eckart}, {Eckert}, {Ettori}, {Forman}, {Galeazzi}, {Gall}, {Gatuzz}, {Hell}, {Hodges-Kluck}, {Jackman}, {Jahromi}, {Jennings}, {Jones}, {Kaaret}, {Kavanagh}, {Kelley}, {Khabibullin}, {Kim}, {Koutroumpa}, {Kovacs}, {Kuntz}, {Lin}, {Lau}, {Lee}, {Leutenegger}, {Lisse}, {Lovisari}, {McCammon}, {McEntee}, {Mernier}, {Miller}, {Nagai}, {Negro}, {Nelson}, {Ness}, {Nulsen}, {Ogorzalek}, {Oppenheimer}, {Oskinova}, {Patnaude}, {Pfeifle}, {Pillepich}, {Plucinsky}, {Pooley}, {Porter}, {Randall}, {Rasia}, {Raymond}, {Ruszkowski}, {Sakai}, {Sarkar}, {Sasaki}, {Sato}, {Schellenberger}, {Schaye}, {Simionescu}, {Smith}, {Steiner}, {Stern}, {Su}, {Sun}, {Tremblay},
  {Truong}, {Tutt}, {Veilleux}, {Vikhlinin}, {Vladutescu-Zopp}, {Vogelsberger}, {Walker}, {Weaver}, {Weigt}, {Werk}, {Werner}, {Wolk}, {Zhang}, {Zhang}, {Zhuravleva}, \& {ZuHone}}]{Kraft2022Line}
{Kraft}, R., {Markevitch}, M., {Kilbourne}, C., {et~al.} 2022, arXiv e-prints, arXiv:2211.09827

\bibitem[{{Kravtsov} \& {Borgani}(2012)}]{Kravtsov2012Formation}
{Kravtsov}, A.~V. \& {Borgani}, S. 2012, \araa, 50, 353

\bibitem[{{Lacey} \& {Cole}(1994)}]{Lacey1994Merger}
{Lacey}, C. \& {Cole}, S. 1994, \mnras, 271, 676

\bibitem[{{Li} {et~al.}(2020){Li}, {Gendron-Marsolais}, {Zhuravleva}, {Xu}, {Simionescu}, {Tremblay}, {Lochhaas}, {Bryan}, {Quataert}, {Murray}, {Boselli}, {Hlavacek-Larrondo}, {Zheng}, {Fossati}, {Li}, {Emsellem}, {Sarzi}, {Arzamasskiy}, \& {Vishniac}}]{Li2020Direct}
{Li}, Y., {Gendron-Marsolais}, M.-L., {Zhuravleva}, I., {et~al.} 2020, \apjl, 889, L1

\bibitem[{{Lochhaas} {et~al.}(2020){Lochhaas}, {Bryan}, {Li}, {Li}, \& {Fielding}}]{Lochhaas2020Properties}
{Lochhaas}, C., {Bryan}, G.~L., {Li}, Y., {Li}, M., \& {Fielding}, D. 2020, \mnras, 493, 1461

\bibitem[{{{\L}okas}(2023)}]{Lokas2023Merging}
{{\L}okas}, E.~L. 2023, \aap, 673, A131

\bibitem[{{Marinacci} {et~al.}(2018){Marinacci}, {Vogelsberger}, {Pakmor}, {Torrey}, {Springel}, {Hernquist}, {Nelson}, {Weinberger}, {Pillepich}, {Naiman}, \& {Genel}}]{marinacci2018first}
{Marinacci}, F., {Vogelsberger}, M., {Pakmor}, R., {et~al.} 2018, \mnras, 480, 5113

\bibitem[{{Markevitch} \& {Vikhlinin}(2007)}]{Markevitch2007Shocks}
{Markevitch}, M. \& {Vikhlinin}, A. 2007, \physrep, 443, 1

\bibitem[{{McNamara} \& {Nulsen}(2007)}]{McNamara2007Heating}
{McNamara}, B.~R. \& {Nulsen}, P.~E.~J. 2007, \araa, 45, 117

\bibitem[{{Mernier} {et~al.}(2023){Mernier}, {Su}, {Markevitch}, {Zhang}, {Simionescu}, {Rasia}, {Lin}, {Zhuravleva}, {Sarkar}, {Kraft}, {Ogorzalek}, {Ayromlou}, {Forman}, {Jones}, {Bregman}, {Ettori}, {Dolag}, {Biffi}, {Churazov}, {Sun}, {ZuHone}, {Bogd{\'a}n}, {Khabibullin}, {Werner}, {Truong}, {Chakraborty}, {Walker}, {Vogelsberger}, {Pillepich}, \& {Mirakhor}}]{Mernier2023Exploring}
{Mernier}, F., {Su}, Y., {Markevitch}, M., {et~al.} 2023, arXiv e-prints, arXiv:2310.04499

\bibitem[{{Mitchell} \& {Schaye}(2022)}]{Mitchell2022Baryonic}
{Mitchell}, P.~D. \& {Schaye}, J. 2022, \mnras, 511, 2600

\bibitem[{Mo {et~al.}(2010)Mo, Van~den Bosch, \& White}]{mo2010galaxy}
Mo, H., Van~den Bosch, F., \& White, S. 2010, Galaxy formation and evolution (Cambridge University Press)

\bibitem[{{Mohapatra} {et~al.}(2022){Mohapatra}, {Jetti}, {Sharma}, \& {Federrath}}]{Mohapatra2022Velocity}
{Mohapatra}, R., {Jetti}, M., {Sharma}, P., \& {Federrath}, C. 2022, \mnras, 510, 2327

\bibitem[{{Mohapatra} \& {Sharma}(2019)}]{Mohapatra2019Turbulence}
{Mohapatra}, R. \& {Sharma}, P. 2019, \mnras, 484, 4881

\bibitem[{{Nagai} {et~al.}(2007){Nagai}, {Kravtsov}, \& {Vikhlinin}}]{Nagai2007Effects}
{Nagai}, D., {Kravtsov}, A.~V., \& {Vikhlinin}, A. 2007, \apj, 668, 1

\bibitem[{{Naiman} {et~al.}(2018){Naiman}, {Pillepich}, {Springel}, {Ramirez-Ruiz}, {Torrey}, {Vogelsberger}, {Pakmor}, {Nelson}, {Marinacci}, {Hernquist}, {Weinberger}, \& {Genel}}]{naiman2018first}
{Naiman}, J.~P., {Pillepich}, A., {Springel}, V., {et~al.} 2018, \mnras, 477, 1206

\bibitem[{{Navarro} {et~al.}(1995){Navarro}, {Frenk}, \& {White}}]{Navarro1995Simulations}
{Navarro}, J.~F., {Frenk}, C.~S., \& {White}, S. D.~M. 1995, \mnras, 275, 720

\bibitem[{{Nelson} {et~al.}(2019){Nelson}, {Pillepich}, {Springel}, {Pakmor}, {Weinberger}, {Genel}, {Torrey}, {Vogelsberger}, {Marinacci}, \& {Hernquist}}]{nelson2019First}
{Nelson}, D., {Pillepich}, A., {Springel}, V., {et~al.} 2019, \mnras, 490, 3234

\bibitem[{{Nelson} {et~al.}(2018){Nelson}, {Pillepich}, {Springel}, {Weinberger}, {Hernquist}, {Pakmor}, {Genel}, {Torrey}, {Vogelsberger}, {Kauffmann}, {Marinacci}, \& {Naiman}}]{nelson18a}
{Nelson}, D., {Pillepich}, A., {Springel}, V., {et~al.} 2018, \mnras, 475, 624

\bibitem[{{Pakmor} {et~al.}(2011){Pakmor}, {Bauer}, \& {Springel}}]{pakmor2011magnetohydrodynamics}
{Pakmor}, R., {Bauer}, A., \& {Springel}, V. 2011, \mnras, 418, 1392

\bibitem[{{Pakmor} \& {Springel}(2013)}]{pakmor2013simulations}
{Pakmor}, R. \& {Springel}, V. 2013, \mnras, 432, 176

\bibitem[{{Peterson} {et~al.}(2003){Peterson}, {Kahn}, {Paerels}, {Kaastra}, {Tamura}, {Bleeker}, {Ferrigno}, \& {Jernigan}}]{Petersen2003HighResolution}
{Peterson}, J.~R., {Kahn}, S.~M., {Paerels}, F.~B.~S., {et~al.} 2003, \apj, 590, 207

\bibitem[{{Pillepich} {et~al.}(2018{\natexlab{a}}){Pillepich}, {Nelson}, {Hernquist}, {Springel}, {Pakmor}, {Torrey}, {Weinberger}, {Genel}, {Naiman}, {Marinacci}, \& {Vogelsberger}}]{pillepich2018First}
{Pillepich}, A., {Nelson}, D., {Hernquist}, L., {et~al.} 2018{\natexlab{a}}, \mnras, 475, 648

\bibitem[{{Pillepich} {et~al.}(2021){Pillepich}, {Nelson}, {Truong}, {Weinberger}, {Martin-Navarro}, {Springel}, {Faber}, \& {Hernquist}}]{Pillepich2021X-ray}
{Pillepich}, A., {Nelson}, D., {Truong}, N., {et~al.} 2021, \mnras, 508, 4667

\bibitem[{{Pillepich} {et~al.}(2018{\natexlab{b}}){Pillepich}, {Springel}, {Nelson}, {Genel}, {Naiman}, {Pakmor}, {Hernquist}, {Torrey}, {Vogelsberger}, {Weinberger}, \& {Marinacci}}]{pillepich2018Simulating}
{Pillepich}, A., {Springel}, V., {Nelson}, D., {et~al.} 2018{\natexlab{b}}, \mnras, 473, 4077

\bibitem[{{Planck Collaboration}(2016)}]{planck2015_xiii}
{Planck Collaboration}. 2016, \aap, 594, A13

\bibitem[{{Ramesh} \& {Nelson}(2023)}]{Ramesh2023Zooming}
{Ramesh}, R. \& {Nelson}, D. 2023, arXiv e-prints, arXiv:2307.11143

\bibitem[{{Sarazin}(2002)}]{Sarazin2002Physics}
{Sarazin}, C.~L. 2002, in Astrophysics and Space Science Library, Vol. 272, Merging Processes in Galaxy Clusters, ed. L.~{Feretti}, I.~M. {Gioia}, \& G.~{Giovannini}, 1--38

\bibitem[{{Simionescu} {et~al.}(2019){Simionescu}, {ZuHone}, {Zhuravleva}, {Churazov}, {Gaspari}, {Nagai}, {Werner}, {Roediger}, {Canning}, {Eckert}, {Gu}, \& {Paerels}}]{Simionescu2019Constraining}
{Simionescu}, A., {ZuHone}, J., {Zhuravleva}, I., {et~al.} 2019, \ssr, 215, 24

\bibitem[{{Spergel} {et~al.}(2003){Spergel}, {Verde}, {Peiris}, {Komatsu}, {Nolta}, {Bennett}, {Halpern}, {Hinshaw}, {Jarosik}, {Kogut}, {Limon}, {Meyer}, {Page}, {Tucker}, {Weiland}, {Wollack}, \& {Wright}}]{Spergel2003First}
{Spergel}, D.~N., {Verde}, L., {Peiris}, H.~V., {et~al.} 2003, \apjs, 148, 175

\bibitem[{{Springel}(2010)}]{springel2010pur}
{Springel}, V. 2010, \mnras, 401, 791

\bibitem[{{Springel} {et~al.}(2018){Springel}, {Pakmor}, {Pillepich}, {Weinberger}, {Nelson}, {Hernquist}, {Vogelsberger}, {Genel}, {Torrey}, {Marinacci}, \& {Naiman}}]{springel2018first}
{Springel}, V., {Pakmor}, R., {Pillepich}, A., {et~al.} 2018, \mnras, 475, 676

\bibitem[{{Springel} {et~al.}(2001){Springel}, {White}, {Tormen}, \& {Kauffmann}}]{springel2001populating}
{Springel}, V., {White}, S. D.~M., {Tormen}, G., \& {Kauffmann}, G. 2001, \mnras, 328, 726

\bibitem[{{Tully}(1987)}]{Tully1987Nearby}
{Tully}, R.~B. 1987, \apj, 321, 280

\bibitem[{{Vazza} {et~al.}(2017){Vazza}, {Jones}, {Br{\"u}ggen}, {Brunetti}, {Gheller}, {Porter}, \& {Ryu}}]{Vazza2017Turbulence}
{Vazza}, F., {Jones}, T.~W., {Br{\"u}ggen}, M., {et~al.} 2017, \mnras, 464, 210

\bibitem[{{Voit}(2005)}]{Voit2005Tracing}
{Voit}, G.~M. 2005, Reviews of Modern Physics, 77, 207

\bibitem[{{Wang} {et~al.}(2021){Wang}, {Ruszkowski}, {Pfrommer}, {Oh}, \& {Yang}}]{Wang2021NonKolmogorov}
{Wang}, C., {Ruszkowski}, M., {Pfrommer}, C., {Oh}, S.~P., \& {Yang}, H. Y.~K. 2021, \mnras, 504, 898

\bibitem[{{Weinberger} {et~al.}(2017){Weinberger}, {Springel}, {Hernquist}, {Pillepich}, {Marinacci}, {Pakmor}, {Nelson}, {Genel}, {Vogelsberger}, {Naiman}, \& {Torrey}}]{weinberger17}
{Weinberger}, R., {Springel}, V., {Hernquist}, L., {et~al.} 2017, \mnras, 465, 3291

\bibitem[{{White} \& {Frenk}(1991)}]{white1991galaxy}
{White}, S. D.~M. \& {Frenk}, C.~S. 1991, \apj, 379, 52

\bibitem[{{White} \& {Rees}(1978)}]{white1978core}
{White}, S.~D.~M. \& {Rees}, M.~J. 1978, \mnras, 183, 341

\bibitem[{{Yang} \& {Reynolds}(2016)}]{Yang2016How}
{Yang}, H. Y.~K. \& {Reynolds}, C.~S. 2016, \apj, 829, 90

\bibitem[{{Zhuravleva} {et~al.}(2018){Zhuravleva}, {Allen}, {Mantz}, \& {Werner}}]{Zhuravleva2018Gas}
{Zhuravleva}, I., {Allen}, S.~W., {Mantz}, A., \& {Werner}, N. 2018, \apj, 865, 53

\bibitem[{{ZuHone} {et~al.}(2016){ZuHone}, {Miller}, {Simionescu}, \& {Bautz}}]{Zuhone2016Simulating}
{ZuHone}, J.~A., {Miller}, E.~D., {Simionescu}, A., \& {Bautz}, M.~W. 2016, \apj, 821, 6

\end{thebibliography}

\begin{appendix}

\section{Halo correlations}
\label{app: halo correlations}
\begin{figure*}
   \centering
   \includegraphics[width=1\textwidth]{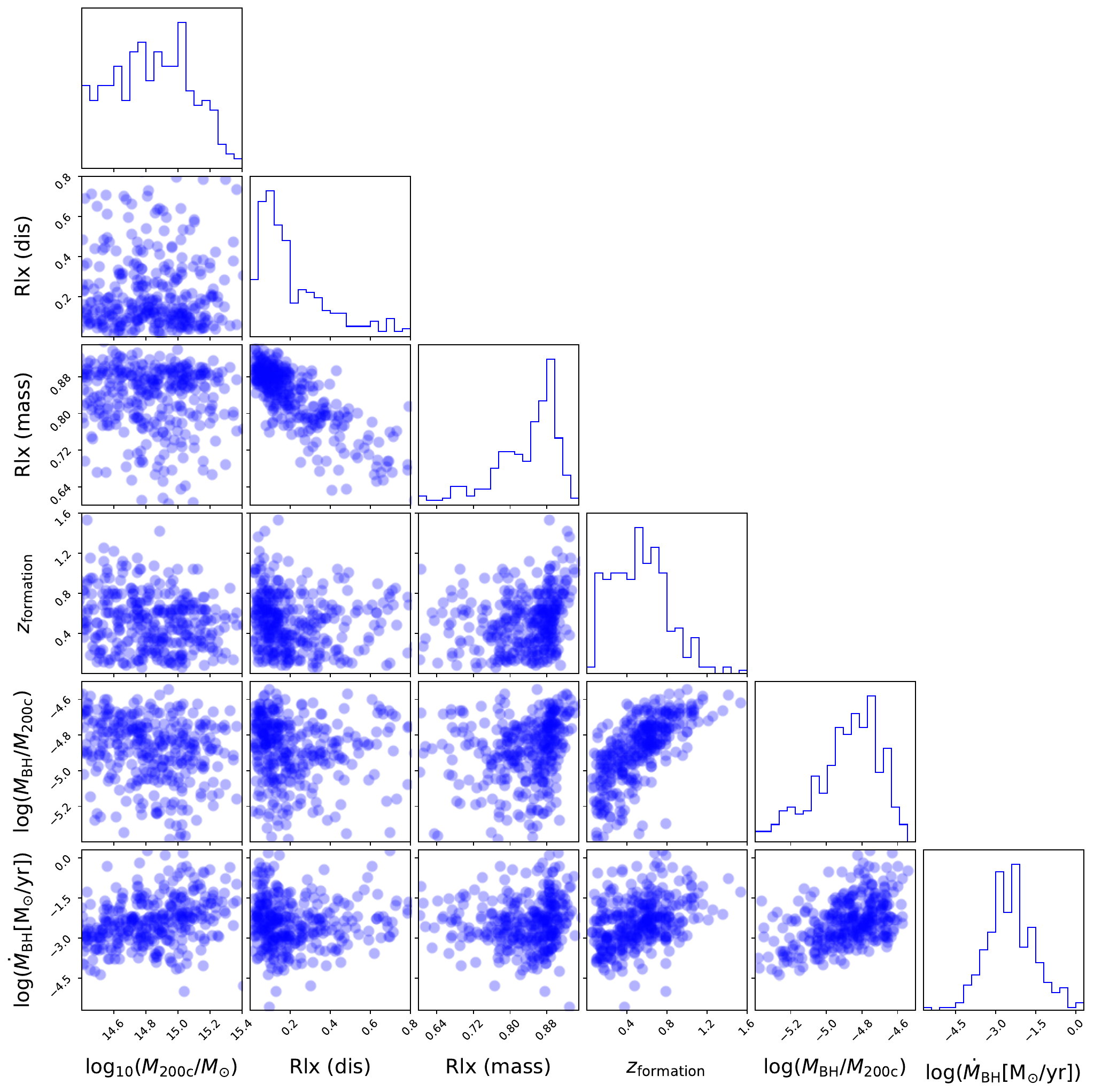}
   \caption{The distribution and cross correlation of several halo properties used in this work. Each row/column corresponds to a specific quantity. The rightmost panel of every row displays the probability distribution of that quantity, while the other panels represent the cross correlation between different pairs of quantities. From left to right: halo mass ($M_{\rm 200c}$), relaxedness parameter based on distance, relaxedness parameter based on mass, formation redshift, SMBH mass, and SMBH accretion rate. Each dot corresponds to an individual galaxy cluster.}
\label{Fig: corner_plot}
\end{figure*}

Fig. \ref{Fig: corner_plot} illustrates the distribution and cross-correlation of several halo properties examined in this study. The rightmost panel in each row presents the distribution of that particular variable, while the remaining panels depict the cross-correlation between different pairs of variables. Our two measures of relaxedness are highly correlated, affirming the robustness of our methodology. Notably, there exists a strong correlation between black hole mass, formation redshift, and relaxedness; more massive black holes are typically found in early-forming and more relaxed halos. We find a similar correlation between black hole 
accretion rate and black hole mass: more massive black holes exhibit higher accretion rates. These cross-correlations are also reflected in the relationship between gas motion and halo properties, as elaborated in Section \ref{subsec: impact of galaxy and halo properties}.

\section{Gas flow rates and velocity dispersion profiles}
\label{app: gas_flow}

\begin{figure*}
    \centering
    \includegraphics[width=0.45\textwidth]{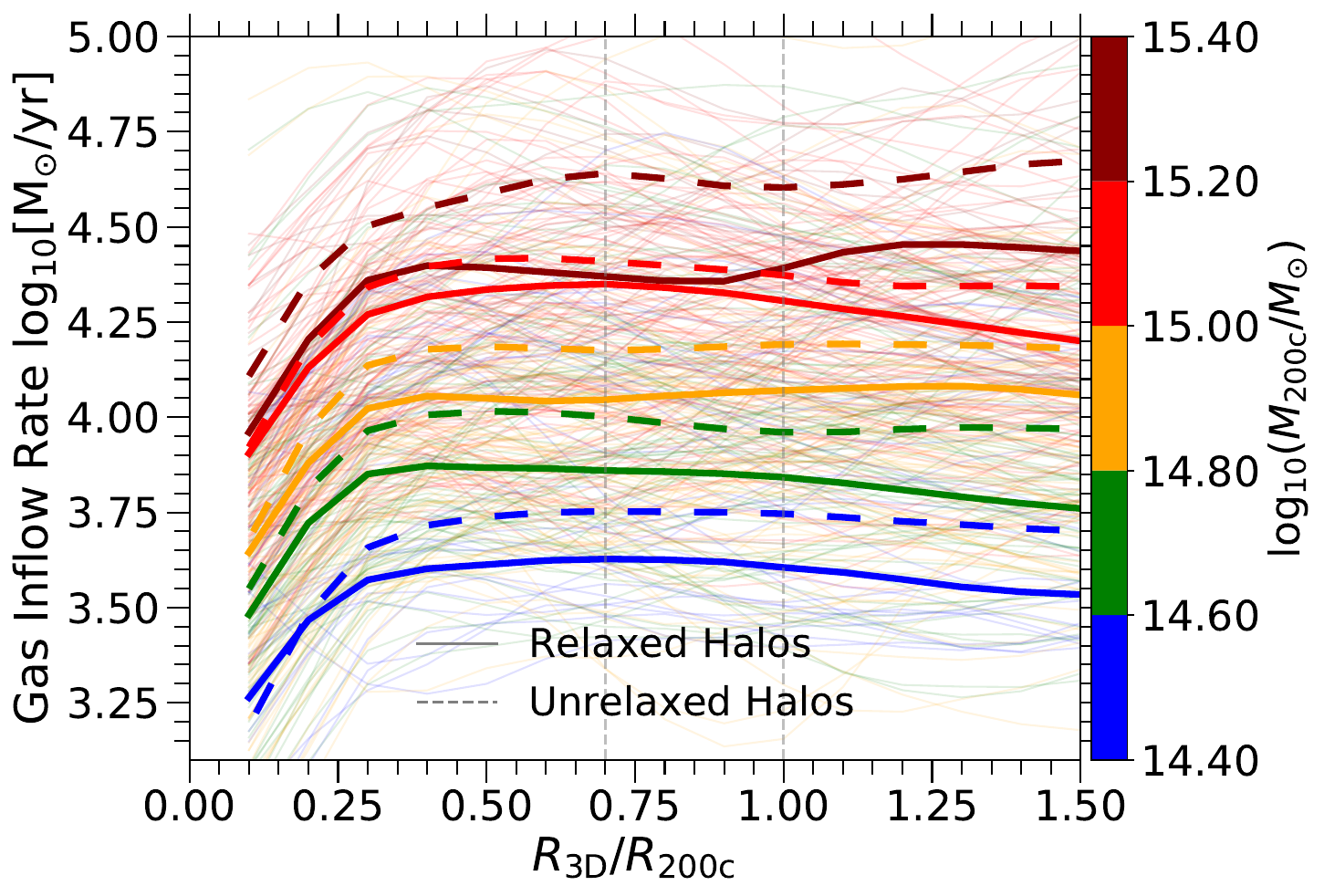}\hspace{5mm}
    \includegraphics[width=0.45\textwidth]{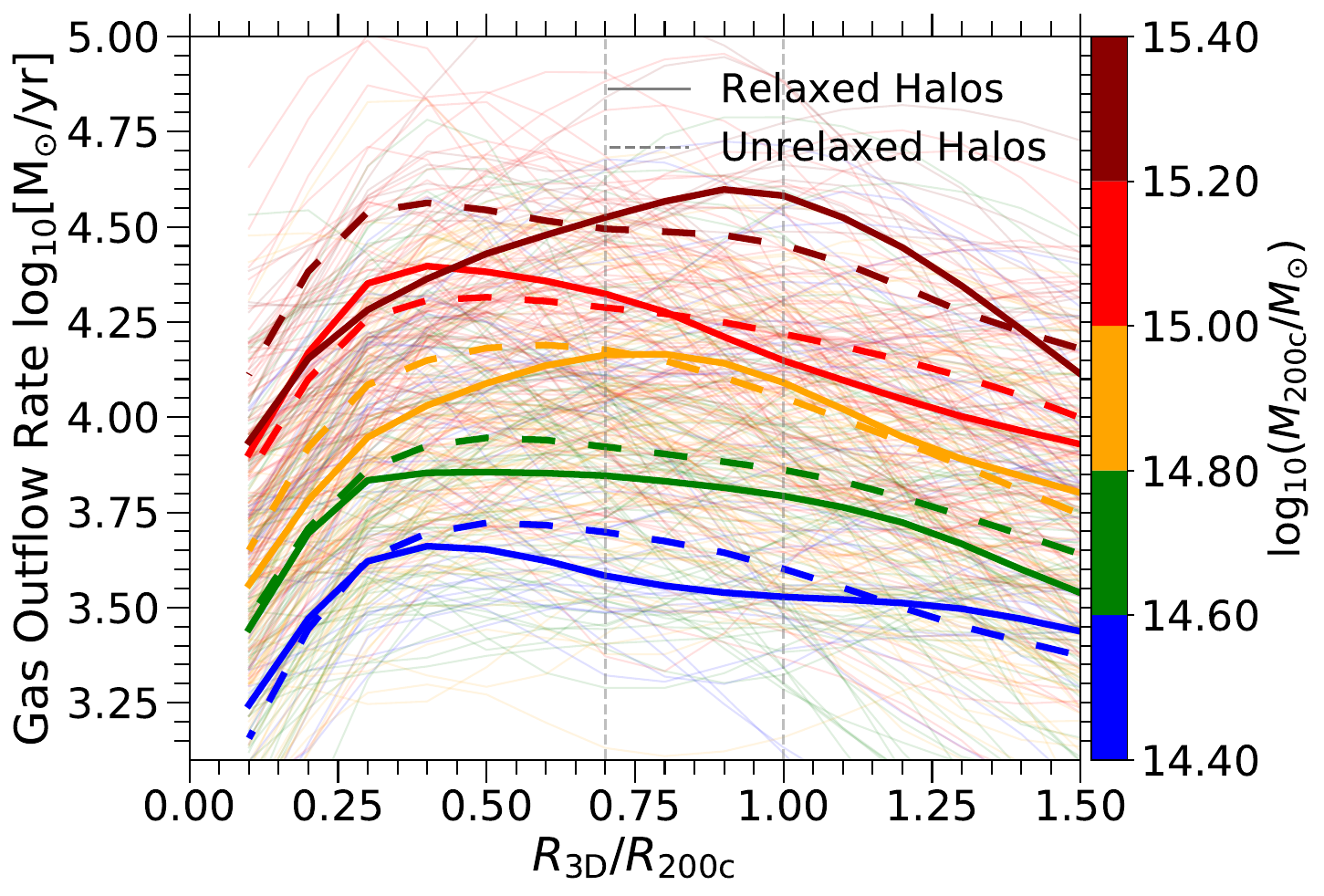}
    \includegraphics[width=0.45\textwidth]{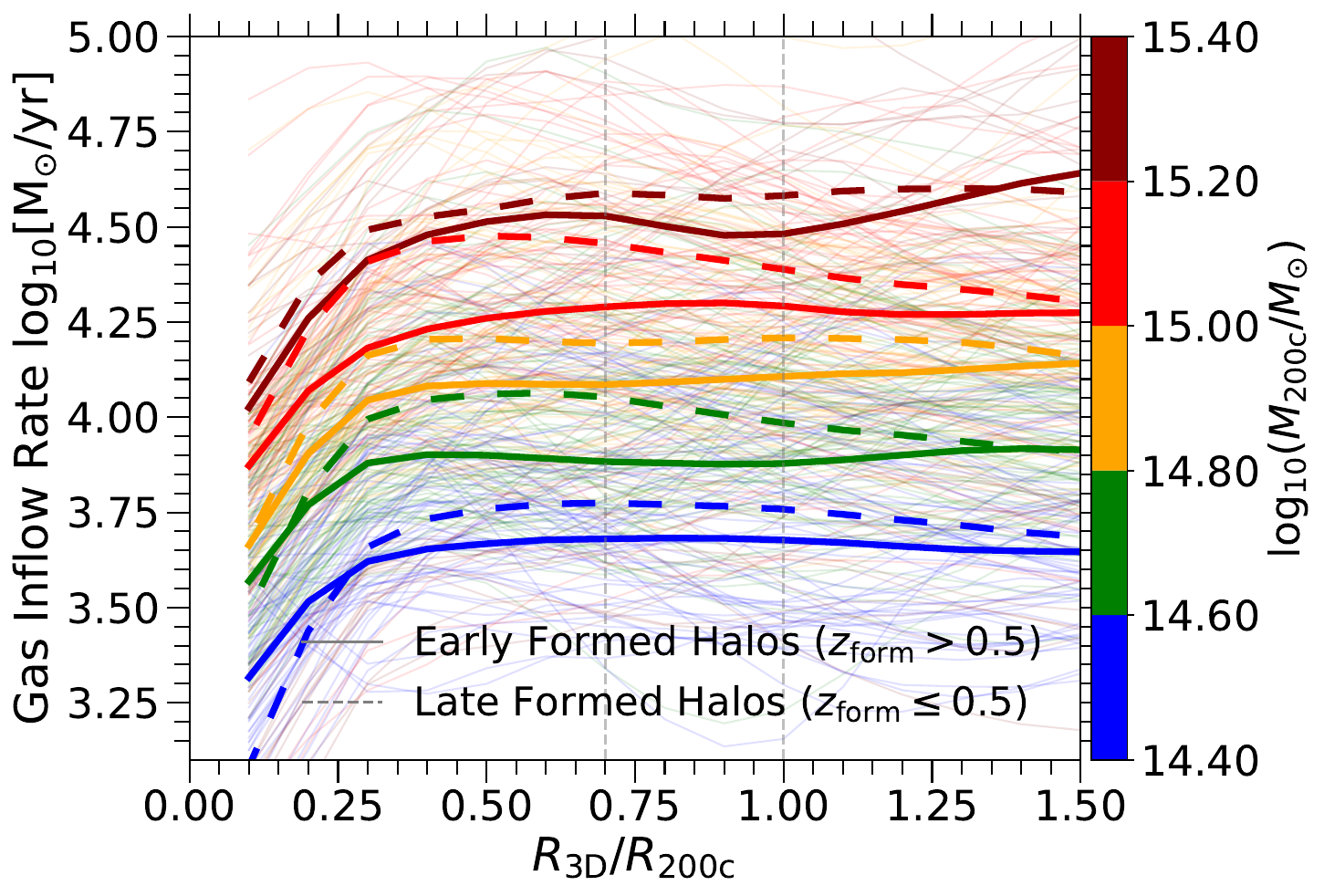}\hspace{5mm}
    \includegraphics[width=0.45\textwidth]{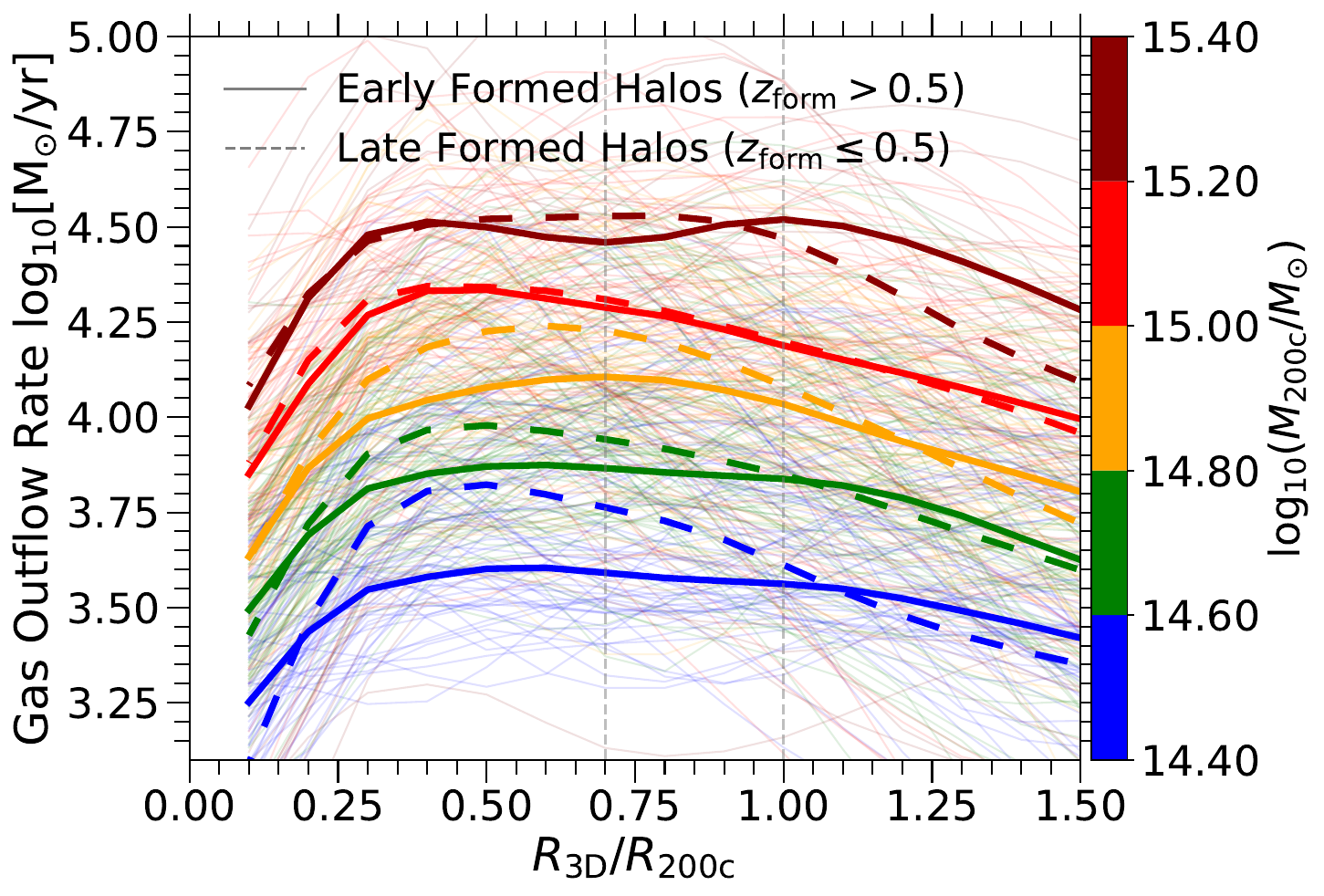}
    \includegraphics[width=0.45\textwidth]{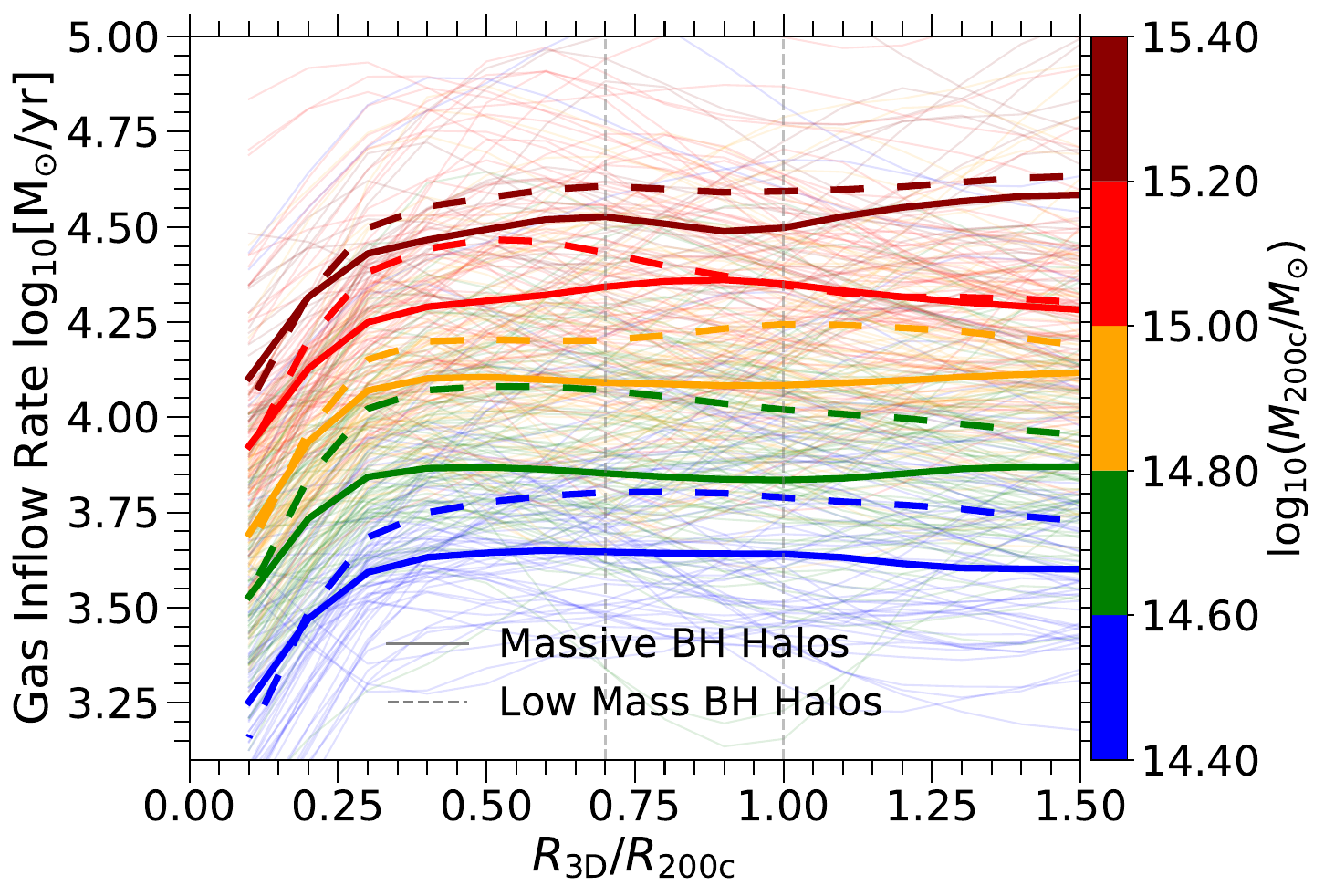}\hspace{5mm}
    \includegraphics[width=0.45\textwidth]{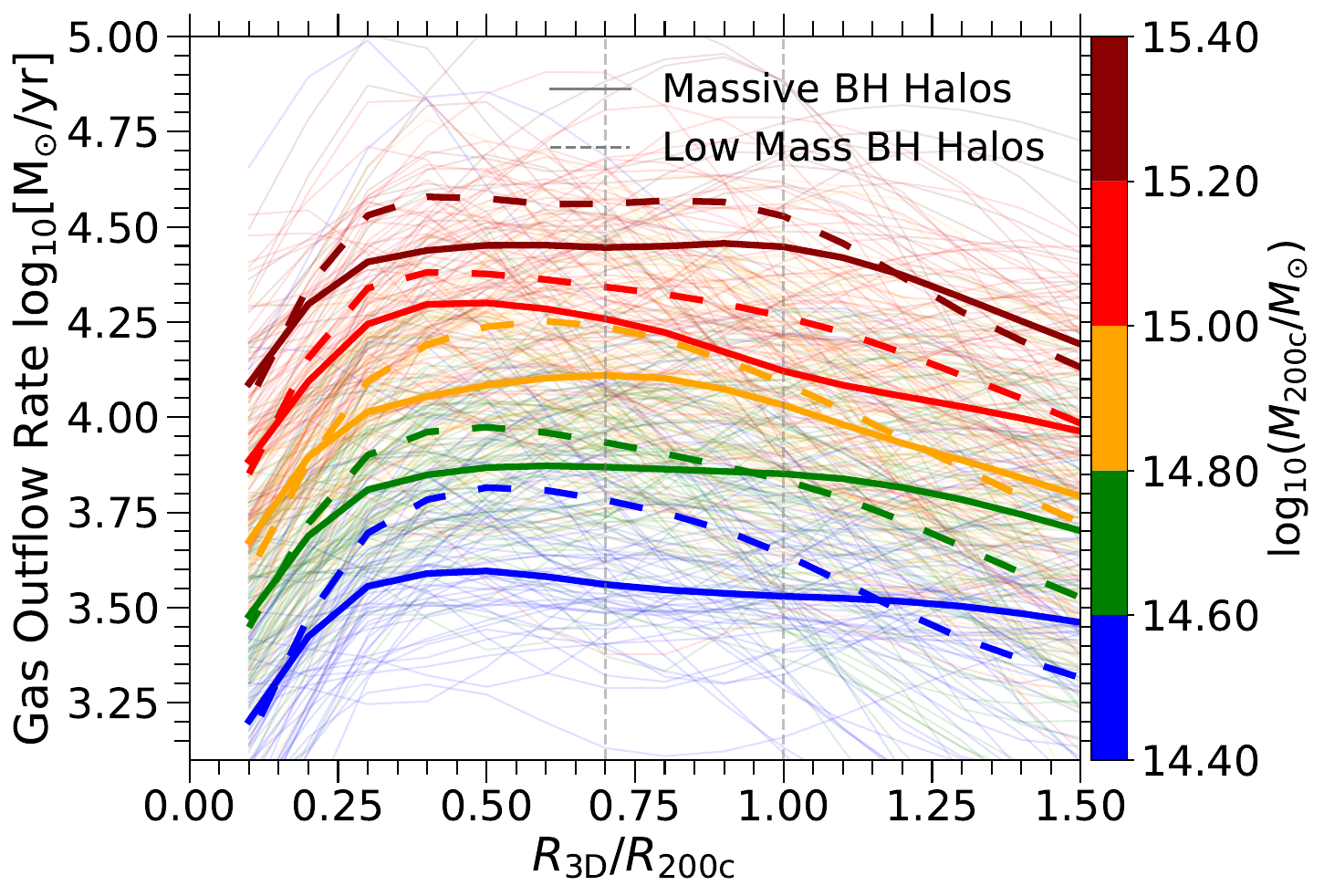}
    \includegraphics[width=0.45\textwidth]{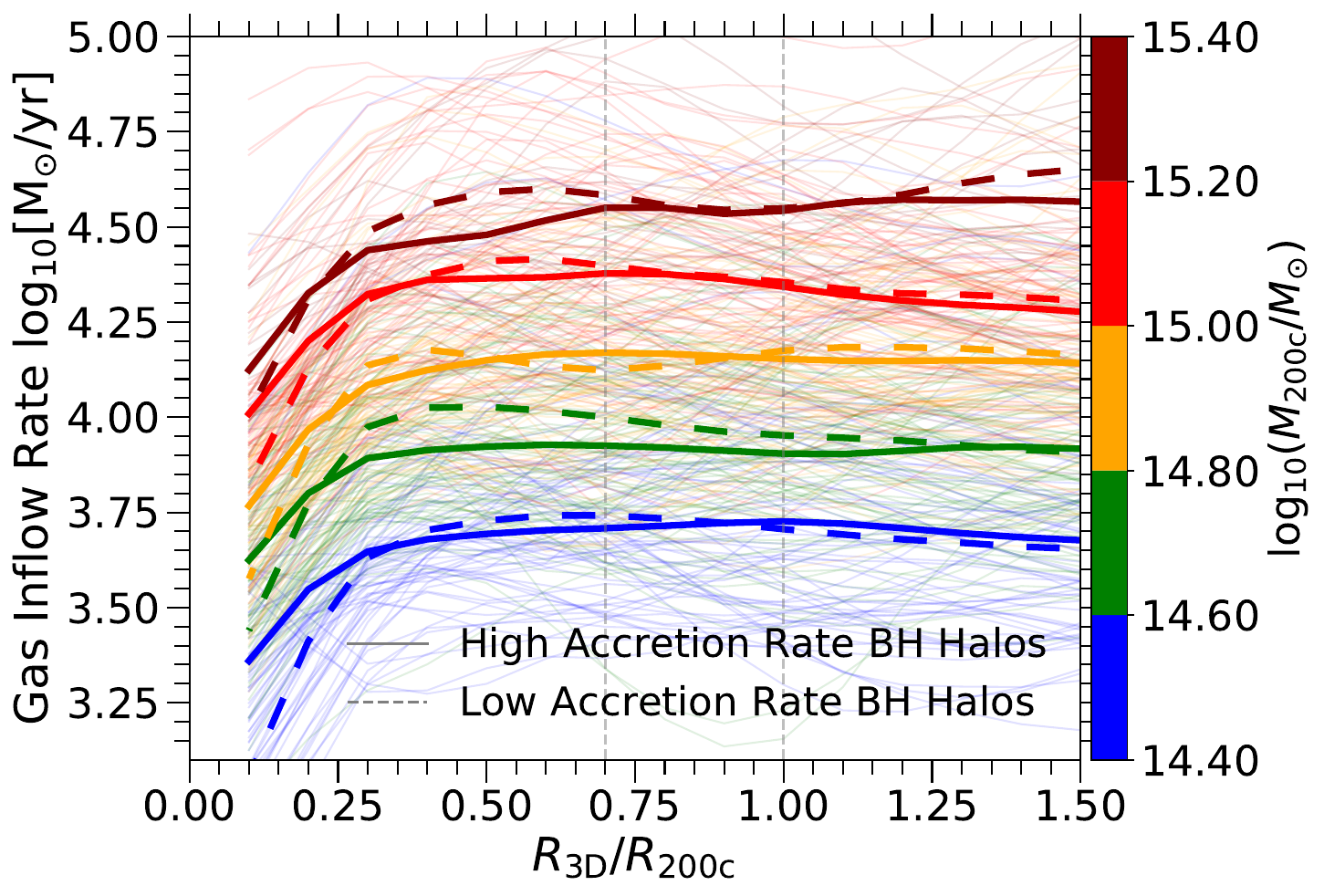}\hspace{5mm}
    \includegraphics[width=0.45\textwidth]{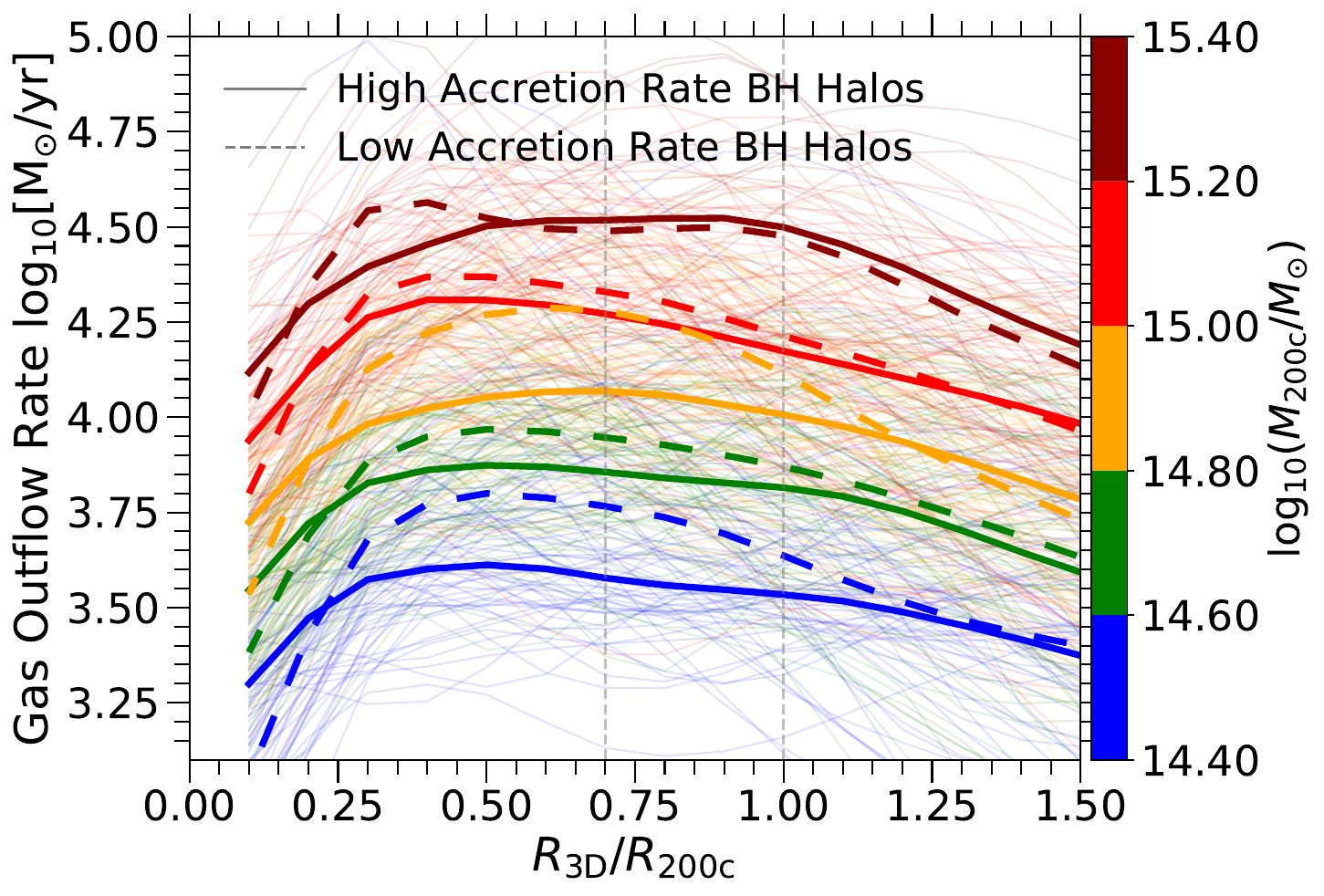}
    \caption{Radial profiles of gas inflow (left column) and gas outflow (right column) for halos classified based on their relaxedness (first row), formation time (second row), SMBH mass (third row), and SMBH accretion rate (fourth row). Each color represents a different halo mass bin. The thick solid lines illustrate the mean profile, while the thin lines represent individual clusters. The figure underscores the significant impact of halo properties on gas inflow and outflow.}
\label{Fig: profiles_gas_flow2}
\end{figure*}

\begin{figure*}
    \centering
    \includegraphics[width=0.46\textwidth]{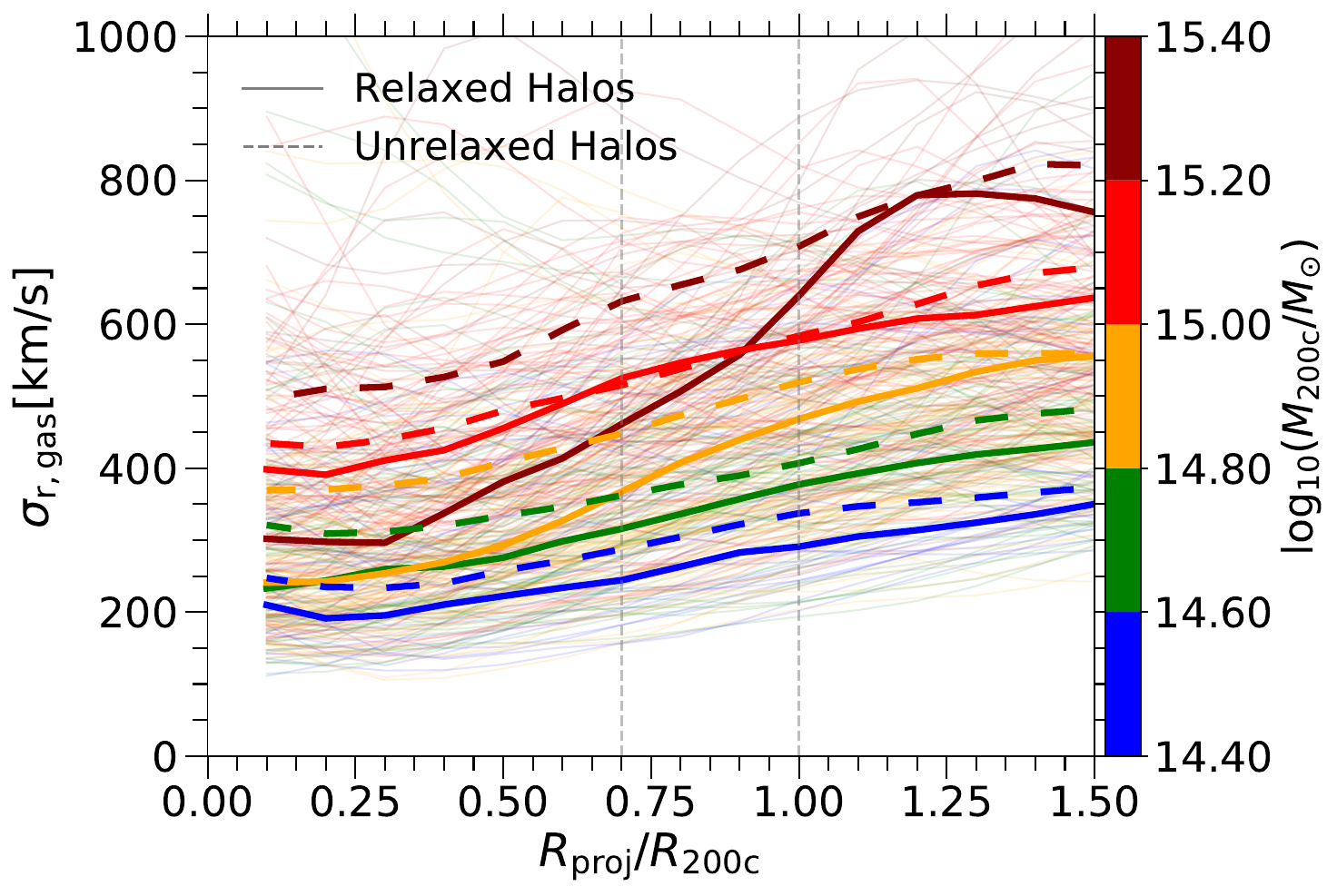}
    \includegraphics[width=0.46\textwidth]{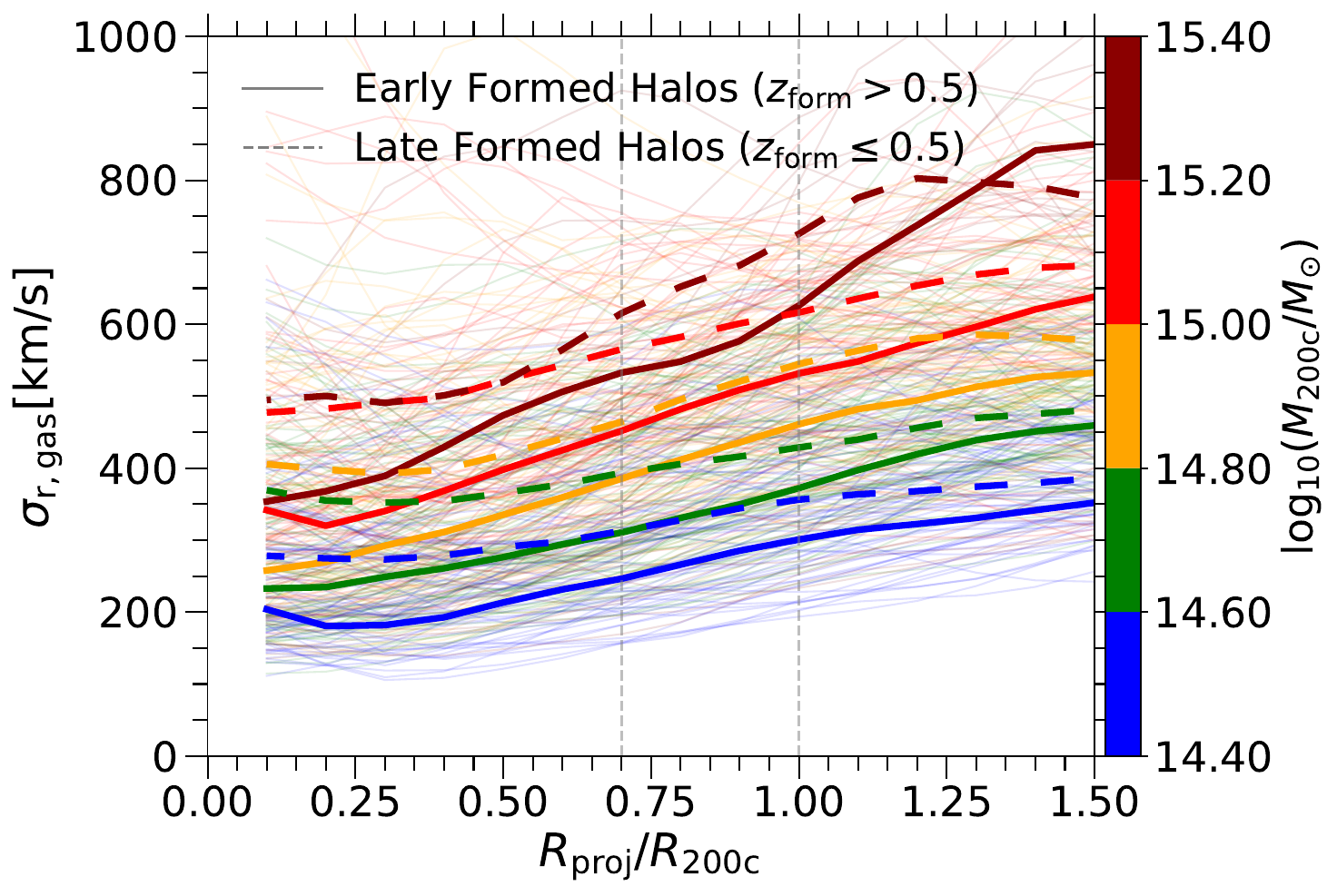}
    \includegraphics[width=0.46\textwidth]{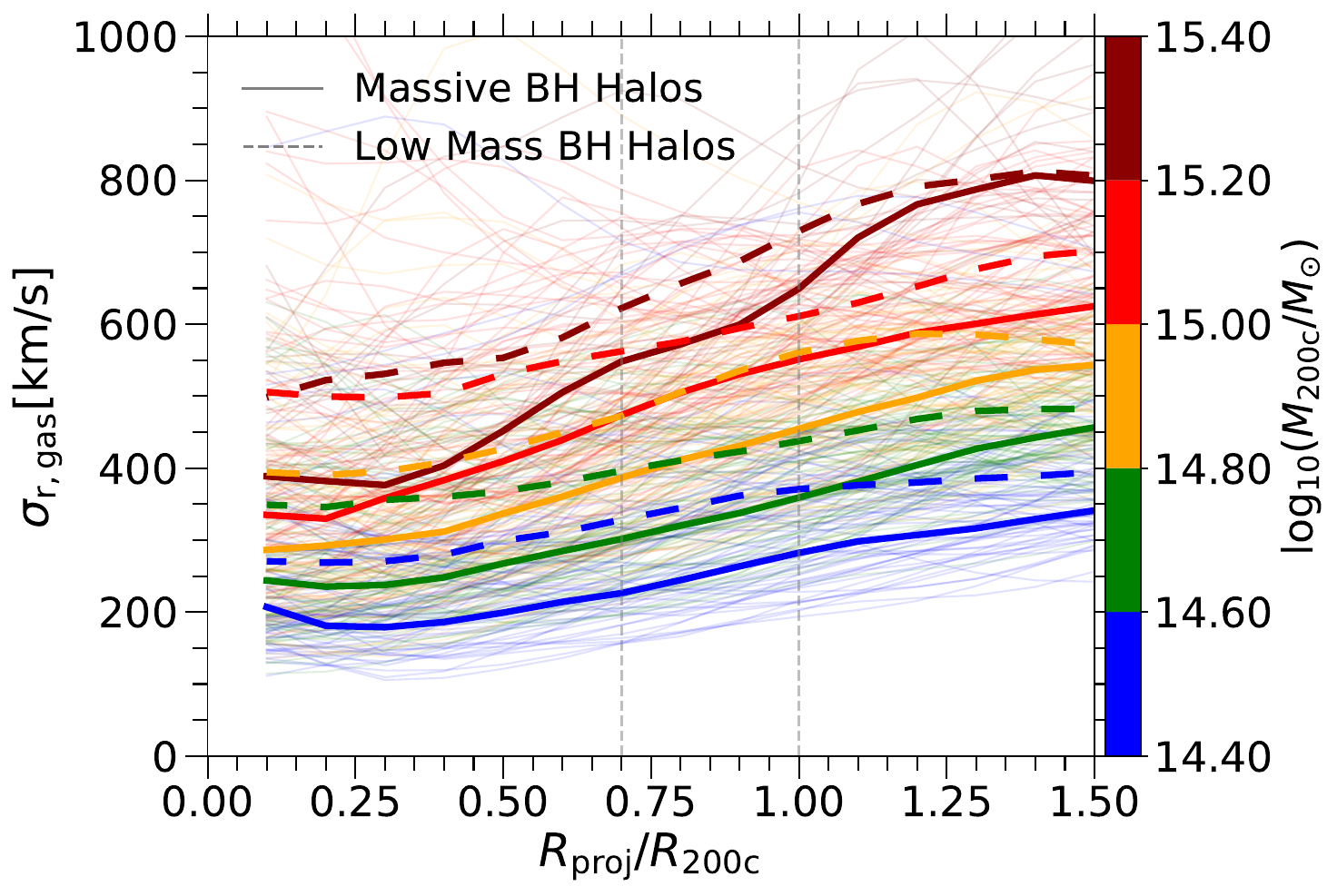}
    \includegraphics[width=0.46\textwidth]{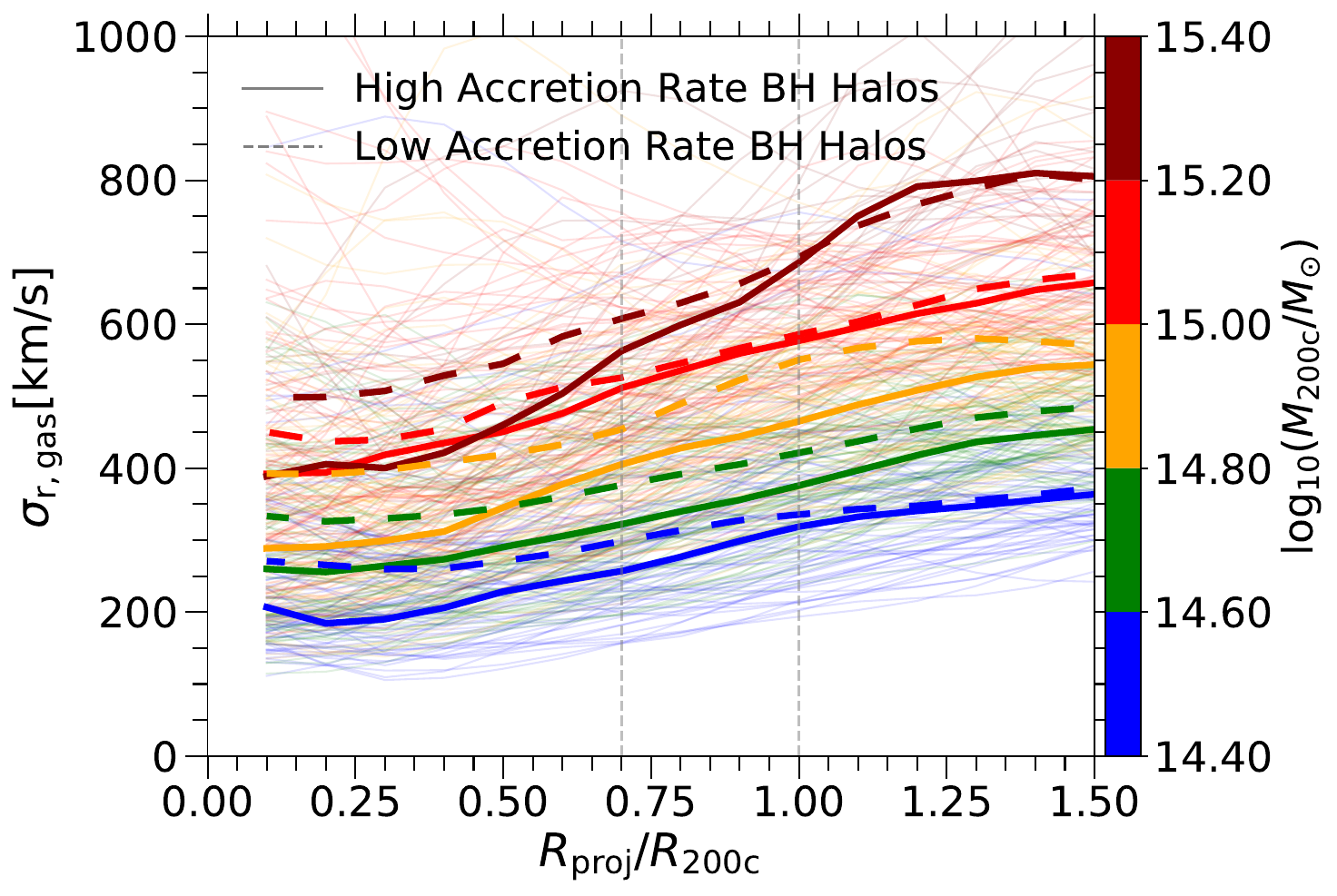}
    \caption{Radial velocity dispersion radial profiles, for halos classified based on their SMBH mass (top left), SMBH accretion rate (top right), relaxedness (bottom left), and formation time (bottom right). Each color corresponds to a different halo mass bin. The thick solid lines depict the mean profile, while the thin lines represent individual clusters. The figure underscores the significant impact of halo properties on the radial velocity dispersion.}
\label{Fig: profiles_vrad_disp}
\end{figure*}

In Fig. \ref{Fig: profiles_gas_flow2}, we present the radial profiles of gas inflow (left column) and gas outflow (right column) for halos, categorized based on their relaxedness (first row), formation time (second row), SMBH mass (third row), and SMBH accretion rate (fourth row). Unrelaxed halos display higher inflow rates and generally elevated outflow rates compared to their relaxed counterparts. Late-forming halos exhibit higher inflow rates, especially near the halo center, whereas early-forming halos show elevated outflow rates. Clusters hosting more massive SMBHs demonstrate lower inflow and outflow rates, while those with higher SMBH accretion rates exhibit increased inflow and outflow rates.

Furthermore, we examine the radial profiles of the 3D gas velocity dispersion for halos, categorized based on their distinct properties in Fig. \ref{Fig: profiles_vrad_disp}. Our findings indicate that unrelaxed and late-forming halos exhibit higher velocity dispersion than relaxed and early-forming halos. Similarly, clusters with more massive SMBHs and higher SMBH accretion rates show higher velocity dispersions. Typically, these trends are stronger near the cluster core and decrease with halocentric distance. These trends align well with the correlations between gas motion and halo properties, as discussed in Section \ref{subsec: impact of galaxy and halo properties}.

\section{Velocity structure function biases}
\label{app: vsf biases}
\begin{figure*}
    \centering
    \includegraphics[width=0.33\textwidth]{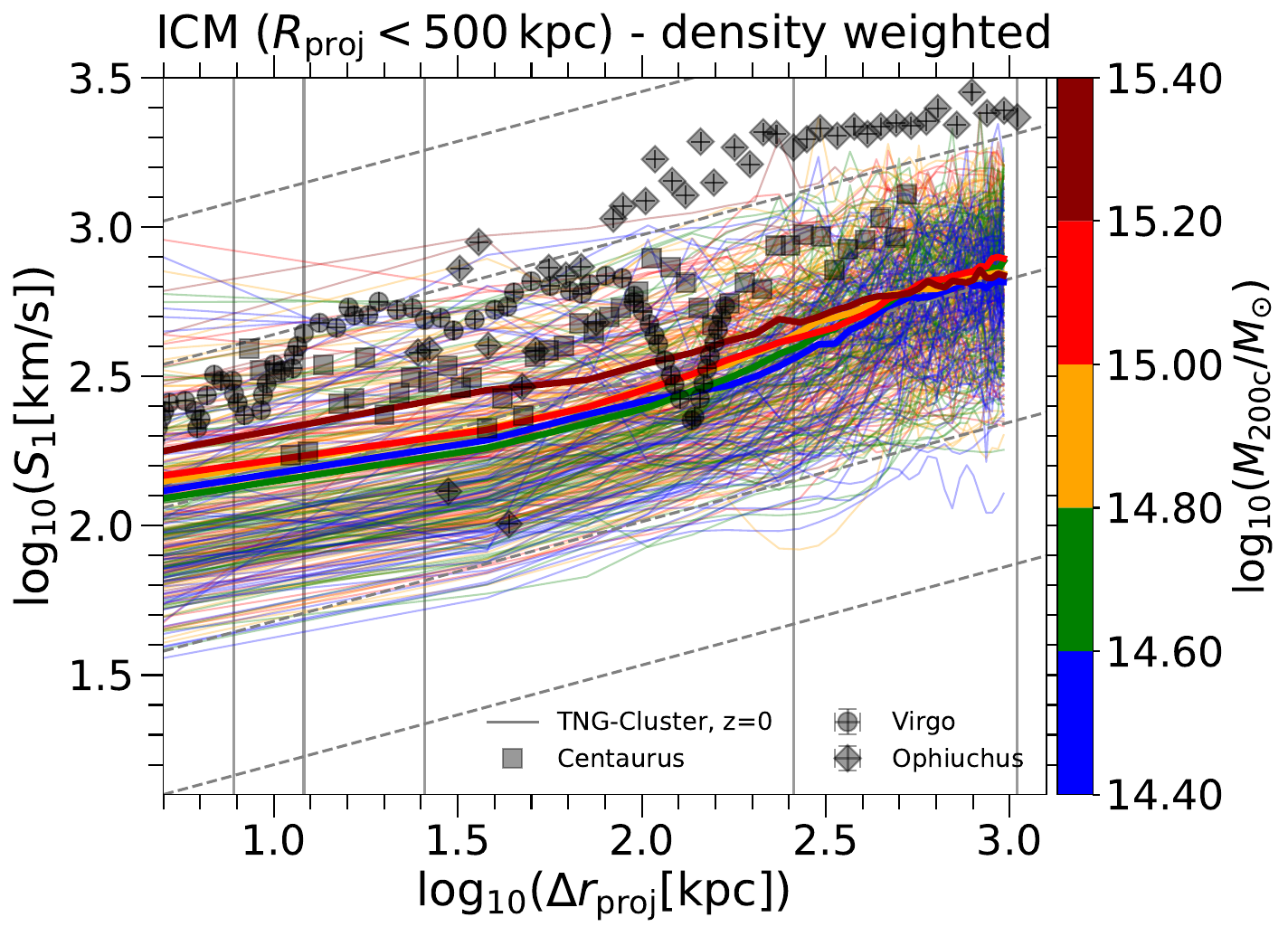}
    \includegraphics[width=0.33\textwidth]{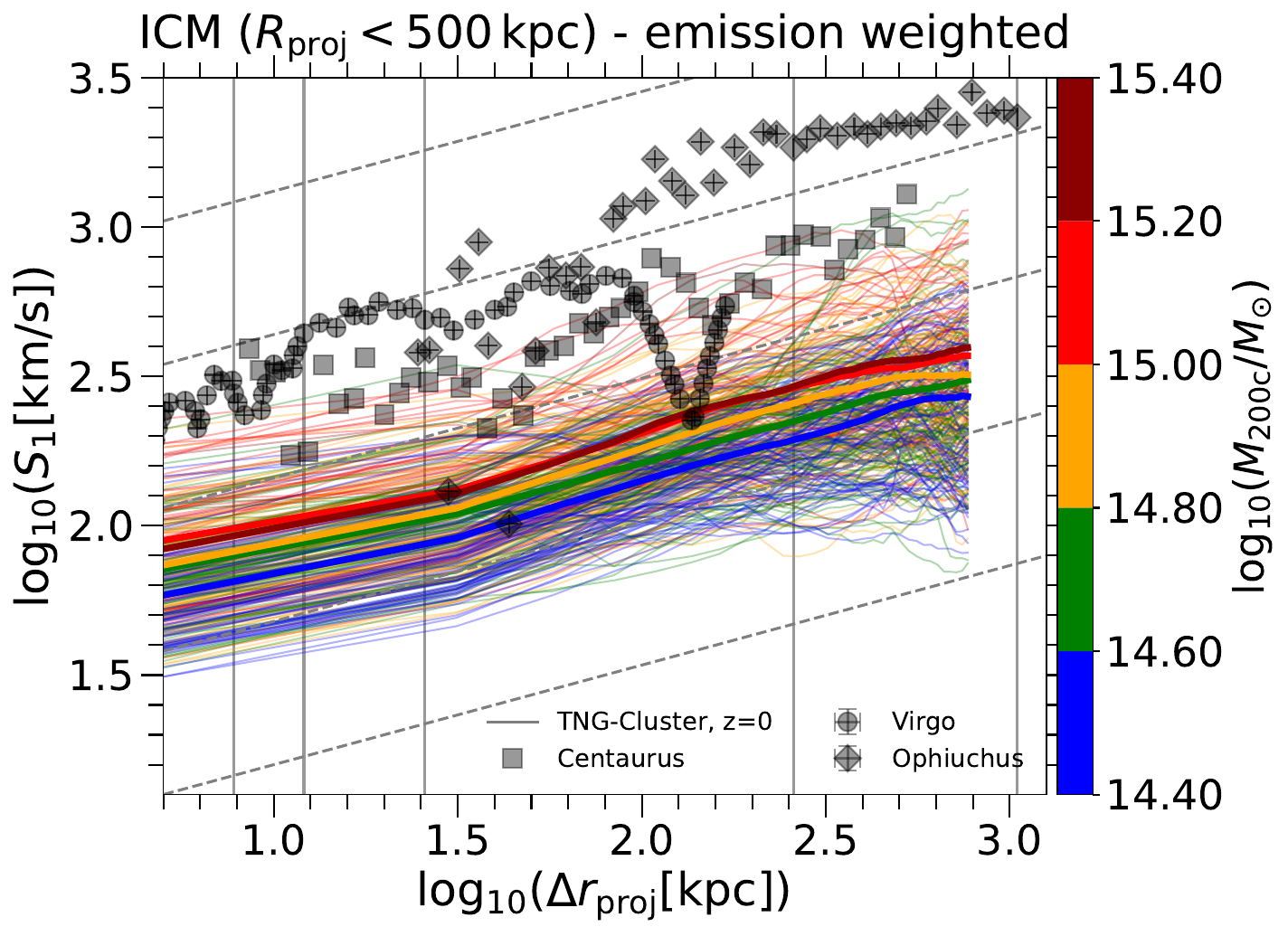}
    \includegraphics[width=0.33\textwidth]{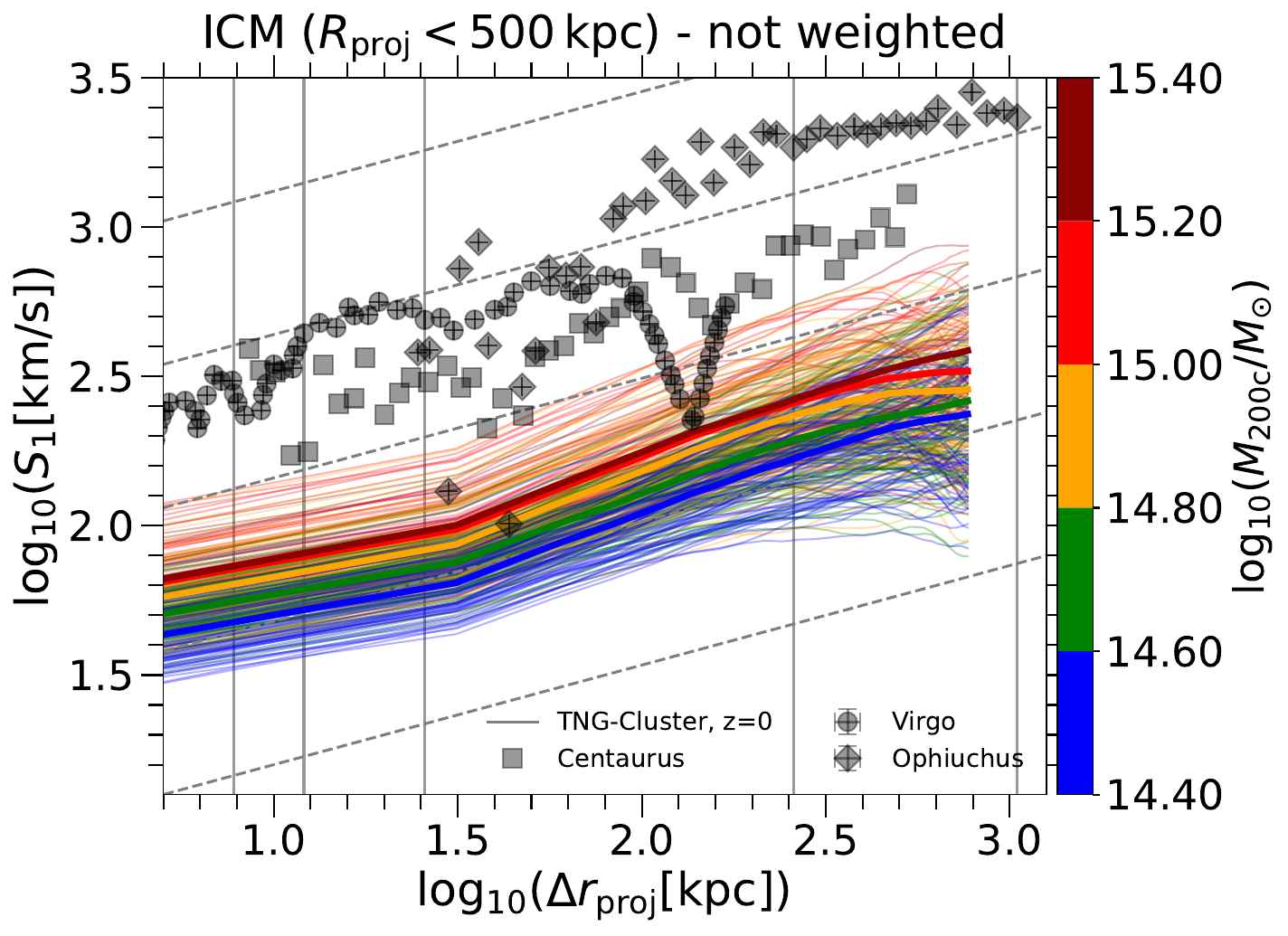}

    \caption{Velocity structure function as a function of projected separation distance. The three panels correspond to the VSF with different weighting schemes: density-weighted (left), bolometric luminosity weighted (middle), and no-weighting (right). The colors represent different halo mass bins. The thick solid lines depict the mean VSF, while the thin lines represent individual clusters. The figure underscores the significant impact of the weighting scheme on the VSF.}
\label{Fig: vsf_app}
\end{figure*}
As discussed in Section \ref{subsec: vsf_obs}, there are many biases in the measurement of the VSF from observational data. In this appendix, we explore the impact of these biases on the VSF measurements from TNG-Cluster, focusing on the large-scale VSF, which is more sensitive to these biases. In Section \ref{subsec: velocity structure function}, we demonstrated that the VSF decreases with halo-centric distance at a fixed separation scale. Therefore, the weighting of different zones of the halo, from cores to outskirts, is critical. One option is to use no weighting, i.e. $w_i = w_j = 1$ in Eq. \ref{eq: VSF}. However, this no-weighting scheme effectively gives more weight to the outskirts due to their significantly larger volume compared to the cluster cores. In Fig. \ref{Fig: vsf_app}, we explored three binning strategies, including those based on density (left panel), emission (middle panel), and no weighting (right panel). We observed that when we weight the bins based on their density, the amplitude of the VSF increases significantly due to the higher weight of the core regions. This alone demonstrates the complexity of the VSF measurement and its sensitivity to the choice of binning and weighting. We note that when weighting our VSF based on density, we achieve significantly better agreement with the measurements of the VSF by \cite{Gatuzz2023Measuring}. Finally, we emphasize that due to the large uncertainties in observational measurements, the complexities of measuring velocities in observations, and the VSF's sensitivity to binning and weighting choices, a careful mock analysis is required to make a fair comparison between simulations and observations—a study we will defer to future work.

\end{appendix}
\end{document}